\documentclass[preprint,12pt]{elsarticle}
\usepackage{graphics,graphicx}
\usepackage{subfigure}
\usepackage{stmaryrd}
\usepackage{wrapfig}
\usepackage{amsfonts,amsmath,amssymb,amsthm}
\usepackage{lineno}
\newtheorem{definition}{Definition}
\newtheorem{theorem}{Theorem}

\newtheorem{example}{Example}
\numberwithin{equation}{section}

\journal{   }
\begin{document}

\begin{frontmatter}
\title{Degenerations of NURBS curves while all of weights approaching infinity}
\author{Yue Zhang}
\ead{zhangyuee@mail.dlut.edu.cn}
\author{Chun-Gang Zhu\corref{cor1}}
\cortext[cor1]{Corresponding author.} \ead{cgzhu@dlut.edu.cn}
\address{School of Mathematical Sciences, Dalian University of Technology, Dalian 116024, China.}

\begin{abstract}
NURBS curve is widely used in Computer Aided Design and Computer Aided Geometric Design. When a single weight approaches infinity, the limit of a NURBS curve tends to the corresponding control point. In this paper, a kind of control structure of a NURBS curve, called regular control curve, is defined. We prove that the limit of the NURBS curve is exactly its regular control curve when all of weights approach infinity, where each weight is multiplied by a certain one-parameter function tending to infinity, different for each control point. Moreover, some representative examples are presented to show this property and indicate its application for shape deformation.
\end{abstract}

\begin{keyword}
NURBS curve \sep weights \sep   regular control curve \sep toric degenerations \sep shape deformation
\MSC[2010] 65D17 \sep  68U07
\end{keyword}

\end{frontmatter}


\section{Introduction}

Non-Uniform Rational B-Spline (NURBS) method, as a  popular  curve and surface modeling technology, is widely used in Computer Aided Geometric Design (CAGD), Computer Aided Design (CAD), and Geometric Modeling. NURBS method is the generalization of B\'{e}zier method,  B-spline method, and rational B\'{e}zier method. It converts the curve and surface fitting tool to a unified representation. The theory of NURBS method can be referred to  books  and literatures by Piegl and Tiller~\cite{Piegl0,Piegl2,Piegl3} and Farin et al.~\cite{Farin0,Farin1}.

\begin{figure}[h!]
\centering
\includegraphics[height=3.5cm,width=3.5cm]{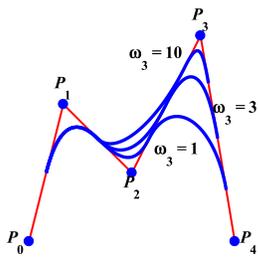}
\caption{The effects of increasing a single weight $\omega_3$.}
\label{Fig1}
\end{figure}

The shape modification of existing objects plays an important role in geometric design
systems. Shape modification of NURBS curves and surfaces can be achieved by means of knot vectors,
control points and weights. It can be easily seen, by changing not only the positions of control
points, but also the values of weights, the shape of curve and surface can be modified.
Piegl and Tiller~\cite{Piegl0,Piegl2} explained the geometric meaning of the NURBS curve when a single weight approaches infinity: the curve tends to the corresponding control point. This property is crucial for interactive shape design. Fig.~\ref{Fig1} shows the effects of increasing a single weight $\omega_3$ of a quadric NURBS curve. Based on this property, Piegl and Tiller~\cite{Piegl2} presented a method to finely tune the shape of a NURBS curve when a weight is allowed to be modified.  Motivated by Piegl's work, scholars investigated how the shape of a NURBS curve changes when the weights of control points are modified. Au and Yuen~\cite{Au}, and S\'{a}nchez-Reyes~\cite{Sanchez-Reyes} introduced approaches to  adjust the shape of  NURBS curves by modifying the weights and location of the control points  simultaneously. Juh\'asz~\cite{Imre} presented a method to adjust the shape of a NURBS curve by modifying some weights. Not only did he study modifying the location of an arbitrary point of a NURBS curve, but also considered the tangent direction. Zhang et al.~\cite{Zhang} studied the effects of the NURBS curve by modifying two weights, he also found that single position and tangent constraints could be realized by a modification of three weights.
The methods above  were presented to finely tune the shape of NURBS curves when one or some weights are allowed to be modified, and they are all based on well-known geometric property of a single weight of NURBS curve. However, seldom work focuses on identifying the meaning  of a NURBS curve when all  weights tend to vary (even to infinity).


Toric geometry is systematically presented and developed gradually in the 1970s. Sturmfels~\cite{Sturmfels} introduced the theory of toric varieties, toric ideals and Gr\"{o}bner bases in detail. In 2002, Krasauskas~\cite{Krasauskas} presented a new kind of multi-sided surface, named toric patches, which is based upon toric variety from algebraic geometry and toric ideals from combinatorics.  The toric patches include the classical rational B\'{e}zier curves and patches, and some multi-sided patches as special cases.
As we know that control points and control nets control the shape of B\'{e}zier patches. But the geometric meaning of control points plus the edges, triangle,  quadrangle formed by the control points is still unknown. Carl de Boor and Ron Goldman proposed a question in geometric modeling: {\em What is the significance for modeling of such control structures (control points plus edges)~\cite{Garcia}?}
It's worth noting that control points plus edges is not necessarily control polygon.  In 2011, Garc\'{\i}a-Puente, Sottile and Zhu~\cite{Garcia} explained the geometric meaning of control structures (surfaces) of toric patches when all  weights tend to infinity by using the theory of toric varieties, toric ideals and toric degenerations, called the toric degenerations of B\'{e}zier patches.  Furthermore, Zhu~\cite{Zhu} presented the toric degenerations of toric varieties and toric ideals induced by regular decomposition. Zhu and Zhao~\cite{Zhao,Zhao2015} also gave the geometric conditions on the control polygon and control points set which guarantees the injectivity (one-to-one property) of rational B\'{e}zier curve and surfaces.

Our purpose of this paper is to explain the geometric meaning of the limit of NURBS curves  when all of weights tend to infinity. We define the regular control curve of the NURBS curve and prove that the limit of NURBS is exactly its regular control curve when all of weights approach infinity.
This property generalizes the geometric meaning of a single weight of NURBS curve and explains the geometric meaning of weights of NURBS curve. Our result provides the possible applications for the injectivity checking of NURBS curve which plays an important role in image warping and morphing, 3D deformation and volume morphing. Furthermore, we also provide an idea for shape modification and deformation of NURBS curve by altering many weights.

The paper is organized as follows. In Section~\ref{relatedwork}, we recall the definition of rational B\'{e}zier curve, toric degenerations of rational B\'{e}zier curves, the definition of NURBS curve and knot insertion algorithm. In Section~\ref{mywork},  we define the \textquotedblleft regular control curve\textquotedblright, a control structure of a NURBS curve by the regular decomposition, and prove that which is the limit of the NURBS curve. Since this property is proved based on toric degeneration of B\'{e}zier curve, we say it is the toric degeneration of NURBS curve. Moreover, we observe that if a curve is the limit of a NURBS curve for the sequence of weights, then this curve must be a regular control curve induced by some regular decomposition. Some representative examples are illustrated to show the degeneration property of NURBS curves and point out its application for shape deformation in Section~\ref{example}.  Finally, Section~\ref{conclusion} concludes the whole paper.

\section{Preliminaries}\label{relatedwork}
\subsection{Rational B\'{e}zier curves}

\begin{definition}[\cite{Farin0,Farin1}]
\label{def1}
\em{
For given control points $\textbf{b}_i\in \mathbb{R}^d$ $(d=2,3)$ and weights $\omega_i$, $i=0,1,\cdots,m$, a {\em rational B\'{e}zier curve} of degree $m$ is defined by
\begin{eqnarray*}\label{eq1.1}
\textbf{F}(v)=\frac{\sum_{i=0}^m\omega_i \textbf{b}_iB_i^m(v)}{\sum_{i=0}^m\omega_iB_i^m(v)},v\in [0,1],
\end{eqnarray*}
where $B_i^m(v)=\binom{m}{i}(1-v)^{m-i}v^i$ are  Bernstein  basis functions and the {\em control polygon} of the curve is the union of segments $\overline{\textbf{b}_{0}\textbf{b}_{1}},\cdots,\overline{\textbf{b}_{m-1}\textbf{b}_{m}}$.
}
\end{definition}

In 2002, a new method to construct the multi-sided surface patches,  toric patches,  was presented by Krasauskas~\cite{Krasauskas}. Following the Krasauskas' toric patches,  the toric B\'{e}zier curve can be defined. Given a set of finite lattice points $\mathcal{A}=\{a_0,a_1,\cdots,a_m\}\subset \mathbb{Z}$,
let  $conv(\mathcal{A})$ be the convex hull of  lattice points of the set $\mathcal{A}$, and
the interval $\Delta_\mathcal{A}=conv(\mathcal{A})$ can be defined by $\{x\in \mathbb{R} \mid 0\leq l_0(x), 0\leq l_1(x)\}$ where $l_0(x)=x-a_0,l_1(x)=a_m-x$ and we assume that $a_i<a_{i+1}, i=0,1,\cdots,m-1$. The following definition we refer to~\cite{Garcia,Zhao}.

\begin{definition}[\cite{Garcia}]
\label{def2}
\em{
Given a set of finite  lattice points  $\mathcal{A}=\{a_0,a_1, \cdots,a_m\}\subset \mathbb{Z}$,  control points $\mathcal{B}=\{\textbf{b}_{a_i}\mid a_i\in \mathcal{A}\}\subset \mathbb{R}^d$, $d=2,3$, and  weights $\omega=\{\omega_{a_i}>0 \mid a_i \in \mathcal{A}\}$, the parametric curve,
\begin{eqnarray*}\label{eq1.4}
\textbf{F}_{\mathcal{A},\omega ,\mathcal{B}}(x)=\frac{\sum_{i=0}^{m}\omega _{a_i} \textbf{b}_{a_i} \beta_{a_i,\mathcal{A}}(x)}{\sum_{i=0}^{m}\omega _{a_i}  \beta_{a_i,\mathcal{A}}(x)} , x \in \Delta_\mathcal{A},
\end{eqnarray*}
is called a {\em toric B\'{e}zier curve}, where the functions $\beta_{a_i,\mathcal{A}}(x)=c_{a_i}l_0(x)^{l_0(a_i)}l_1(x)^{l_1(a_i)}$ are called  toric B\'{e}zier basis functions, the coefficient  $c_{a_i}>0$ and the union of segments $\overline{\textbf{b}_{a_0}\textbf{b}_{a_1}},\cdots,\overline{\textbf{b}_{a_{m-1}}\textbf{b}_{a_m}}$ is called the {\em control polygon} of the curve.
}
\end{definition}

Write $\textbf{F}_{\mathcal{A},\omega ,\mathcal{B}}$ for the image $\textbf{F}_{\mathcal{A},\omega ,\mathcal{B}}(\Delta_{\mathcal{A}})$.
Note that when we set $\mathcal{A}=\{0,1,\cdots,m\}$,  $\Delta_\mathcal{A}=[0,m]$, $l_0(x)=x$, $l_1(x)=m-x$ is the boundary equations of $\Delta_\mathcal{A}$, then toric B\'{e}zier basis is $\beta_{i,\mathcal{A}}(x)=c_il_0(x)^{l_0(i)}l_1(x)^{l_1(i)}=c_ix^{i}(m-x)^{m-i}$, for any $i\in \mathcal{A}$. We assume $x=nv$ and $c_i=\binom{m}{i}m^{-m}$, then $\beta_{i,\mathcal{A}}(x)$ becomes the classical Bernstein basis $B_i^m(v)$ and the toric B\'{e}zier curve can be parameter transformed as the rational B\'{e}zier curve.

Since parameter transformation of curve does not affect the shape of curve, the definition of toric B\'{e}zier curve is equivalent to the definition of rational B\'{e}zier curve. Moreover, because of the concept of the set of finite lattice points $\mathcal{A}$ is introduced in toric B\'{e}zier curve, the weights and control points of curve can been indexed by the lattice points of  $\mathcal{A}$. For convenience of analysis, we will use Definition~\ref{def2} to represent a rational B\'{e}zier curve.

\subsection{Toric degenerations of rational B\'{e}zier curves}\label{2.2}

In~\cite{Garcia}, for explaining the geometric meaning of control surfaces of toric patches, Garc\'{\i}a-Puente, Sottile and Zhu defined the regular decomposition of a set of finite lattice points. And following this work, Zhu and Zhao \cite{Zhao} proposed the definition of regular decomposition of $\mathcal{A}$ for dealing with the self-intersections of rational B\'{e}zier curves. In this section, we first recall some notations from \cite{Garcia} and \cite{Zhao}.

Let $\mathcal{A}\subset \mathbb{Z}$ be a set of finite lattice points and suppose that $\lambda:\mathcal{A}\rightarrow \mathbb{R}$ is a function. We use a {\em lifting function} $\lambda$ to lift all the lattice points of $\mathcal{A}$ into $\mathbb{R}^2$. Let $P_\lambda$ be the convex hull of the lifted points of the lattice points,
\begin{eqnarray*}\label{eq1.5}
P_\lambda =conv\{\left(a_i,\lambda(a_i)\right)\mid a_i\in \mathcal{A}\}\subset \mathbb{R}^2.
\end{eqnarray*}
Each face of $P_\lambda$ has an outward pointing normal vector, and its{ \em upper edges} are those whose normal vector has positive last coordinate.  The union of the upper edges is the upper hull of $P_\lambda$. If we project each of these upper edges back to $\mathbb{R}$, then we get a set of closed intervals of $\Delta_\mathcal{A}$ and the union of them covers $\Delta_\mathcal{A}$. These closed intervals  induce a  {\em regular domain decomposition} $\mathcal{T}_\lambda$ of $\Delta_\mathcal{A}$. All the lattice points of $\mathcal{A}$ which belong to the same closed interval and whose lifted points lie on a common \textquotedblleft upper edge\textquotedblright, we get subset $s_j$ of $\mathcal{A}$. Then the union of these subsets of $\mathcal{A}$ is called a {\em regular decomposition} $S_\lambda$ of $\mathcal{A}$ induced by  $\lambda$.

Suppose that a lifting function $\lambda$  induces a regular decomposition $S_\lambda$ of $\mathcal{A}$. If $s_j \in S_\lambda$  is a subset, then the weights $\omega{\mid}_{ s_j}=\{\omega_{a_i} \mid a_i \in s_j\}$ and the control points $\mathcal{B}{\mid}_{ s_j}= \{\textbf{b}_{a_i}\mid a_i \in s_j\}$ indexed by elements of $s_j$  can construct a rational B\'{e}zier curve by Definition~\ref{def2}, denoted by $\textbf{F}_{s_j,\omega{\mid}_{ s_j},\mathcal{B}{\mid}_{ s_j}}$.  The {\em regular control curve} of the rational B\'{e}zier curve $\textbf{F}_{\mathcal{A},\omega ,\mathcal{B}}$ induced by the regular decomposition $S_\lambda$ is the union of those curves~\cite{Garcia},
\begin{eqnarray*}\label{eq1.6}
\textbf{{F}}_{\mathcal{A},\omega ,\mathcal{B}}(S_\lambda):=\bigcup_{s_j\in S_\lambda }\textbf{{F}}_{s_j,\omega{\mid}_{ s_j}, \mathcal{B} {\mid}_{ s_j} }.
\end{eqnarray*}

\begin{definition}[\cite{Garcia}]
\label{def3}
\em{
Given a set of finite integers $\mathcal{A}=\{a_0,\cdots,a_m\}\subset \mathbb{Z}$, control points $\mathcal{B}=\{\textbf{b}_{a_i} \mid a_i \in \mathcal{A} \}\subset \mathbb{R}^d$, $d=2,3$, and weights $\omega=\{\omega_{a_i}>0 \mid a_i \in \mathcal{A}\}$, the curve,
\begin{eqnarray*}\label{eq1.7}
\textbf{F}_{\mathcal{A},\omega_\lambda(t) ,\mathcal{B}}(x;t):=\frac{\sum_{i=0}^{m}t^{\lambda(a_i)}\omega_{a_i} \textbf{b}_{a_i} \beta _{a_i,\mathcal{A}}(x)}{\sum_{i=0}^{m}t^{\lambda(a_i)}\omega _{a_i} \beta _{a_i,\mathcal{A}}(x)},x \in \Delta_\mathcal{A}
\end{eqnarray*}
is called a rational B\'{e}zier curve parameterized by $t$, where $\omega_\lambda(t):=\{t^{\lambda(a_i)}\omega_{a_i} \mid a_i\in \mathcal{A}\}$.
}
\end{definition}

\begin{theorem}[\cite{Garcia}]
\label{th1.1}
\begin{eqnarray*}
\lim_{t\rightarrow \infty }\textbf{F}_{\mathcal{A},\omega _\lambda (t),\mathcal{B}}=\textbf{F}_{\mathcal{A},\omega ,\mathcal{B}}(S_\lambda)
\end{eqnarray*}
\end{theorem}

By Theorem~\ref{th1.1}, we know that if the control points $\mathcal{B}$ are fixed but the parameter $t \rightarrow \infty$, the regular control curve induced by the lifting function $\lambda$ is exactly the limit of rational B\'{e}zier curve $F_{\mathcal{A},\omega_\lambda(t) ,\mathcal{B}}$. Garc\'{\i}a-Puente et al.~\cite{Garcia} proved  this property and which is called  {\em toric degeneration of a rational  B\'{e}zier curve}. Furthermore, they also proved the following result, which is converse to Theorem~\ref{th1.1}.

\begin{theorem}[\cite{Garcia}]
\label{th1.2}
If $\textbf{F}\subset \mathbb{R}^3$  is a set for which there is a sequence $\omega^{(1)},\omega^{(2)},\cdots$ of weights so that
\begin{eqnarray*}
\lim_{\tau\rightarrow \infty }\textbf{F}_{\mathcal{A},\omega^{(\tau)},\mathcal{B}}=\textbf{F},
\end{eqnarray*}
then there exist a lifting function $\lambda:\mathcal{A} \rightarrow \mathbb{R}$ and weights $\omega$ such that $\textbf{F}=\textbf{F}_{\mathcal{A},\omega,\mathcal{B}}(S_\lambda)$ is a regular control curve.
\end{theorem}

\begin{example}
\label{example2.1}
\em{
Let $\mathcal{A}=\{0,1,2,3,4\}$ and $\Delta_\mathcal{A}=[0,4]$. For given control points $\mathcal{B}=\{\textbf{b}_0,\textbf{b}_1,\textbf{b}_2, \textbf{b}_3,\textbf{b}_4\}$ and weights $\omega=\{3,4,2,1.5,1\}$, the quartic rational B\'{e}zier curve $\textbf{F}_{\mathcal{A},\omega ,\mathcal{B}}$ is shown in Fig.~$\ref{Fig2a}$.
Suppose a lifting function $\lambda_1$ take the values  $\{2,3,4,3,2\}$ at the  lattice points of $\mathcal{A}$ (see Fig.~$\ref{Fig3a}$), which induces a regular domain decomposition of $\Delta_\mathcal{A}$,  $\mathcal{T}_{\lambda_1}=\{[0,2],[2,4]\}$, and a regular decomposition of $\mathcal{A}$, $S_{\lambda_1}=\{\{0,1,2\},\{2,3,4\}\}$.
Another lifting function $\lambda_2$ take the values $\{2,3,4,2,3\}$ at the  lattice points of $\mathcal{A}$ (see Fig.~$\ref{Fig3b}$), which induces the same regular domain decomposition with $\lambda_1$ and a different regular decomposition with $\lambda_1$, that is $S_{\lambda_2}=\{\{0,1,2\},\{2,4\}\}$,
since the lifted point $(3,\lambda_2(3))$ does not lie on any upper edge of the $P_{\lambda_2}$ (see Fig.~$\ref{Fig3b}$).

For the subset $s_1=\{0,1,2\} \in S_{\lambda_2}$, a rational quadratic B\'{e}zier curve $\textbf{{F}}_{s_1,\omega{\mid}_{ s_1},\mathcal{B}{\mid}_{ s_1}}$ can be constructed by the corresponding control points $\mathcal{B}{\mid}_{ s_1}=\{\textbf{b}_0,\textbf{b}_1,\textbf{b}_2\}$ and weights $\omega{\mid}_{ s_1}=\{3,4,2\}$, and parametric domain of the curve is $[0,2]$. For the subset $s_2=\{2,4\}$ of $S_{\lambda_2}$, a linear B\'{e}zier curve $\textbf{{F}}_{s_2,\omega{\mid}_{ s_2},\mathcal{B}{\mid}_{ s_2}}$ can be constructed by the corresponding control points $\mathcal{B}{\mid}_{ s_2}=\{\textbf{b}_2,\textbf{b}_4\}$ and weights $\omega{\mid}_{ s_2}=\{2,1\}$,  and parametric domain of the curve is $[2,4]$. Then the union of those two curves $\textbf{{F}}_{s_1,\omega{\mid}_{s_1},\mathcal{B}{\mid}_{s_1}}\cup \textbf{{F}}_{s_2,\omega{\mid}_ {s_2},\mathcal{B}{\mid}_{ s_2}}$  is the regular control curve of  $\textbf{F}_{\mathcal{A},\omega ,\mathcal{B}}$ (as Fig.~$\ref{Fig2b}$ shown). By Theorem $\ref{th1.1}$, the limit of a rational B\'{e}zier curve $\textbf{F}_{\mathcal{A},\omega _\lambda (t),\mathcal{B}}$ is the regular control curve shown in Fig.~$\ref{Fig2b}$ while the parameter $t \rightarrow \infty$, where $\omega_{\lambda_2}(t)=\{3t^2,4t^3,2t^4,1.5t^2,t^3\}$.
}
\begin{figure}[h!]
\centering
\subfigure[Rational B\'{e}zier curve] {\includegraphics[height=3cm,width=3cm]{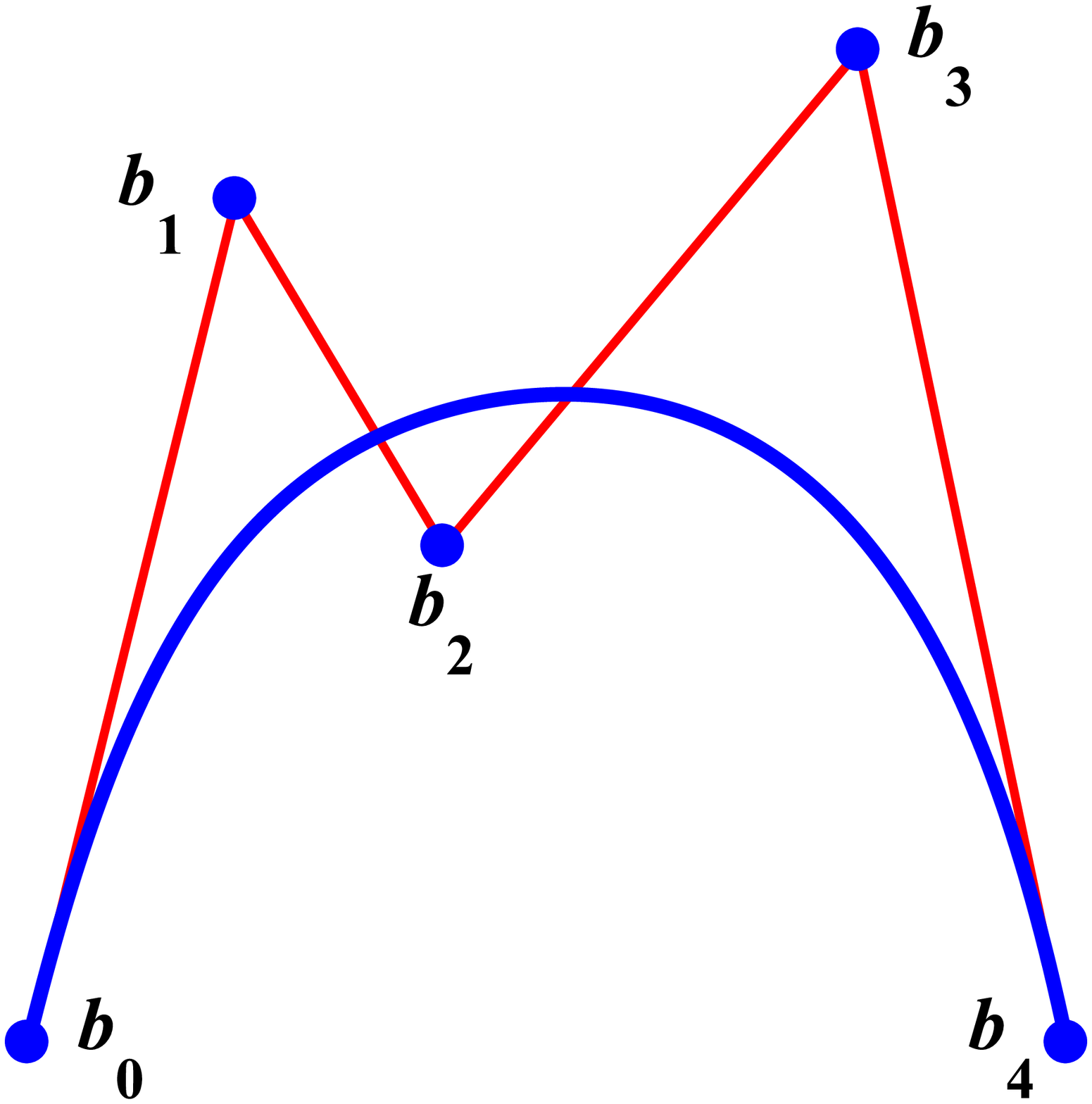}\label{Fig2a}}\hspace{5ex}
\subfigure[Regular control curve] {\includegraphics[height=3cm,width=3cm]{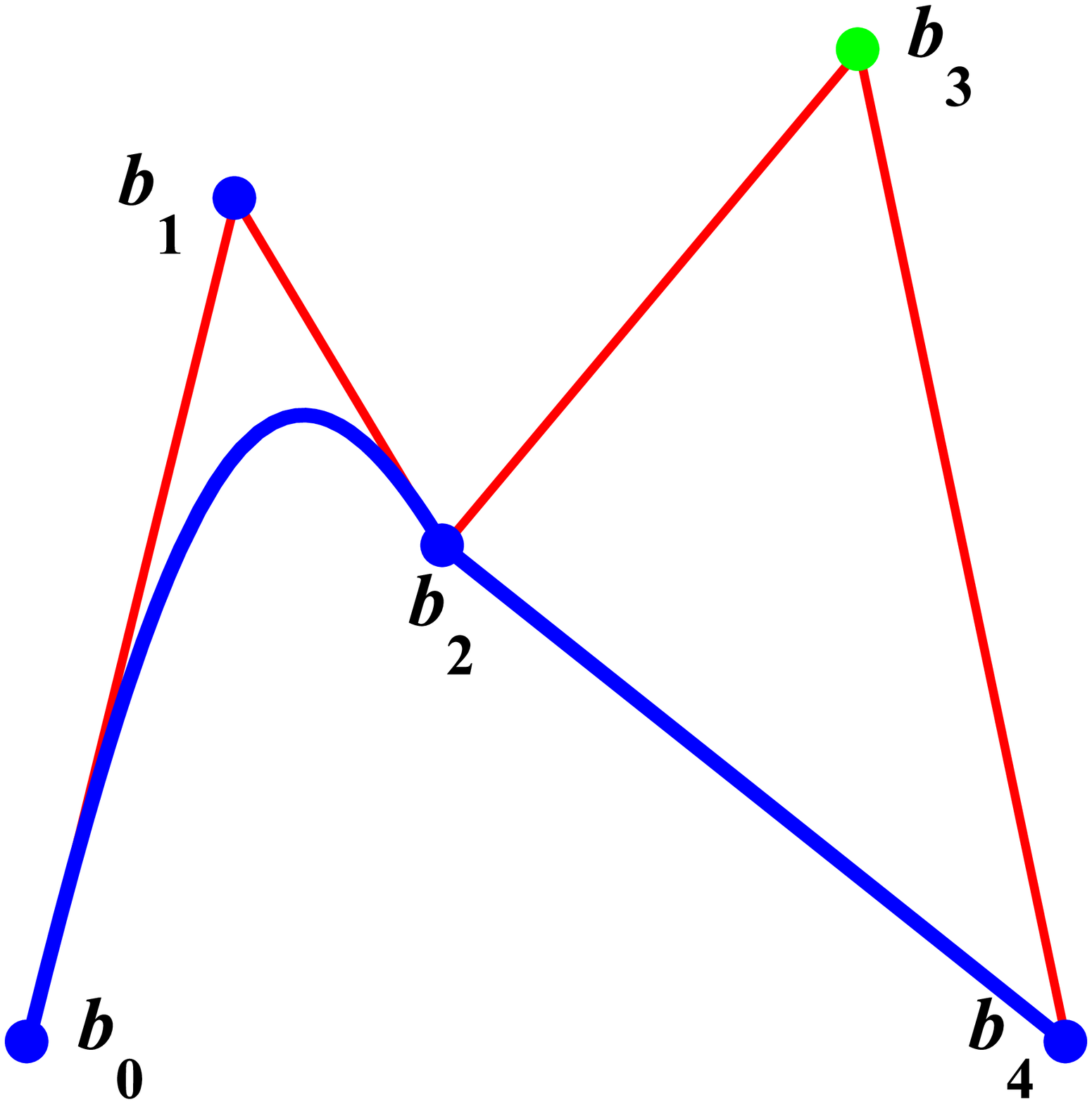}\label{Fig2b}}
\caption{Rational B\'{e}zier curve and its regular control curve.}\label{Fig2}
\end{figure}
\begin{figure}[h!]
\centering
\subfigure[$\mathcal{T}_{\lambda_1}$ and $S_{\lambda_1}$] {\includegraphics[height=3cm,width=3cm]{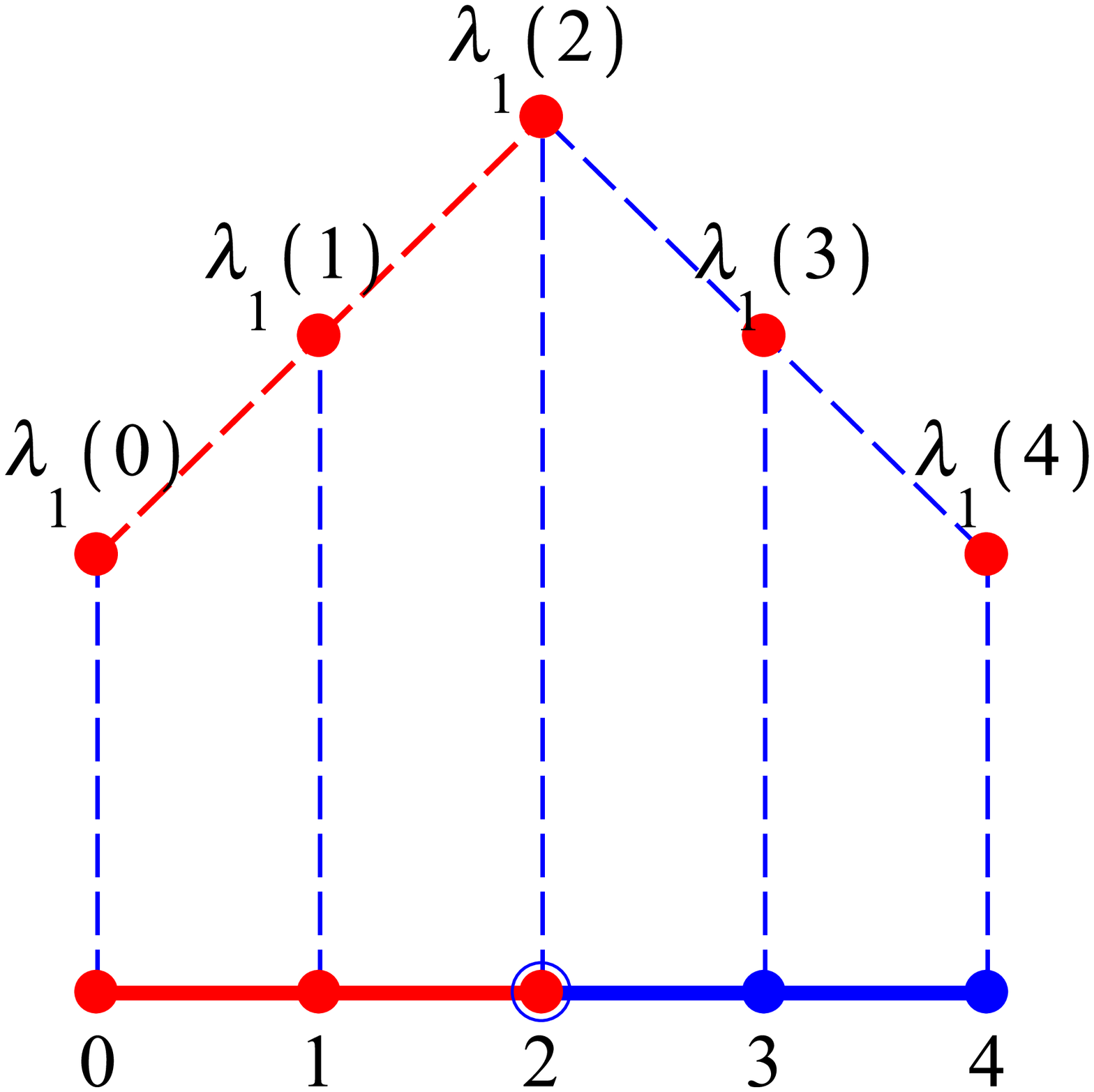}\label{Fig3a}}\hspace{5ex}
\subfigure[$\mathcal{T}_{\lambda_2}$ and $S_{\lambda_2}$] {\includegraphics[height=3cm,width=3cm]{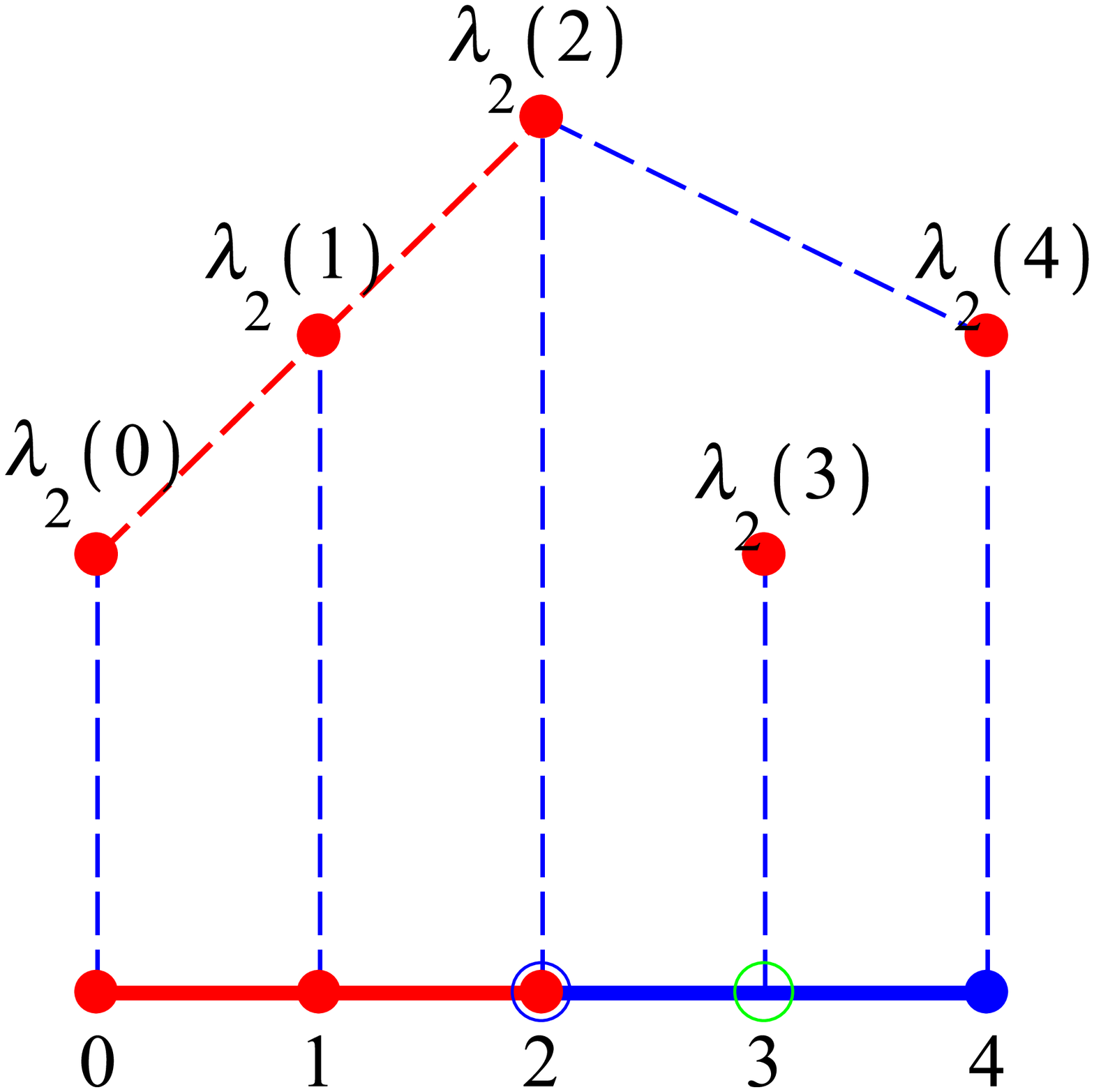}\label{Fig3b}}
\caption{Regular decompositions of $\mathcal{A}$ induced by $\lambda_1$ and  $\lambda_2$.}\label{Fig3}
\end{figure}
\end{example}
\subsection{NURBS curves and knot insertion}
\begin{definition}[\cite{Piegl0,Farin0,Farin1}]
\label{def4}
\em{
A $p$th degree {\em NURBS curve} with the control points $\textbf{P}=\{\textbf{P}_0,\textbf{P}_1,\cdots, \textbf{P}_{n+p-1}\}$ and weights $\omega=\{\omega_0,\omega_1,\cdots,\omega_{n+p-1}\}$ are defined by
\begin{eqnarray}\label{eq2.1}
\textbf{R}(u)=\frac{\sum_{i=0}^{n+p-1}\omega_i\textbf{P}_iN_{i,p}(u)}{\sum_{i=0}^{n+p-1}\omega_iN_{i,p}(u)},u \in [0,1]
\end{eqnarray}
where $\{N_{i,p}(u)\}$ are the  B-spline basis functions of degree $p$ defined on knot vector
\begin{eqnarray}\label{eq2.3}
U=\{\underbrace{0,\cdots ,0}_{p+1},u_1,u_2,\cdots ,u_{n-1},\underbrace{1,\cdots ,1}_{p+1}\}.
\end{eqnarray}
The union of segments $\overline{\textbf{P}_0\textbf{P}_1},\cdots,\overline{\textbf{P}_{n+p-2}\textbf{P}_{n+p-1}}$ is called the {\em control polygon} of the curve.
}
\end{definition}

Obviously, the NURBS curve $\textbf{R}(u)$ defined on the knot vector Eq.~(\ref{eq2.3})   satisfies the endpoint interpolation property. The geometric meaning of a single weight of NURBS curve can be explained as: within the influence range $(u_\alpha,u_\beta)$ of the control point $\textbf{P}_i$,
\begin{eqnarray*}\label{eq2.6}
\lim_{\omega _i\rightarrow + \infty} \textbf{R}(u)=\textbf{P}_i, \forall u\in (u_\alpha,u_\beta),
\end{eqnarray*}
where $u_\alpha,u_\beta \in  U$ and $\beta-\alpha=p+1$. Fig.~\ref{Fig1} shows the NURBS curve moves toward or away from the control points when one weight increases or decreases.

\begin{figure*}
\centering
\includegraphics[height=3cm]{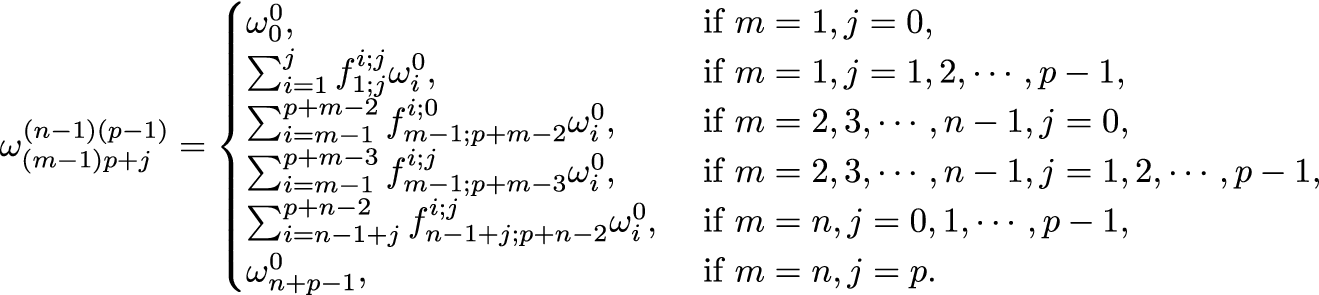}
\includegraphics[height=4cm]{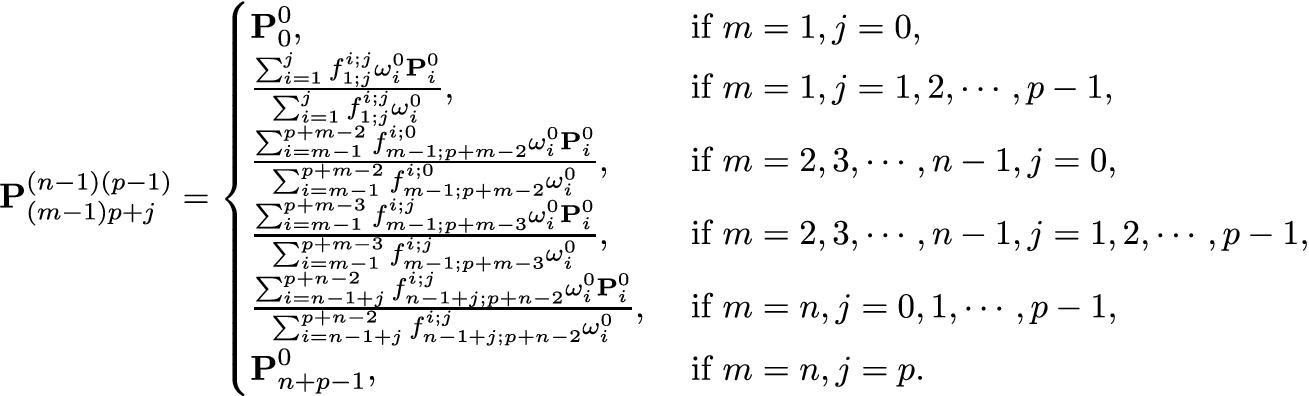}
\caption{The Pyramid relationships between the weights $\bar{\omega}$ and control points $\bar{\mathcal{B}}$ with the original weights $\omega$ and control points $\mathcal{B}$.}
\label{figth8}
\end{figure*}

In order to study toric degeneration of NURBS curve, we recall the conversion from NURBS curves  to B\'{e}zier
form by knot insertion  \cite{Piegl0,Farin0,Farin1,Boehm1980}. Without loses of generality, in the rest of paper we assume each of the interior knot $u_i$ of $U$ for NURBS curve $\textbf{R}(u)$ of degree $p$ defined as Eq.~(\ref{eq2.1}) is  with multiplicity $1$ and $u_i<u_{i+1}$, $(i=1,2,\cdots,n-2)$. For NURBS curve $\textbf{R}(u)$, we can insert each existing knot $p-1$ times to make all of interior knots  $u_i$~$(i=1,2,\cdots,n-1)$ with multiplicity $p$. Note that the result of knot insertion is independent of the ordering  of inserting knots. If a interior knot is inserted, weights and control points of  NURBS curve will be recomputed. When $(n-1)(p-1)$ interior knots are inserted, the weights
\begin{eqnarray}\label{eq2.77}
\omega^{(n-1)(p-1)}=\{\omega^{(n-1)(p-1)}_0,\omega^{(n-1)(p-1)}_1,\cdots,\omega^{(n-1)(p-1)}_{np}\},
\end{eqnarray}
and control points
\begin{eqnarray}\label{eq2.88}
\textbf{P}^{(n-1)(p-1)}=\{\textbf{P}^{(n-1)(p-1)}_0,\textbf{P}^{(n-1)(p-1)}_1,\cdots,\textbf{P}^{(n-1)(p-1)}_{np}\}
\end{eqnarray}
of NURBS curve $\textbf{R}(u)$ are generated, which defined on the new knot vector $U^{(n-1)(p-1)}$
\begin{eqnarray*}
\begin{array}{l}
=\{\underbrace{0,\cdots ,0}_{p+1},\underbrace{u_1,\cdots ,u_1}_{p},\cdots , \underbrace{u_{n-1},\cdots,u_{n-1}}_{p},\underbrace{1,\cdots ,1}_{p+1}\}.
\end{array}
\end{eqnarray*}

Let  $u_0=0,u_n=1$. By the parameter transformation $v=\frac{u-u_i}{u_{i+1}-u_i}$, the NURBS curve $\textbf{R}(u)$ is a $p$th degree rational B\'{e}zier curve $\textbf{F}^m(v)$ in every interval $[u_{m-1},u_{m}]$, $m=1,2,\cdots,n$, with the weights
\begin{equation}\label{eq2.221}
\begin{array}{c}
\{\omega^{(n-1)(p-1)}_0,\omega^{(n-1)(p-1)}_1,\cdots ,\omega^{(n-1)(p-1)}_p\},\\
\{\omega^{(n-1)(p-1)}_p,\omega^{(n-1)(p-1)}_{p+1},\cdots ,\omega^{(n-1)(p-1)}_{2p}\}, \\ \cdots ,\\
\{\omega^{(n-1)(p-1)}_{(n-1)p},\omega^{(n-1)(p-1)}_{(n-1)p+1},\cdots ,\omega^{(n-1)(p-1)}_{np}\},
\end{array}
\end{equation}
and control points
\begin{equation}\label{eq2.22}
\begin{array}{c}
\{\textbf{P}^{(n-1)(p-1)}_0,\textbf{P}^{(n-1)(p-1)}_1,\cdots ,\textbf{P}^{(n-1)(p-1)}_p\},\\
\{\textbf{P}^{(n-1)(p-1)}_p,\textbf{P}^{(n-1)(p-1)}_{p+1},\cdots ,\textbf{P}^{(n-1)(p-1)}_{2p}\},\\ \cdots ,\\
\{\textbf{P}^{(n-1)(p-1)}_{(n-1)p},\textbf{P}^{(n-1)(p-1)}_{(n-1)p+1},\cdots ,\textbf{P}^{(n-1)(p-1)}_{np}\}.
\end{array}
\end{equation}
Then the NURBS curve $\textbf{R}(u)$ is transformed into the union of those $n$ pieces of rational B\'{e}zier curves.

The representations of the weights in Eq.~(\ref{eq2.221}) and control points in Eq.~(\ref{eq2.22}) are discussed in Theorem \ref{th2}, which relate to the original weights and control points. Theorem \ref{th2} can be proved step by step via knot insertion \cite{Piegl0,Farin0,Farin1,Boehm1980}, which will be omitted here.

\begin{theorem}
\label{th2}
Let the NURBS curve $\textbf{R}(u)$ of degree $p$ defined in Eq.~$(\ref{eq2.1})$ with the control points $\mathcal{B}=\{\textbf{P}_0^0,\textbf{P}_1^0,\cdots,\textbf{P}_{n+p-1}^0\}$ and the weights $\omega=\{\omega_0^0,\omega_1^0,\cdots,\omega_{n+p-1}^0\}$. After inserting knots to  make all interior knots $u_i$~ $(i=1,2,\cdots,n-1)$ in the knot vector
$
U=\{\underbrace{0,\cdots ,0}_{p+1},u_1,u_2,\cdots ,u_{n-1},\underbrace{1,\cdots ,1}_{p+1}\}
$
with multiplicity $p$, the generated weights $\omega^{(n-1)(p-1)}$ in Eq.~$(\ref{eq2.77})$ denoted by $\bar{\omega}$ and the generated control points $\textbf{P}^{(n-1)(p-1)}$ in Eq.~$(\ref{eq2.88})$ denoted by $\bar{\mathcal{B}}$ which satisfy a certain relationship with the original weights $\omega$ and the control points $\mathcal{B}$, as Fig.~$\ref{figth8}$  shown, where coefficients $f_{a;b}^{i;j}$~$(i=a,a+1,\cdots,b)$  are relevant to $\omega^0_i$, $\textbf{P}^0_i$, computed via knot insertion, and $\sum_{i=a}^{b} f_{a;b}^{i;j}=1, (a\leq b)$ in every element of $\bar{\omega}$ and  $\bar{\mathcal{B}}$.
\end{theorem}

\section{Toric degenerations of NURBS curves}\label{mywork}

This paper focuses on what happens when all of weights of a NURBS curve assume extreme values. Since the curve is pulled towards the corresponding control point when a single weight increases, furthermore, does the NURBS curve approximate all of control points simultaneously when all of weights approach infinity.
In this section, by defining a kind of control structure of a NURBS curve, we present the toric degeneration of NURBS curve by using the toric degeneration of rational B\'{e}zier curve and indicate that the NURBS curve approximates to this control structure when all of weights approach infinity.

For the convenience, we will use the following representation to represent a NURBS curve $\textbf{R}(u)$ in the rest of paper, which is similar with Definition~\ref{def2} and equivalent to Definition~\ref{def4}. Given a set of finite lattice points $\mathcal{A} =\{0,1,\cdots,n+p-1\}\subset \mathbb{Z}$, $\Delta_{\mathcal{A}}=conv(\mathcal{A})=[0,n+p-1]$, control points $\mathcal{B}=\{\textbf{P}_i^0 \mid i\in \mathcal{A}\}\subset \mathbb{R}^{d}$, $d=2,3$, and weights $\omega=\{\omega_i^0>0  \mid i\in \mathcal{A}\}$, the parametric curve,
\begin{eqnarray}\label{eq2.42}
\textbf{R}_{\mathcal{A},\omega,\mathcal{B}}(u):=\frac{\sum_{i=0}^{n+p-1}\omega_i^0 \textbf{P}_i^0N_{i,p}(u)}{\sum_{i=0}^{n+p-1}\omega_i^0 N_{i,p}(u)},u\in[0,1]
\end{eqnarray}
is called a {\em NURBS curve} of degree $p$, where the  B-spline basis functions $N_{i,p}(u)$ are defined on the knot vector
\begin{eqnarray}\label{eqvector}
U^0=\{\underbrace{0,\cdots ,0}_{p+1},u_1,u_2,\cdots ,u_{n-1},\underbrace{1,\cdots ,1}_{p+1}\}.
\end{eqnarray}

We set $\bar{\mathcal{A}}=\{0,1,\cdots,np-1,np\}$ and $\Delta_{\bar{\mathcal{A}}}=[0,np]$. The weights $\bar{\omega}=\{\omega_i^{(n-1)(p-1)} \mid i\in \bar{\mathcal{A}}\}$ and control points $\bar{\mathcal{B}}=\{\textbf{P}_i^{(n-1)(p-1)} \mid i \in \bar{\mathcal{A}} \}$ can be computed by Theorem \ref{th2}, Eq.~(\ref{eq2.221}) and Eq.~(\ref{eq2.22}). For NURBS curve $\textbf{R}_{\mathcal{A},\omega,\mathcal{B}}$, we transform the curve into the union of $n$ pieces of rational B\'{e}zier curves via knot insertion, and denote it by  $\textbf{R}_{\bar{\mathcal{A}},\bar{\omega},\bar{\mathcal{B}}}$.
Let $\mathcal{A}^m=\{(m-1)p,(m-1)p+1,\cdots,mp\}\subset \bar{\mathcal{A}}$ and $\Delta_{\mathcal{A}^m}=[(m-1)p,mp]\subset \Delta_{\bar{\mathcal{A}}}$. Then $\bigcup_{m=1}^{n}\mathcal{A}^m=\bar{\mathcal{A}}$. By Definition~\ref{def2}, the $m$th piece rational B\'{e}zier curve of NURBS curve $\textbf{R}_{\bar{\mathcal{A}},\bar{\omega},\bar{\mathcal{B}}}$ is denoted by $\textbf{F}_{\mathcal{A}^m,\omega^m,\mathcal{B}^m}$,
where the weights $\omega^m=\{\omega^{(n-1)(p-1)}_i \mid i\in \mathcal{A}^m\}$ and control points $\mathcal{B}^m=\{\textbf{P}^{(n-1)(p-1)}_i \mid i\in \mathcal{A}^m\}$  are indexed by the lattice points of $\mathcal{A}^m$. Then we have
\begin{eqnarray}\label{eq2.461}
\textbf{R}_{\mathcal{A},\omega,\mathcal{B}}=\textbf{R}_{\bar{\mathcal{A}},\bar{\omega},\bar{\mathcal{B}}}=\bigcup_{m=1}^{n}\textbf{F}_{\mathcal{A}^m,\omega^m,\mathcal{B}^m}.
\end{eqnarray}

\begin{figure}[h!]\centering
\includegraphics[height=3cm,width=7cm]{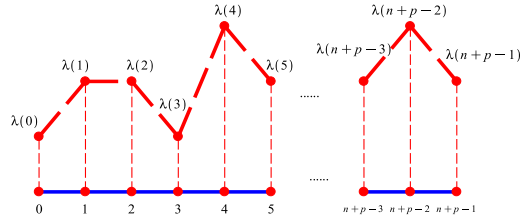}
\caption{A lifting function $\lambda$.}
\label{Fig4}
\end{figure}

We study the change of a NURBS curve when the weights of all control points approach infinite. Since the speed of weight of each control point tends to infinite may be different, we introduce the concept of the value of a lifted point $\lambda(i)$ associated with the lattice point $i$. Assume the speed of weight $\omega_i^0$ tends to infinite is $t^{\lambda(i)}$, then we replace the weight $\omega_i^0$ with $t^{\lambda(i)}\omega_i^0$. If the value of a lifted point $\lambda(i)$ has a larger value, the weight $\omega_i^0 $ of control point $\textbf{P}_i^0$ goes faster to infinity.
In this paper, we assume a lifting function $\lambda: i\rightarrow (i,\lambda(i))$ to lift all the lattice points $i$ of $\mathcal{A}$ into $\mathbb{R}^2$ (as Fig.~\ref{Fig4} shown).
If we set $\omega_{\lambda}(t):=\{t^{\lambda(i)}\omega _i^0 \mid i\in \mathcal{A}\}$, then the curve,
\begin{eqnarray*}\label{eq2.43}
\textbf{R}_{\mathcal{A},\omega_{\lambda}(t),\mathcal{B}}(u;t):=\frac{\sum_{i=0}^{n+p-1} t^{\lambda(i)}\omega_i^0 \textbf{P}_i^0N_{i,p}(u)}{\sum_{i=0}^{n+p-1} t^{\lambda(i)}\omega_i^0 N_{i,p}(u)},u\in[0,1]
\end{eqnarray*}
is called NURBS curve $\textbf{R}_{\mathcal{A},\omega,\mathcal{B}}(u)$ of degree $p$ defined in Eq.~(\ref{eq2.42}) parameterized by $t$.

Suppose that the weights $\omega=\{\omega^0_0,\omega^0_1,\cdots,\omega^0_{n+p-1}\}$  is parameterized by $t$, $\omega_{\lambda }(t):=\{t^{\lambda(i)}\omega _i^0 \mid i\in \mathcal{A}\}$. By Theorem \ref{th2}, we take $\omega_{\lambda}(t)$ as the new weights and the control points $\mathcal{B}$ stay the same. After inserting knots, the  weights $\bar{\omega}_{\lambda}(t) =\{\omega_i^{(n-1)(p-1)} \mid i\in \bar{\mathcal{A}}\}$  and control points $\bar{\mathcal{B}}_{\lambda}(t)=\{\textbf{P}_i^{(n-1)(p-1)} \mid i \in \bar{\mathcal{A}} \}$ with parameter $t$ can be computed.   For NURBS curve $\textbf{R}_{\mathcal{A},\omega_{\lambda}(t),\mathcal{B}}$, we transform the curve into the union of $n$ pieces of rational B\'{e}zier  curves, and denote it by  $\textbf{R}_{\bar{\mathcal{A}},\bar{\omega}_{\lambda}(t),\bar{\mathcal{B}}_{\lambda}(t)}$.
By Definition~\ref{def2}, the $m$th piece rational B\'{e}zier curve of NURBS curve $\textbf{R}_{\bar{\mathcal{A}},\bar{\omega}_{\lambda}(t),\bar{\mathcal{B}}_{\lambda}(t)}$ is denoted by $\textbf{F}_{\mathcal{A}^m,\omega^m_{\lambda^m}(t),\mathcal{B}^m_{\lambda^m}(t)}$,
where the weights $\omega^m_{\lambda^m}(t)=\{\omega^{(n-1)(p-1)}_i \mid i\in \mathcal{A}^m\}\subset \bar{\omega}_{\lambda}(t)$ and control points $\mathcal{B}^m_{\lambda^m}(t)=\{\textbf{P}^{(n-1)(p-1)}_i \mid i\in \mathcal{A}^m\}\subset \bar{\mathcal{B}}_{\lambda}(t)$  are indexed by the lattice points of $\mathcal{A}^m$. Then we have
\begin{eqnarray}\label{eq2.4611}
\textbf{R}_{\mathcal{A},\omega_{\lambda}(t),\mathcal{B}}=\textbf{R}_{\bar{\mathcal{A}},\bar{\omega}_{\lambda}(t),\bar{\mathcal{B}}_{\lambda}(t)}=\bigcup_{m=1}^{n}\textbf{F}_{\mathcal{A}^m,\omega^m_{\lambda^m}(t),\mathcal{B}^m_{\lambda^m}(t)}.
\end{eqnarray}

We study the $n$ pieces of rational B\'{e}zier curves $\textbf{F}_{\mathcal{A}^m,\omega^m_{\lambda^m}(t),\mathcal{B}^m_{\lambda^m}(t)}$, $m=1,2,\cdots,n$. For the first piece rational B\'{e}zier curve $\textbf{F}_{\mathcal{A}^1,\omega^1_{\lambda^1}(t),\mathcal{B}^1_{\lambda^1}(t)}$, the control points
\begin{eqnarray*}
& &\mathcal{B}^1_{\lambda^1}(t)=\{\textbf{P}^{(n-1)(p-1)}_0,\textbf{P}^{(n-1)(p-1)}_1,\cdots,\textbf{P}^{(n-1)(p-1)}_p\}
\\& &=\{\textbf{P} _0^0,\textbf{P}_1^0,\frac{ \sum_{i=1}^{2}f_{1;2}^{i;2}t^{\lambda(i)}\omega _i^0\textbf{P}_i^0}{\sum_{i=1}^{2}f_{1;2}^{i;2}t^{\lambda(i)}\omega _i^0},\cdots,\frac{\sum_{i=1}^{p}f_{1;p}^{i;0}t^{\lambda(i)}\omega _i^0\textbf{P}_i^0}{ \sum_{i=1}^{p}f_{1;p}^{i;0}t^{\lambda(i)}\omega _i^0}\},
\end{eqnarray*}
and  weights  are
\begin{eqnarray*}
& &\omega^1_{\lambda^1}(t)=\{\omega^{(n-1)(p-1)}_0,\omega^{(n-1)(p-1)}_1,\cdots,\omega^{(n-1)(p-1)}_p\}\\& &=\{t^{\lambda (0)}\omega _0^0,t^{\lambda (1)}\omega _1^0, \sum_{i=1}^{2}f_{1;2}^{i;2}t^{\lambda(i)}\omega _i^0,\cdots,\sum_{i=1}^{p}f_{1;p}^{i;0}t^{\lambda(i)}\omega _i^0\}.
\end{eqnarray*}
For $m=2,3,\cdots,n-1$, the control points and weights of the $m$th piece rational B\'{e}zier curve $\textbf{F}_{\mathcal{A}^m,\omega^m_{\lambda^m}(t),\mathcal{B}^m_{\lambda^m}(t)}$ can be computed by
\begin{eqnarray*}
& &\mathcal{B}^m_{\lambda^m}(t)=\{\textbf{P}^{(n-1)(p-1)}_{(m-1)p},\textbf{P}^{(n-1)(p-1)}_{(m-1)p+1},\cdots,\textbf{P}^{(n-1)(p-1)}_{mp}\} \\& &=\{\frac{ \sum_{i=m-1}^{p+m-2}f_{m-1;p+m-2}^{i;0}t^{\lambda(i)}\omega _i^0\textbf{P}_i^0}{ \sum_{i=m-1}^{p+m-2}f_{m-1;p+m-2}^{i;0}t^{\lambda(i)}\omega _i^0},\frac{ \sum_{i=a}^{b}f_{a;b}^{i;1}t^{\lambda(i)}\omega _i^0\textbf{P}_i^0}{ \sum_{i=a}^{b}f_{a;b}^{i;1}t^{\lambda(i)}\omega _i^0},
\\& &\cdots ,\frac{ \sum_{i=a}^{b}f_{a;b}^{i;p-1}t^{\lambda(i)}\omega _i^0\textbf{P}_i^0}{ \sum_{i=a}^{b}f_{a;b}^{i;p-1}t^{\lambda(i)}\omega _i^0},\frac{\sum_{i=m}^{p+m-1}f_{m;p+m-1}^{i;0}t^{\lambda(i)}\omega _i^0\textbf{P}_i^0}{\sum_{i=m}^{p+m-1}f_{m;p+m-1}^{i;0}t^{\lambda(i)}\omega _i^0}\},
\end{eqnarray*}
\begin{eqnarray*}
& &\omega^m_{\lambda^m}(t)=\{\omega^{(n-1)(p-1)}_{(m-1)p},\omega^{(n-1)(p-1)}_{(m-1)p+1},\cdots,\omega^{(n-1)(p-1)}_{mp}\}
\\& &=\{\sum_{i=m-1}^{p+m-2}f_{m-1;p+m-2}^{i;0}t^{\lambda(i)}\omega _i^0,\sum_{i=a}^{b}f_{a;b}^{i;1}t^{\lambda(i)}\omega _i^0,
\\& &\cdots ,\sum_{i=a}^{b}f_{a;b}^{i;p-1}t^{\lambda(i)}\omega _i^0, \sum_{i=m}^{p+m-1}f_{m;p+m-1}^{i;0}t^{\lambda(i)}\omega _i^0\},
\end{eqnarray*}
where $a=m,b=p+m-2$. For the last piece rational B\'{e}zier curve $\textbf{F}_{\mathcal{A}^n,\omega^n_{\lambda^n}(t),\mathcal{B}^n_{\lambda^n}(t)}$, the control points and weights are
\begin{eqnarray*}
& &\mathcal{B}^{n}_{\lambda^n}(t)=\{\textbf{P}^{(n-1)(p-1)}_{(n-1)p},\textbf{P}^{(n-1)(p-1)}_{(n-1)p+1},\cdots,\textbf{P}^{(n-1)(p-1)}_{np}\} \\& &=\{ \frac{ \sum_{i=n-1}^{p+n-2}f_{n-1;p+n-2}^{i;0}t^{\lambda(i)}\omega _i^0\textbf{P}_i^0}{ \sum_{i=n-1}^{p+n-2}f_{n-1;p+n-2}^{i;0}t^{\lambda(i)}\omega _i^0},\frac{ \sum_{i=n}^{p+n-2}f_{n;p+n-2}^{i;1}t^{\lambda(i)}\omega _i^0\textbf{P}_i^0}{ \sum_{i=n}^{p+n-2}f_{n;p+n-2}^{i;1}t^{\lambda(i)}\omega _i^0},\\& &\cdots,\textbf{P}_{n+p-2}^0,\textbf{P}_{n+p-1}^0\},
\end{eqnarray*}
\begin{eqnarray*}
& &\omega^n_{\lambda^n}(t)=\{\omega^{(n-1)(p-1)}_{(n-1)p},\omega^{(n-1)(p-1)}_{(n-1)p+1},\cdots,\omega^{(n-1)(p-1)}_{np}\} \\& &=\{\sum_{i=n-1}^{p+n-2}f_{n-1;p+n-2}^{i;0}t^{\lambda(i)}\omega _i^0,\sum_{i=n}^{p+n-2}f_{n;p+n-2}^{i;1}t^{\lambda(i)}\omega _i^0,
\\& &\cdots ,t^{\lambda (n+p-2)}\omega_{n+p-2}^0,t^{\lambda (n+p-1)}\omega_{n+p-1}^0\}.
\end{eqnarray*}

Consider the  $m$th piece rational B\'{e}zier curve $\textbf{F}_{\mathcal{A}^m,\omega^m_{\lambda^m}(t),\mathcal{B}^m_{\lambda^m}(t)}$, we discuss the weights $\omega^m_{\lambda^m}(t)$ and the location of  control points  $\mathcal{B}^m_{\lambda^m}(t)$ when $t\rightarrow \infty $. Suppose that the control point $\textbf{P}_{(m-1)p+j}^{(n-1)(p-1)}\in \mathcal{B}^m_{\lambda^m}(t)$, $j=0,1,\cdots,p$, is formed by the convex combination of the original \ control points  $\textbf{P}_a^0,\textbf{P}_{a+1}^0,\cdots,\textbf{P}_b^0$ of $\mathcal{B}$, where $a,b \in \mathcal{A}=\{0,1,\cdots,n+p-1\}$ and $a \leq  b$, corresponding to the values of the lifted points, $\lambda(a),\lambda(a+1),\cdots,\lambda(b) $. Then by Theorem \ref{th2}, we have
\begin{eqnarray}
\textbf{P}_{(m-1)p+j}^{(n-1)(p-1)}=\frac{\sum_{i=a}^{b}f_{a;b}^{i;j}t^{\lambda(i)}\omega _i^0 \textbf{P}_i^0}{\sum_{i=a}^{b}f_{a;b}^{i;j}t^{\lambda(i)}\omega _i^0},j=0,1,\cdots,p.
\end{eqnarray}
Suppose that the set $\psi$ be the set of lattice points of $\mathcal{A}$ corresponding to the largest value of
$\{\lambda (a),\lambda(a+1),\cdots,\lambda(b)\}$, 
then we have
\begin{eqnarray}
\centering
\lim_{t\rightarrow \infty }\textbf{P}_{(m-1)p+j}^{(n-1)(p-1)}=\frac{\sum_{i\in \psi}f_{a;b}^{i;j}\omega_{i}^0\textbf{P}_{i}^0}{\sum_{i\in \psi}f_{a;b}^{i;j}\omega_{i}^0},
\end{eqnarray}
and its corresponding weight $\lim_{t\rightarrow \infty }\omega_{(m-1)p+j}^{(n-1)(p-1)}=\sum_{i\in \psi}f_{a;b}^{i;j}\omega_{i}^0$.
 Using this method, we can get the geometric position of every control point in $\mathcal{B}^m_{\lambda^m}(t)$ and the value of its corresponding weight. We set $\bar{\mathcal{B}}^m=\lim_{t\rightarrow \infty }\mathcal{B}^m_{\lambda^m}(t)$ and $\bar{\omega}^m=\lim_{t\rightarrow \infty }\omega^m_{\lambda^m}(t)$ denote the collections of the control points of $\mathcal{B}^m_{\lambda^m}(t)$ and weights of $\omega^m_{\lambda^m}(t)$ when $t\rightarrow \infty$, respectively.

According to regular decomposition of rational B\'{e}zier curve presented in Section \ref{2.2}, we can define a regular decomposition $S_\lambda^m$ of the $m$th piece rational B\'{e}zier curve $\textbf{F}_{\mathcal{A}^m,\omega^m_{\lambda^m}(t),\mathcal{B}^m_{\lambda^m}(t)}$ for $m=1,\cdots,n$. In order to get regular decompositions of rational B\'{e}zier curves, we will discuss the values of the lifted points of every rational B\'{e}zier curve. Consider the function $\lambda: i\rightarrow (i,\lambda(i))$, $i \in \mathcal{A}=\{0,1,\cdots,n+p-1\}$ induces a regular decomposition of $\mathcal{A}^m$ by lifting $\mathcal{A}^m$ into $\mathbb{R}^2$ as follows, where $m=1,\cdots,n$. For $m=1$, the values of the lifted points associated with the lattice points of $\mathcal{A}^1$ are assigned as
\begin{eqnarray*}
\lambda^1=\{\lambda(0),\lambda(1),max\{\lambda(1),\lambda(2)\},\cdots,max\{\lambda(1),\cdots,\lambda(p)\}\}.
\end{eqnarray*}
For $m=2,\cdots,n-1$, the values of the lifted points associated with the lattice points of $\mathcal{A}^m$  are assigned as
\begin{eqnarray*}
\lambda^m=\{max\{\lambda(m-1),\cdots,\lambda(p+m-2)\},\underbrace{\chi,\chi,\cdots,\chi}_{p-1},max\{\lambda(m),\cdots,\lambda(p+m-1)\}\},
\end{eqnarray*}
where $\chi=max\{\lambda(m),\cdots,\lambda(p+m-2)\}.$
 For $m=n$, the values of the lifted points associated with the lattice points of $\mathcal{A}^n$  are assigned as
\begin{eqnarray*}
\lambda^n&&=\{max\{\lambda(n-1),\cdots,\lambda(p+n-2)\},  max\{\lambda(n),\cdots,\lambda(p+n-2)\},
\cdots,\\&&max\{\lambda(n+p-3),\lambda(p+n-2)\},  \lambda(p+n-2),\lambda(p+n-1)\}.
\end{eqnarray*}
According to the above values of the lifted points, the  regular decomposition $S_\lambda^m$ of $\mathcal{A}^m$ induced by $\lambda^m$ can be obtained directly. The union of $S_\lambda^m$ for $m=1,\cdots,n$ is called the \emph{regular decomposition of $\bar{\mathcal{A}}$}, denoted by $\bar{S}_{{\lambda}}$.

Let $s_j^m$ be the subset of $S^m_\lambda$, $\overline{\omega}^m{\mid }_{s_j^m}=\{\omega^{(n-1)(p-1)}_{i} \mid i\in s_j^m\}$ and $\overline{\mathcal{B}}^m{\mid}_{ s_j^m}=\{\textbf{P}^{(n-1)(p-1)}_{i} \mid i\in s_j^m\}$ be the weights and control points indexed by elements of $s_j^m$. $\overline{\omega}^m{\mid }_{s_j^m}$ and $\overline{\mathcal{B}}^m{\mid}_{ s_j^m}$ can construct a rational B\'{e}zier curve, denoted by $\textbf{F}_{s_j^m,\overline{\omega}^m{\mid }_{s_j^m},\overline{\mathcal{B}}^m{\mid}_{ s_j^m}}$.
By the regular decomposition $S^m_\lambda$ of $\mathcal{A}^m$ induced by $\lambda^m$ , control points $\bar{\mathcal{B}}^m$ and weights $\bar{\omega}^m$, we get the {regular control curve}
$$\textbf{F}_{\mathcal{A}^m,\overline{\omega}^m,\overline{\mathcal{B}}^m}(S_\lambda^m)=\bigcup_{s_j^m\in S_{\lambda}^m}\textbf{F}_{s_j^m,\overline{\omega}^m{\mid }_{s_j^m},\overline{\mathcal{B}}^m{\mid}_{ s_j^m}}$$
of $\textbf{F}_{\mathcal{A}^m,\overline{\omega}^m,\overline{\mathcal{B}}^m}$ induced by $S_\lambda^m$.

\begin{definition}
\label{def5}
\em{
Given a set of finite  lattice points  $\mathcal{A} =\{0,1,\cdots,n+p-1\}\subset \mathbb{Z}$, the control points $\mathcal{B}=\{\textbf{P}_i^0 \mid i\in \mathcal{A}\}\subset \mathbb{R}^{d}$, $d=2,3$,  and  weights $\omega=\{\omega_i^0>0 \mid i\in \mathcal{A}\}$, suppose that we have a regular decomposition $\bar{S}_{{\lambda}}$ of $\bar{\mathcal{A}}$ induced by a lifting function $\lambda$, then the curve
\begin{eqnarray*}\label{eq2.48}
\textbf{R}_{\mathcal{A},\omega,\mathcal{B}}(\bar{S}_\lambda)
&&=\textbf{R}_{\bar{\mathcal{A}},\bar{\omega},\bar{\mathcal{B}}}(\bar{S}_\lambda)\\
&&=\bigcup_{m=1}^{n}\textbf{F}^{m}_{\mathcal{A}^m,\overline{\omega}^m,\overline{\mathcal{B}}^m}(S_\lambda^m)\\
&&=\bigcup_{m=1}^{n}\bigcup_{s_j^m\in S_{\lambda}^m}\textbf{F}^m_{s_j^m,\overline{\omega}^m{\mid}_{ s_j^m},\overline{\mathcal{B}}^m{\mid}_{ s_j^m}}
\end{eqnarray*}
 is called the {\em regular control curve} of NURBS curve $\textbf{R}_{\mathcal{A},\omega,\mathcal{B}}$ induced by the regular decomposition $\bar{S}_\lambda$.
 }
\end{definition}

Note that the regular control curve $\textbf{R}_{\mathcal{A},w,\mathcal{B}}(\bar{S}_\lambda)$ is a $C^0$ spline curve and we will show that it is the limit of the NURBS curve $\textbf{R}_{\mathcal{A},\omega_{\lambda}(t),\mathcal{B}}$ when $t\rightarrow\infty$.

\begin{example}
\label{example5}
\begin{figure*}
\centering
\subfigure[NURBS curve] {\includegraphics[height=3cm,width=3cm]{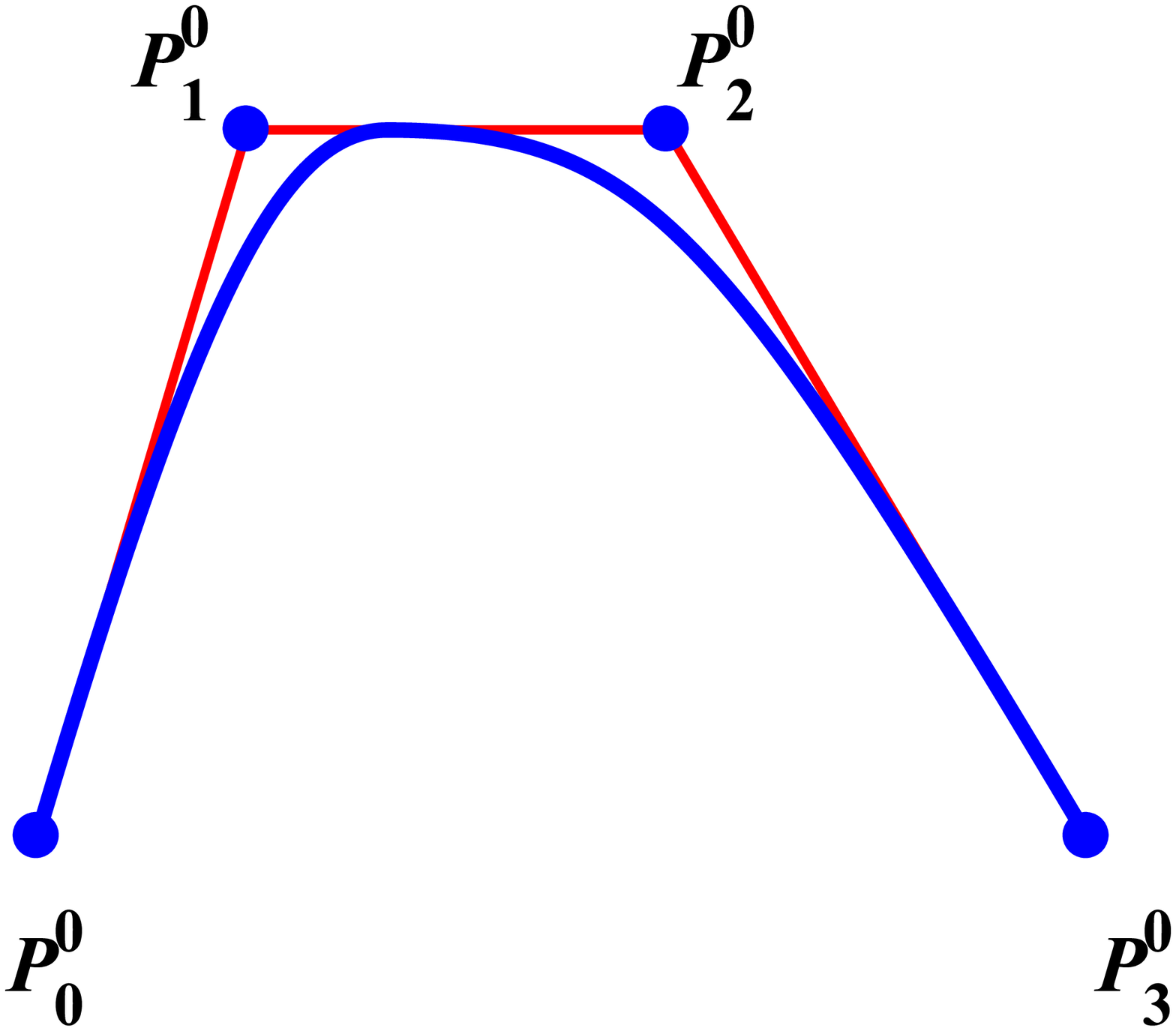}\label{Fig5a}}\hspace{3ex}
\subfigure[Curve after knot insertion] {\includegraphics[height=3cm,width=3cm]{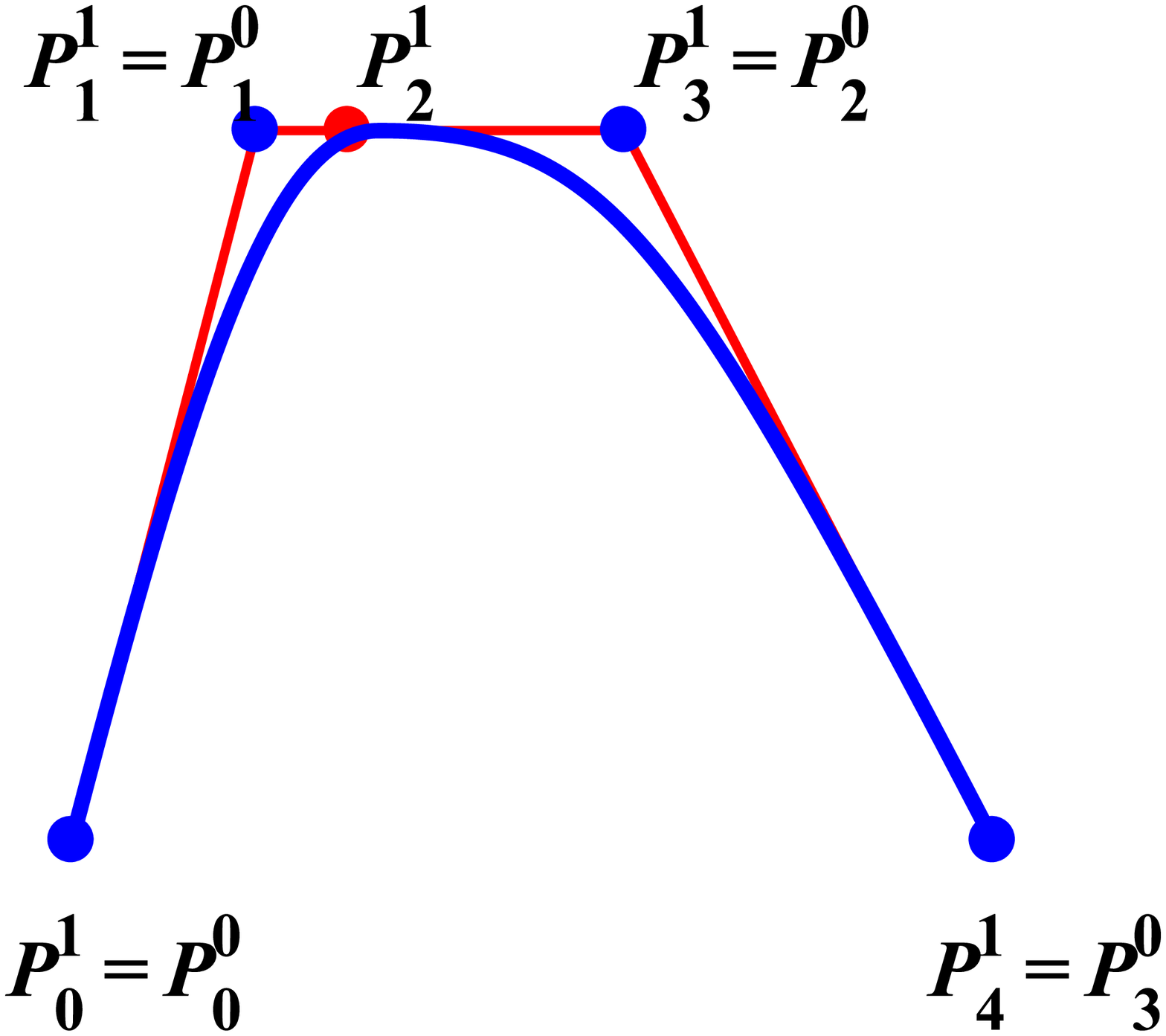}\label{Fig5d}}\hspace{3ex}
\subfigure[Regular control curve induced by $\lambda_{1}$] {\includegraphics[height=3cm,width=3cm]{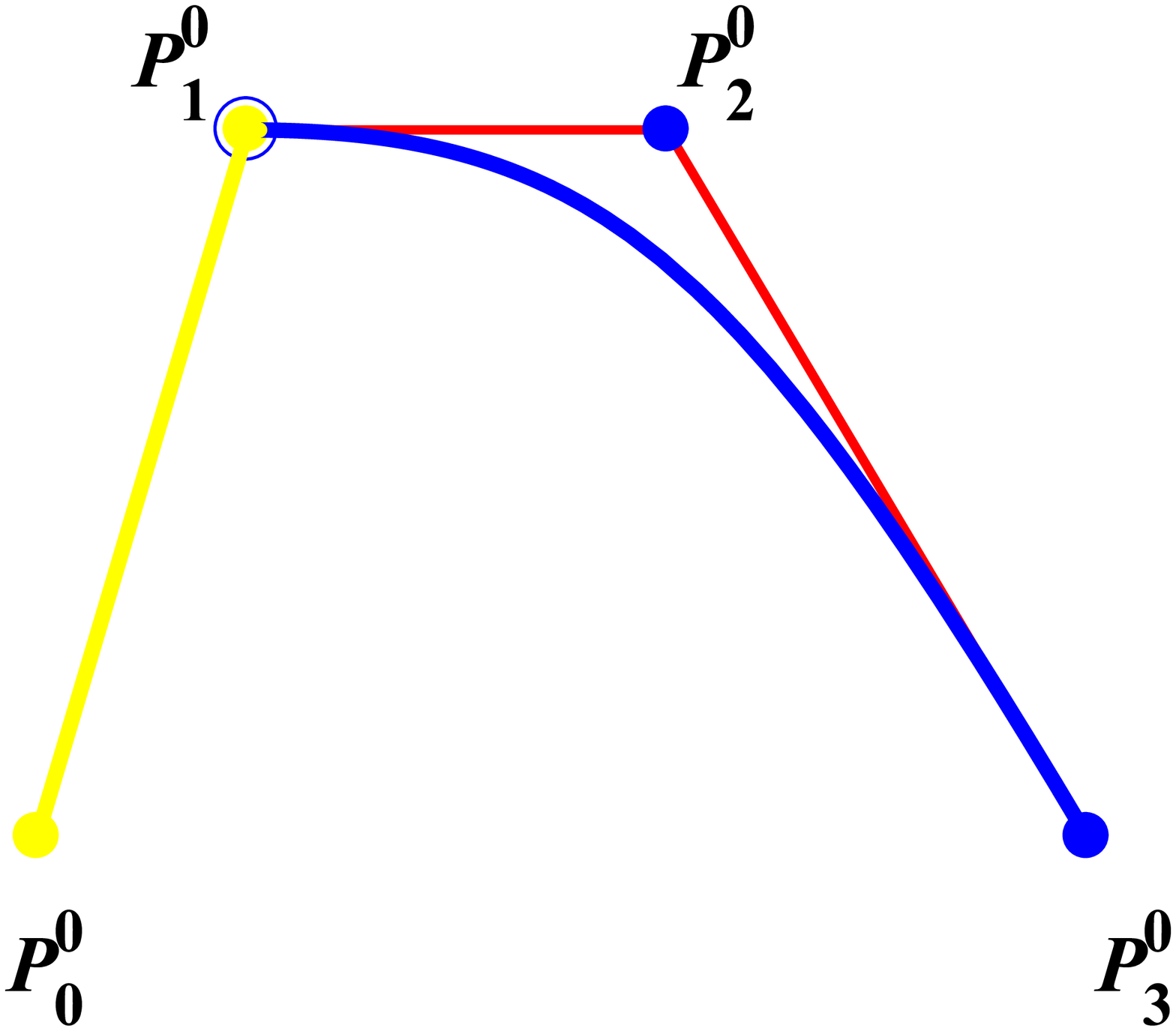}\label{Fig5b}}\hspace{3ex}
\subfigure[Regular control curve induced by $\lambda_{2}$] {\includegraphics[height=3cm,width=3cm]{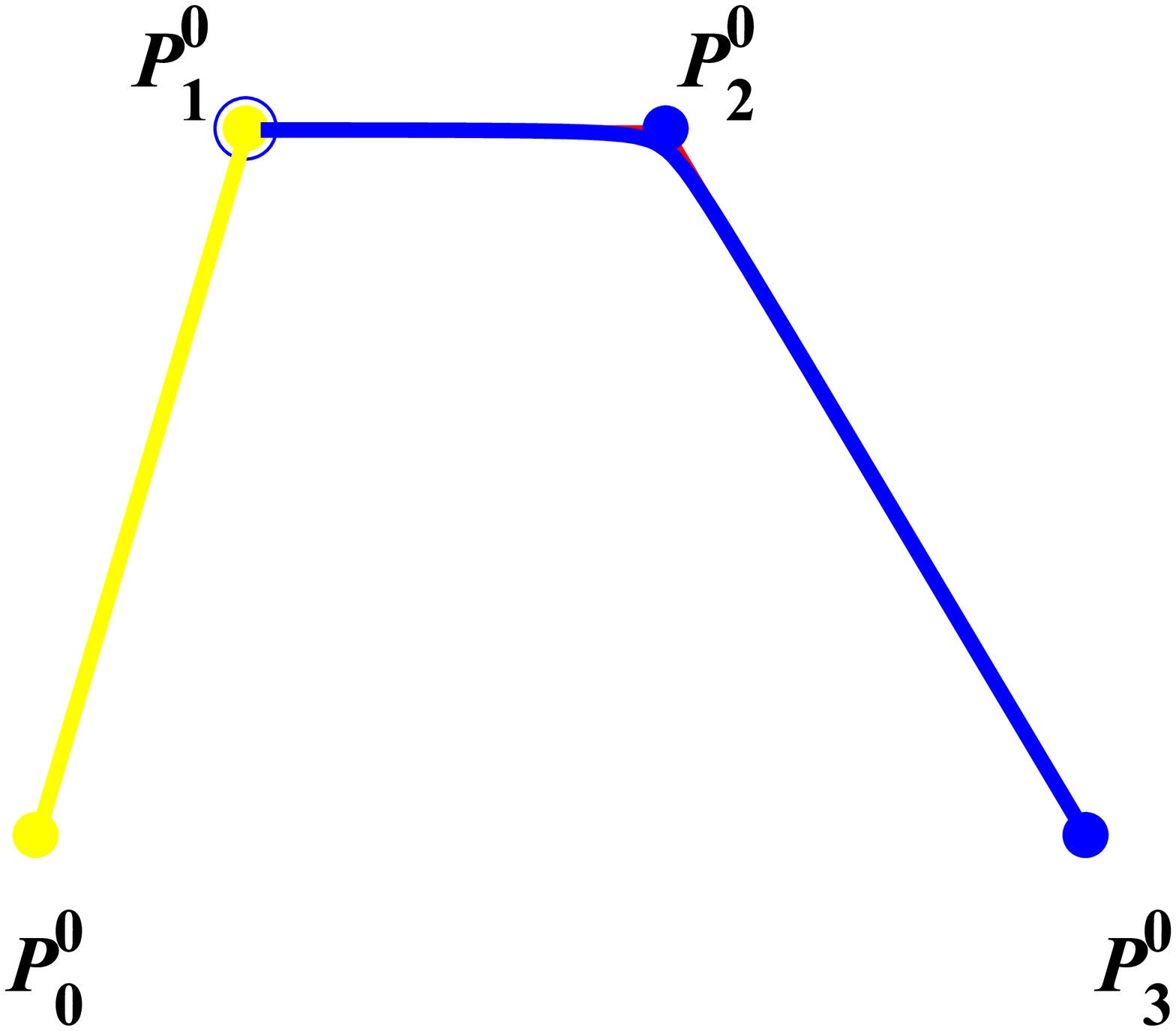}\label{Fig5c}}
\caption{The quadric NURBS curve and its regular control curve induced by $\lambda_{1}$ and by $\lambda_{2}$.}
\label{Fig5}
\end{figure*}
\em{
Let
\begin{eqnarray*}
\textbf{R}_{\mathcal{A},\omega,\mathcal{B}}(u)=\frac{\sum_{i=0}^{3}\omega_i^0 \textbf{P}_i^0N_{i,2}(u)}{\sum_{i=0}^{3}\omega_i^0N_{i,2}(u)},u\in[0,1]
\end{eqnarray*}
be a quadratic   NURBS curve defined on knot vector $U^0=\{0,0,0,\frac{1}{4},1,1,1\}$ with the control points $\mathcal{B}=\{\textbf{P}^0_0,\textbf{P}^0_1,\textbf{P}^0_2,\textbf{P}^0_3\}$ and  weights $\omega=\{\omega^0_0,\omega^0_1,\omega^0_2,\omega^0_3\}=\{3,1,2,2\}$, where $\mathcal{A}=\{0,1,2,3\}$ (see Fig.~$\ref{Fig5a}$). The curve $\textbf{R}_{\mathcal{A},\omega,\mathcal{B}}$ consists of two pieces of rational B\'ezier curves after knot insertion (see Fig.~$\ref{Fig5d}$).

$(1)$ Suppose that the lifting function $\lambda_{1}$ has the assignments $\{1,3,2,1\}$ at the lattice points of $\mathcal{A}$, then we get the regular decomposition $\bar{S}_{{\lambda}_{1}}=\{\{\{0,1\},\{1,2\}\},\{\{2,3,4\}\}\}$ of $\bar{\mathcal{A}}=\{0,1,2,3,4\}$. The regular control curve $\textbf{R}_{\mathcal{A},\omega,\mathcal{B}}(\bar{S}_\lambda)$ is the union of two parts. The first one is the union of linear B\'ezier curves formed by control points $\{\textbf{P}^1_0=\textbf{P}^0_0,\textbf{P}^1_1=\textbf{P}^0_1\}$, $\{\textbf{P}^1_1,\textbf{P}^1_2\}$, and the second part is  a rational quadratic  B\'{e}zier curve formed by control points $\{\textbf{P}^1_2,\textbf{P}^1_3=\textbf{P}^0_2,\textbf{P}^1_4=\textbf{P}^0_3\}$ and their corresponding weights. Since the control points $\textbf{P}^1_2$ goes to $\textbf{P}^1_1=\textbf{P}^0_1$ while $t$ goes to infinity, then $\textbf{R}_{\mathcal{A},\omega,\mathcal{B}}(\bar{S}_\lambda)$  degenerates into the union of a line segment $\overline{\textbf{P}^0_0\textbf{P}^0_1}$ and a rational quadratic  B\'{e}zier curves  by control points $\{\textbf{P}^0_1,\textbf{P}^0_2,\textbf{P}^0_3\}$ and their corresponding weights (shown in Fig.~$\ref{Fig5b}$).

$(2)$ Suppose that the lifting function $\lambda_{2}$ has the assignments $\{1,3,2,0\}$ at the lattice points of $\mathcal{A}$, we get the regular decomposition $\bar{S}_{{\lambda}_{2}}=\{\{\{0,1\},\{1,2\}\},\{\{2,3\},\{3,4\}\}\}$.  The regular control curve $\textbf{R}_{\mathcal{A},\omega,\mathcal{B}}(\bar{S}_\lambda)$ is the union of two parts. The first one is the union of linear B\'ezier curves formed by control points $\{\textbf{P}^1_0,\textbf{P}^1_1\}$ and  $\{\textbf{P}^1_1,\textbf{P}^1_2\}$, where as the second part consists of two linear B\'ezier curves formed by  $\{\textbf{P}^1_2,\textbf{P}^1_3\}$ and $\{\textbf{P}^1_3,\textbf{P}^1_4\}$. Since the control points $\textbf{P}^1_2$ goes to $\textbf{P}^1_1=\textbf{P}^0_1$ while $t$ goes to infinity, then $\textbf{R}_{\mathcal{A},\omega,\mathcal{B}}(\bar{S}_\lambda)$  degenerates into the union of three line segments $\overline{\textbf{P}^0_0\textbf{P}^0_1}\cup\overline{\textbf{P}^0_1\textbf{P}^0_2}\cup\overline{\textbf{P}^0_2\textbf{P}^0_3}$,  which is exactly the control polygon of the NURBS curve $\textbf{R}_{\mathcal{A},\omega,\mathcal{B}}$ (see Fig.~$\ref{Fig5c}$).
}

\end{example}

We will explain how to get the regular control curve of a NURBS curve in Section $\ref{example}$ in detail. The following result shows that the regular control curve of the NURBS curve $\textbf{R}_{\mathcal{A},\omega,\mathcal{B}}$ is just the limit of $\textbf{R}_{\mathcal{A},\omega_{\lambda}(t),\mathcal{B}}$ when $t$ goes to infinity.

\begin{theorem}
\label{th4}
Let $\textbf{R}_{\mathcal{A},\omega,\mathcal{B}}$ be the NURBS curve of degree $p$  with the control points $\mathcal{B}=\{\textbf{P}_i \mid i\in \mathcal{A}\}\subset \mathbb{R}^{d}$, $d=2,3$,  and weights $\omega=\{\omega_i>0 \mid i\in \mathcal{A}\}$ defined on the knot vector Eq.~$(\ref{eqvector})$, where $\mathcal{A} =\{0,1,\cdots,n+p-1\}\subset \mathbb{Z}$.
Suppose  that $\textbf{R}_{\mathcal{A},\omega,\mathcal{B}}(\bar{S}_\lambda)$ is the regular control curve of $\textbf{R}_{\mathcal{A},\omega,\mathcal{B}}$ induced by the regular decomposition $\bar{S}_\lambda$ of $\bar{\mathcal{A}}$ and lifting function $\lambda$, then
\begin{eqnarray}\label{eq2.49}
\lim_{t\rightarrow \infty }\textbf{R}_{\mathcal{A},\omega_\lambda(t),\mathcal{B}}=\textbf{R}_{\mathcal{A},\omega,\mathcal{B}}(\bar{S}_\lambda).
\end{eqnarray}
\end{theorem}

\noindent {\bf Proof :}
By knot insertion, the NURBS curve $\textbf{R}_{\mathcal{A},\omega_{\lambda}(t),\mathcal{B}}$ can be converted into $n$ pieces of rational B\'{e}zier curves $\textbf{F}_{\mathcal{A}^m,\omega^m_{\lambda^m}(t),\mathcal{B}^m_{\lambda^m}(t)}$, $m=1,2,\cdots,n$.

Let $S_\lambda^m$ be the regular decomposition of $\mathcal{A}^m$ by $\lambda^m$. Consider the rational B\'{e}zier curve $\textbf{F}_{\mathcal{A}^m,\omega^m_{\lambda^m}(t),\mathcal{B}^m_{\lambda^m}(t)}$, we have
\begin{eqnarray*}
\left \| \textbf{F}_{\mathcal{A}^m,\omega^m_{\lambda^m}(t),\mathcal{B}^m_{\lambda^m}(t)}-\textbf{F}_{\mathcal{A}^m,\overline{\omega}^m,\overline{\mathcal{B}}^m}(S_\lambda^m) \right \|
&\leq \left \|\textbf{F}_{\mathcal{A}^m,\omega^m_{\lambda^m}(t),\mathcal{B}^m_{\lambda^m}(t)}-\textbf{F}_{\mathcal{A}^m,{\omega^m_{\lambda^m}(t)},\overline{\mathcal{B}}^m} \right \|
\\&
+\left \| \textbf{F}_{\mathcal{A}^m,\omega^m_{\lambda^m}(t),\overline{\mathcal{B}}^m}-\textbf{F}_{\mathcal{A}^m,\overline{\omega}^m,\overline{\mathcal{B}}^m}(S_\lambda^m) \right \|.
\end{eqnarray*}
where $\|\cdot\|$ is the Hausdorff distance between two subsets of $\mathbb{R}^{3}$~\cite{Garcia}.
Since $\lim_{t\rightarrow \infty }\mathcal{B}^m_{\lambda^m}(t)=\overline{\mathcal{B}}^m$, we have
\begin{eqnarray*}
\lim_{t\rightarrow \infty }\left \| \textbf{F}_{\mathcal{A}^m,\omega^m_{\lambda^m}(t),\mathcal{B}^m_{\lambda^m}(t)}-\textbf{F}_{\mathcal{A}^m,{\omega^m_{\lambda^m}(t)},\overline{\mathcal{B}}^m} \right \| =0.
\end{eqnarray*}
By Theorem \ref{th1.1} and $\lim_{t\rightarrow \infty }{\omega}^m_{\lambda^m}(t)=\overline{\omega}^m$,  when the control points $\overline{\mathcal{B}}^m$ are fixed but the parameter $t \rightarrow \infty$, the regular control curve induced by the regular decomposition $S_\lambda^m$ of $\mathcal{A}^m$ is exactly the limit of rational B\'{e}zier curve, that is
\begin{eqnarray*}
\lim_{t\rightarrow \infty}\textbf{F}_{\mathcal{A}^m,\omega^m_{\lambda^m}(t),\bar{\mathcal{B}}^m}=\textbf{F}_{\mathcal{A}^m,\overline{\omega}^m,\overline{\mathcal{B}}^m}(S_\lambda^m),
\end{eqnarray*}
then we get
\begin{eqnarray*}
\lim_{t\rightarrow \infty}\left \| \textbf{F}_{\mathcal{A}^m,\omega^m_{\lambda^m}(t),\bar{\mathcal{B}}^m}-\textbf{F}_{\mathcal{A}^m,\overline{\omega}^m,\overline{\mathcal{B}}^m}(S_\lambda^m)\right \|=0.
\end{eqnarray*}
This means
$$
\lim_{t\rightarrow \infty} \textbf{F}_{\mathcal{A}^m,\omega^m_{\lambda^m}(t),\mathcal{B}^m_{\lambda^m}(t)}=\textbf{F}_{\mathcal{A}^m,\overline{\omega}^m,\overline{\mathcal{B}}^m}(S_\lambda^m).
$$

Note that the NURBS curve $\textbf{R}_{\mathcal{A},\omega_\lambda(t),\mathcal{B}}$ can be convert into $n$ pieces of rational B\'{e}zier curves after knot insertion,
\begin{equation*}
\textbf{R}_{\mathcal{A},\omega_\lambda(t),\mathcal{B}}=\bigcup_{m=1}^{n}\textbf{F}_{\mathcal{A}^m,\omega^m_{\lambda^m}(t),\mathcal{B}^m_{\lambda^m}(t)},
\end{equation*}
then the  limit of NURBS curve $\textbf{R}_{\mathcal{A},\omega_\lambda(t),\mathcal{B}}$ can be written as the limit of the union of those rational B\'{e}zier curves $\textbf{F}_{\mathcal{A}^m,\omega^m_{\lambda^m}(t),\mathcal{B}^m_{\lambda^m}(t)}$, that is
\begin{eqnarray*}
\lim_{t\rightarrow \infty }\textbf{R}_{\mathcal{A},\omega_\lambda(t),\mathcal{B}}
&&=\lim_{t\rightarrow \infty }\bigcup_{m=1}^{n}\textbf{F}_{\mathcal{A}^m,\omega^m_{\lambda^m}(t),\mathcal{B}^m_{\lambda^m}(t)}\\
&&=\bigcup_{m=1}^{n}\lim_{t\rightarrow \infty }\textbf{F}_{\mathcal{A}^m,\omega^m_{\lambda^m}(t),\mathcal{B}^m_{\lambda^m}(t)}
\\&&=\bigcup_{m=1}^{n}\textbf{F}_{\mathcal{A}^m,\overline{\omega}^m,\overline{\mathcal{B}}^m}(S_\lambda^m).
\end{eqnarray*}
By Definition~\ref{def5}, we get
$$\lim_{t\rightarrow \infty }\textbf{R}_{\mathcal{A},\omega_\lambda(t),\mathcal{B}}=\textbf{R}_{\mathcal{A},\omega,\mathcal{B}}(\bar{S}_\lambda),$$
and this ends the proof.

This property of NURBS curve  is called the \emph{toric degeneration of NURBS curve} following the toric degenerations of B\'{e}zier curves and surfaces. The following result is converse to Theorem \ref{th4}.

\begin{theorem}
\label{th5}
Let $\mathcal{A}=\{0,1,\cdots,n+p-1\}\subset \mathbb{Z}$  and $\mathcal{B}=\{\textbf{P}_i^0 \mid i\in \mathcal{A}\}\subset \mathbb{R}^3$  be control points. If $\textbf{R}\subset \mathbb{R}^3$ is a set for which there is a sequence $\omega^{(1)},\omega^{(2)},\cdots$ of weights so that
\begin{eqnarray}
\lim_{\tau\rightarrow \infty }\textbf{R}_{\mathcal{A},\omega^{(\tau)},\mathcal{B}}=\textbf{R},
\end{eqnarray}
then there are a regular decomposition $\bar{S}_\lambda$ of $\bar{\mathcal{A}}$ induced by a lifting function $\lambda$ and weights $\omega=\{\omega_i^0>0 \mid i\in \mathcal{A}\}$, such that $\textbf{R}$ is a regular control curve of NURBS curve, $\textbf{R}=\textbf{R}_{\mathcal{A},\omega,\mathcal{B}}(\bar{S}_\lambda)$.

\end{theorem}

\noindent {\bf Proof :}
Let $\mathcal{A}^m=\{(m-1)p,(m-1)p+1,\cdots,mp\}\subset \bar{\mathcal{A}}$, $\Delta_{\mathcal{A}^m}=[(m-1)p,mp]$, $\mathcal{B}^m=\{\textbf{P}_i^{(n-1)(p-1)} \mid i \in \mathcal{A}^m \}$ be control points and $\omega^{m^{(\tau)}}=\{\omega_i^{(n-1)(p-1)} \mid i\in \mathcal{A}^m\}$ be weights for the $m$th piece rational B\'{e}zier curve  $\textbf{F}_{\mathcal{A}^m,\omega^{m^{(\tau)}},\mathcal{B}^m}$ $(m=1,\cdots,n)$ after  knot insertion for NURBS curve $\textbf{R}_{\mathcal{A},\omega^{(\tau)},\mathcal{B}}$.

From the assumption,
\begin{eqnarray*}\label{eq2.51}
\lim_{\tau\rightarrow \infty }\textbf{F}_{\mathcal{A}^m,\omega^{m^{(\tau)}},\mathcal{B}^m}=\textbf{F}^{m},
\end{eqnarray*}
where $\textbf{F}^{m}$ is a set of $\mathbb{R}^3$.
By Theorem \ref{th1.2}, there is regular decomposition of  $\mathcal{A}^m$ induced by a lifting function $\lambda^m$, weights $\omega_*^{m}$ and control points $\mathcal{B}_*^{m}$, such that $\textbf{F}^{m}=\textbf{F}_{\mathcal{A}^m,{\omega_*^m},{{\mathcal{B}}_*^m}}(S_\lambda^m)$ is a regular control curve.

Note that the NURBS curve $\textbf{R}_{\mathcal{A},\omega^{(\tau)},\mathcal{B}}$  is coincident with  $\textbf{R}_{\bar{\mathcal{A}},\bar{\omega}^{(\tau)},\bar{\mathcal{B}}}$, then $\lim_{\tau\rightarrow \infty }\textbf{R}_{\mathcal{A},\omega^{(\tau)},\mathcal{B}}=\lim_{\tau\rightarrow \infty }\textbf{R}_{\bar{\mathcal{A}},\bar{\omega}^{(\tau)},\bar{\mathcal{B}}}$. We set $\mathbf{R}=\bigcup_{m=1}^{n}\textbf{F}^{m}$,
then
\begin{eqnarray*}
\lim_{\tau \rightarrow \infty }\textbf{R}_{\mathcal{A},\omega^{(\tau)},\mathcal{B}}&&=\lim_{\tau \rightarrow \infty }\textbf{R}_{\bar{\mathcal{A}},\bar{\omega}^{(\tau)},\bar{\mathcal{B}}}=\lim_{\tau\rightarrow \infty}\bigcup_{m=1}^{n}\textbf{F}^{m}_{\mathcal{A}^m,\omega^{{m^{(\tau)}}},\mathcal{B}^m}=\bigcup_{m=1}^{n}\lim_{\tau \rightarrow \infty}\textbf{F}^{m}_{\mathcal{A}^m,\omega^{{m^{(\tau)}}},\mathcal{B}^m}\\
& &=\bigcup_{m=1}^{n}\textbf{F}^{m}=\bigcup_{m=1}^{n}\textbf{F}^{m}_{\mathcal{A}^m,{\omega_*^m},{{\mathcal{B}}_*^m}}(S_\lambda^m).
\end{eqnarray*}
We set the lifting function $\lambda$ of $\mathcal{A}$ taking the same value at the lattice points of $\mathcal{A}^m$, the weights $\omega=\{\omega_{i}^0 \mid i\in \mathcal{A} \}$ satisfying $\bar{\omega}=\bigcup_{m=1}^{n}\omega_*^m$ and control points $\mathcal{B}=\{\textbf{P}_{i}^0 \mid i\in \mathcal{A} \}$ satisfying $\bar{\mathcal{B}}=\bigcup_{m=1}^{n}\mathcal{B}_*^m$ after knot insertion. Let $\bar{S}_\lambda$ be the regular decomposition of $\bar{\mathcal{A}}$. By Definition \ref{def5},  we get
$$\bigcup_{m=1}^{n}\textbf{F}_{\mathcal{A}^m,{\omega_*^m},{{\mathcal{B}}_*^m}}(S_\lambda^m)=\textbf{R}_{\mathcal{A},\omega,\mathcal{B}}(\bar{S}_\lambda)$$
is a regular control curve and  this completes the proof.

\section{Examples}\label{example}

\begin{example}
\label{example1}

\em{
Let \begin{eqnarray*}
\textbf{R}_{\mathcal{A},\omega,\mathcal{B}}(u)=\frac{\sum_{i=0}^{4}\omega_i^0 \textbf{P}_i^0N_{i,2}(u)}{\sum_{i=0}^{4}\omega_i^0N_{i,2}(u)}, u\in [0,1],
\end{eqnarray*}
 be a quadratic  NURBS curve defined on knot vector $U^0=\{0,0,0,\frac{1}{4},\frac{3}{4},1,1,1\}$ with the weights $\omega=\{3,2,3,2,5\}$ and control points $\mathcal{B}=\{\textbf{P}^0_0,\textbf{P}^0_1,\textbf{P}^0_2,\textbf{P}^0_3,\textbf{P}^0_4\}$ (see Fig.~$\ref{Fig6a}$).
Fig.~$\ref{Fig6b}$ shows the NURBS curve after inserting the knots $\frac{1}{4}$ and $\frac{3}{4}$. Suppose that the lifting function $\lambda$ has the assignments $\{1,2,3,2,1\}$ at the lattice points  of $\mathcal{A}=\{0,1,2,3,4\}$. We can define the NURBS curve  $\textbf{R}_{\mathcal{A},\omega_\lambda(t),\mathcal{B}}$ with parameter $t$ by a family of weights $\omega_\lambda(t)=\{3t,2t^2,3t^3,2t^2,5t\}$. After inserting the knots $\frac{1}{4}$ and $\frac{3}{4}$,  the NURBS curve $\textbf{R}_{\mathcal{A},\omega_\lambda(t),\mathcal{B}}$ is converted to the union of three pieces of rational B\'ezier curves, $\textbf{R}_{\bar{\mathcal{A}},\bar{\omega}_\lambda(t),\bar{\mathcal{B}}_\lambda(t)}=\bigcup_{m=1}^{3}\textbf{F}_{\mathcal{A}^m,\omega^m_{\lambda^m}(t),\mathcal{B}^m_{\lambda^m}(t)}$. The weights and control points of NURBS curve $\textbf{R}_{\bar{\mathcal{A}},\bar{\omega}_\lambda(t),\bar{\mathcal{B}}_\lambda(t)}$ can be obtained by Theorem $\ref{th2}$,
\begin{eqnarray*}
\bar{\omega}_\lambda(t)
& &=\{\omega^2_0,\omega^2_1,\omega^2_2,\omega^2_3,\omega^2_4,\omega^2_5,\omega^2_6\}\\
& &=\{t^{\lambda(0)}\omega^0_0,t^{\lambda(1)}\omega^0_1,\sum_{i=1}^{2}f_{1;2}^{i;0}t^{\lambda(i)}\omega _i^0,t^{\lambda(2)}\omega^0_2,\sum_{i=2}^{3}f_{2;3}^{i;0}t^{\lambda(i)}\omega _i^0,t^{\lambda(3)}\omega^0_3,t^{\lambda(4)}\omega^0_4\},
\end{eqnarray*}
\begin{eqnarray*}
\bar{\mathcal{B}}_\lambda(t)&&=\{\textbf{P}^2_0,\textbf{P}^2_1,\textbf{P}^2_2,\textbf{P}^2_3,\textbf{P}^2_4,\textbf{P}^2_5,\textbf{P}^2_6\}\\
& &=\{\textbf{P}^0_0,\textbf{P}^0_1,\frac{ \sum_{i=1}^{2}f_{1;2}^{i;0}t^{\lambda(i)}\omega _i^0\textbf{P}_i^0}{\sum_{i=1}^{2}f_{1;2}^{i;0}t^{\lambda(i)}\omega _i^0},\textbf{P}^0_2,\frac{ \sum_{i=2}^{3}f_{2;3}^{i;0}t^{\lambda(i)}\omega _i^0\textbf{P}_i^0}{\sum_{i=2}^{3}f_{2;3}^{i;0}t^{\lambda(i)}\omega _i^0},\textbf{P}^0_3,\textbf{P}^0_4\},
\end{eqnarray*}
where $f_{1;2}^{1;0}=\frac{2}{3}, f_{1;2}^{2;0}=\frac{1}{3}, f_{2;3}^{2;0}= \frac{1}{3}, f_{2;3}^{3;0}=\frac{2}{3}$.

The lifting function ${\lambda}$ induces the assignments on $\bar{\mathcal{A}}$ by $\{\{1,2,3\},\{3,3,3\},\{3,2,1\}\}$ and  derives a regular decomposition $\bar{S}_{\lambda}=\{\{\{0,1,2\}\},\{\{2,3,4\}\},\{\{4,5,6\}\}\}$ of $\bar{\mathcal{A}}$.

Consider the regular control curve of the first rational B\'ezier curve $\textbf{F}_{\mathcal{A}^1,\omega^1_{\lambda^1}(t),\mathcal{B}^1_{\lambda^1}(t)}$ with the control points $\{\textbf{P}^2_0,\textbf{P}^2_1,\textbf{P}^2_2\}$,  weights $\{\omega^2_0,\omega^2_1,\omega^2_2\}$ and lifting function $\lambda^1=\{1,2,3\}$. Let $\mathcal{A}^1=\{0,1,2\}$ and $\Delta_{\mathcal{A}^1}=[0,2]$.
Since $\lambda(1)=2<\lambda(2)=3$,  the weights $\omega^2_0=\omega^0_0=3,\omega^2_1=\omega^0_1=2,\lim_{t\rightarrow \infty }\omega^2_2=\frac{1}{3}\omega^0_2=1$ and the control points $\textbf{P}^2_0=\textbf{P}^0_0,\textbf{P}^2_1=\textbf{P}^0_1,\lim_{t\rightarrow \infty }\textbf{P}^2_2=\textbf{P}^0_2$.  The lifting function $\lambda^1=\{1,2,3\}$
induces a regular decomposition $S_{\lambda}^1=\{\{0,1,2\}\}$ of $\mathcal{A}^1$.
For the subset $\{0,1,2\}$, we can construct a rational quadratic   B\'{e}zier curve by the control points $\{\textbf{P}^2_0,\textbf{P}^2_1,\textbf{P}^2_2\}=\{\textbf{P}^0_0,\textbf{P}^0_1,\textbf{P}^0_2\}$ and weights $\{\omega^2_0,\omega^2_1,\omega^2_2\}=\{3,2,1\}$.  Then the regular control curve of the first rational B\'{e}zier curve is the rational quadratic  B\'{e}zier  curve.

Consider the regular control curve of  the second rational B\'{e}zier curve $\textbf{F}_{\mathcal{A}^2,\omega^2_{\lambda^2}(t),\mathcal{B}^2_{\lambda^2}(t)}$ with the control points $\{\textbf{P}^2_2,\textbf{P}^2_3,\textbf{P}^2_4\}$, weights $\{\omega^2_2,\omega^2_3,\omega^2_4\}$ and lifting function $\lambda^2=\{3,3,3\}$.
 Let $\mathcal{A}^2=\{2,3,4\}$ and $\Delta_{\mathcal{A}^2}=[2,4]$.
Since $\lambda(1)=2<\lambda(2)=3$ and $\lambda(2)=3>\lambda(3)=2$, the weights $\lim_{t\rightarrow \infty }\omega^2_2=\frac{1}{3}\omega^0_2=1$, $\omega^2_3=\omega^0_2=3$, $\lim_{t\rightarrow \infty }\omega^2_4=\frac{1}{3}\omega^0_2=1$ and the control points $\lim_{t\rightarrow \infty }\textbf{P}^2_2=\textbf{P}^0_2$, $\textbf{P}^2_3=\textbf{P}^0_2$, $\lim_{t\rightarrow \infty }\textbf{P}^2_4=\textbf{P}^0_2$. The lifting function $\lambda^2=\{3,3,3\}$
induces a regular decomposition $S_{\lambda}^2=\{\{2,3,4\}\}$ of $\mathcal{A}^2$. For the subset $\{2,3,4\}$, we can construct a rational quadratic  B\'{e}zier curve by the control points $\{\textbf{P}^2_2,\textbf{P}^2_3,\textbf{P}^2_4\}$ and weights $\{\omega^2_2,\omega^2_3,\omega^2_4\}$.  Then the regular control curve of the second rational B\'{e}zier curve is the rational quadratic  B\'{e}zier curve. Since the control points $\textbf{P}^2_2$ and $\textbf{P}^2_4$ goes to $\textbf{P}^2_3=\textbf{P}^0_2$ while $t$ goes to infinity, then the regular control curve degenerates into a point $\textbf{P}^0_2$.

Consider the regular control curve of  the third rational B\'{e}zier curve $\textbf{F}_{\mathcal{A}^3,\omega^3_{\lambda^3}(t),\mathcal{B}^3_{\lambda^3}(t)}$ with the control points $\{\textbf{P}^2_4,\textbf{P}^2_5,\textbf{P}^2_6\}$, weights $\{\omega^2_4,\omega^2_5,\omega^2_6\}$ and lifting function $\lambda^3=\{3,2,1\}$.
 Let $\mathcal{A}^3=\{4,5,6\}$ and $\Delta_{\mathcal{A}^3}=[4,6]$.
Since $\lambda(2)=3>\lambda(3)=2$,  the weights $\lim_{t\rightarrow \infty }\omega^2_4=\frac{1}{3}\omega_2^0=1,\omega^2_5=\omega_3^0=2,\omega^6_2=\omega_4^0=5$ and the control points $\lim_{t\rightarrow \infty }\textbf{P}^2_4=\textbf{P}^0_2,\textbf{P}^2_5=\textbf{P}^0_3,\textbf{P}^2_6=\textbf{P}^0_4$.  The lifting function $\lambda^3=\{3,2,1\}$
induces a regular decomposition $S_{\lambda}^3=\{\{4,5,6\}\}$ of $\mathcal{A}^3$.
For the subset $\{4,5,6\}$, we can construct a rational quadratic B\'{e}zier curve by  the control points $\{\textbf{P}^2_4,\textbf{P}^2_5,\textbf{P}^2_6\}=\{\textbf{P}^0_2,\textbf{P}^0_3,\textbf{P}^0_4\}$ and weights $\{\omega^2_4,\omega^2_5,\omega^2_6\}=\{1,2,5\}$. Then the regular control curve of the third rational B\'{e}zier curve is the rational quadratic  B\'{e}zier  curve.

Then the regular control curve of the quadratic  NURBS curve $\textbf{R}_{\mathcal{A},\omega,\mathcal{B}}$ is the union of two pieces of rational quadratic  B\'{e}zier curves by control points $\{\textbf{P}^0_0,\textbf{P}^0_1,\textbf{P}^0_2\}$ and weights $\{3,2,1\}$, and control points  $\{\textbf{P}^0_2,\textbf{P}^0_3,\textbf{P}^0_4\}$  and weights $\{1,2,5\}$, respectively (see Fig.~$\ref{Fig6c}$). Fig.~$\ref{Fig7}$ shows the degeneration process of the curve with $t=2,3,5,10$, respectively.
}
\begin{figure*}
\centering
\subfigure[NURBS curve] {\includegraphics[height=3cm,width=3cm]{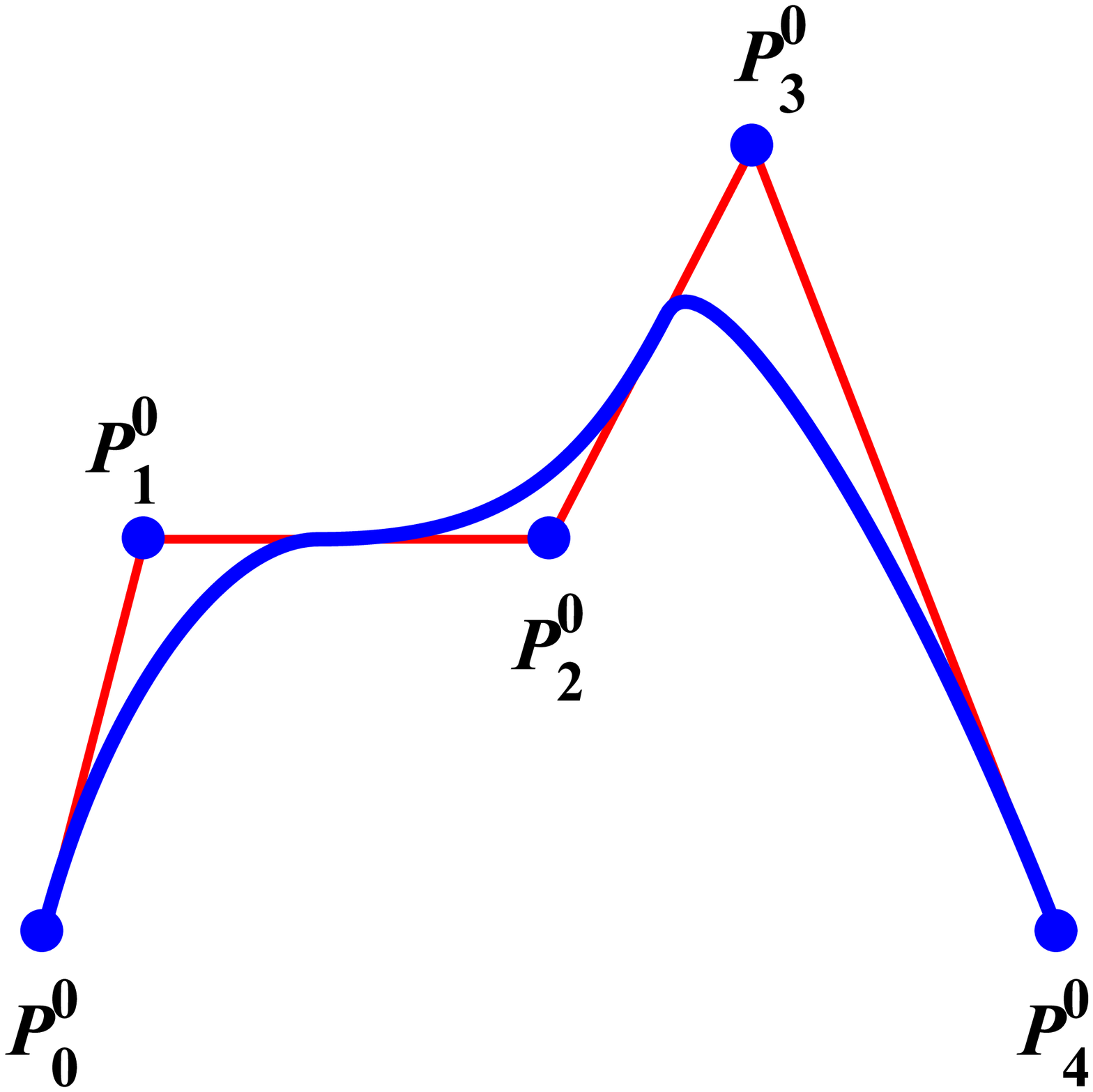}\label{Fig6a}}\hspace{5ex}
\subfigure[Curve after knot insertion] {\includegraphics[height=3cm,width=3.2cm]{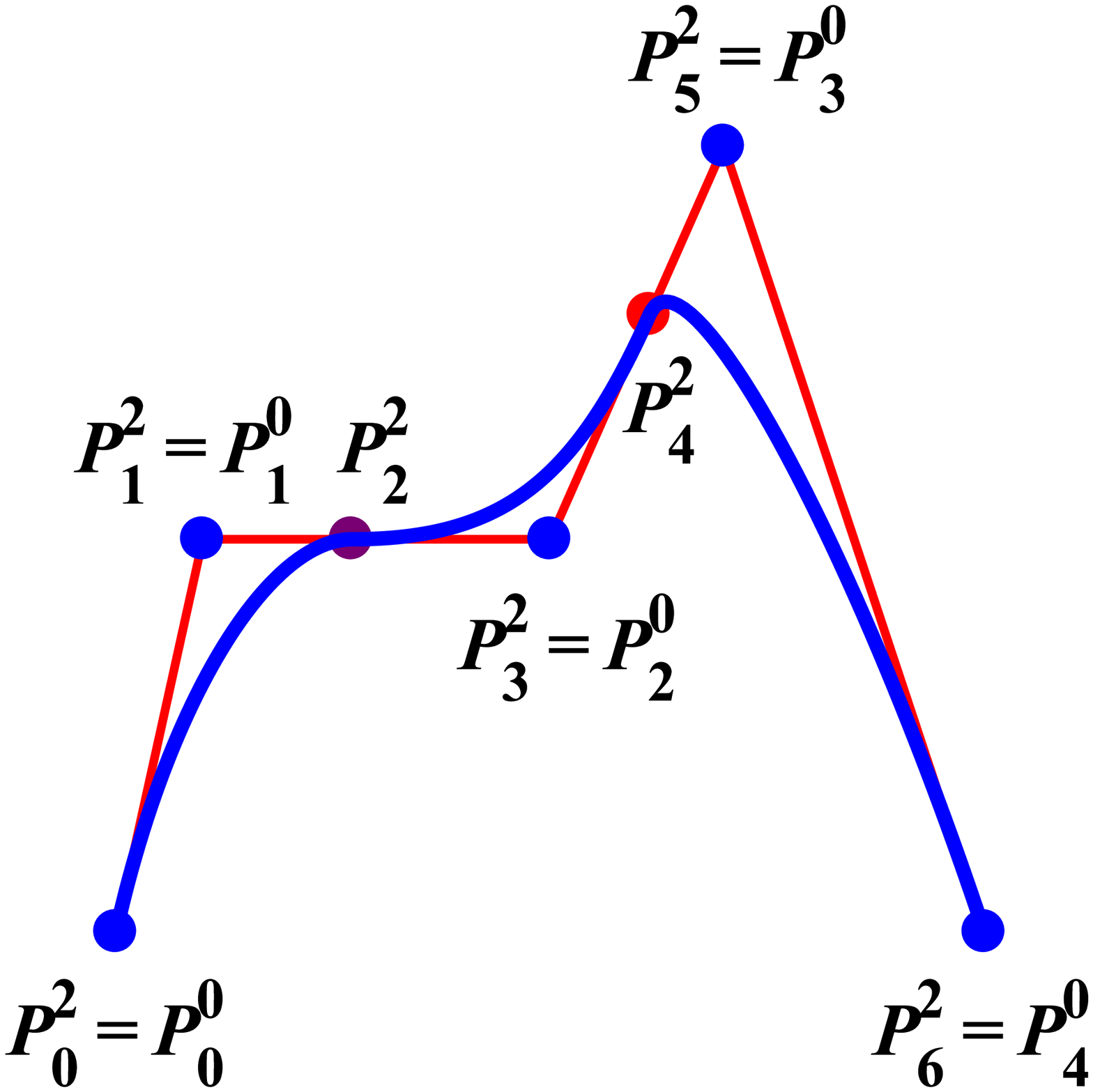}\label{Fig6b}}\hspace{5ex}
\subfigure[The regular control curve] {\includegraphics[height=3cm,width=3cm]{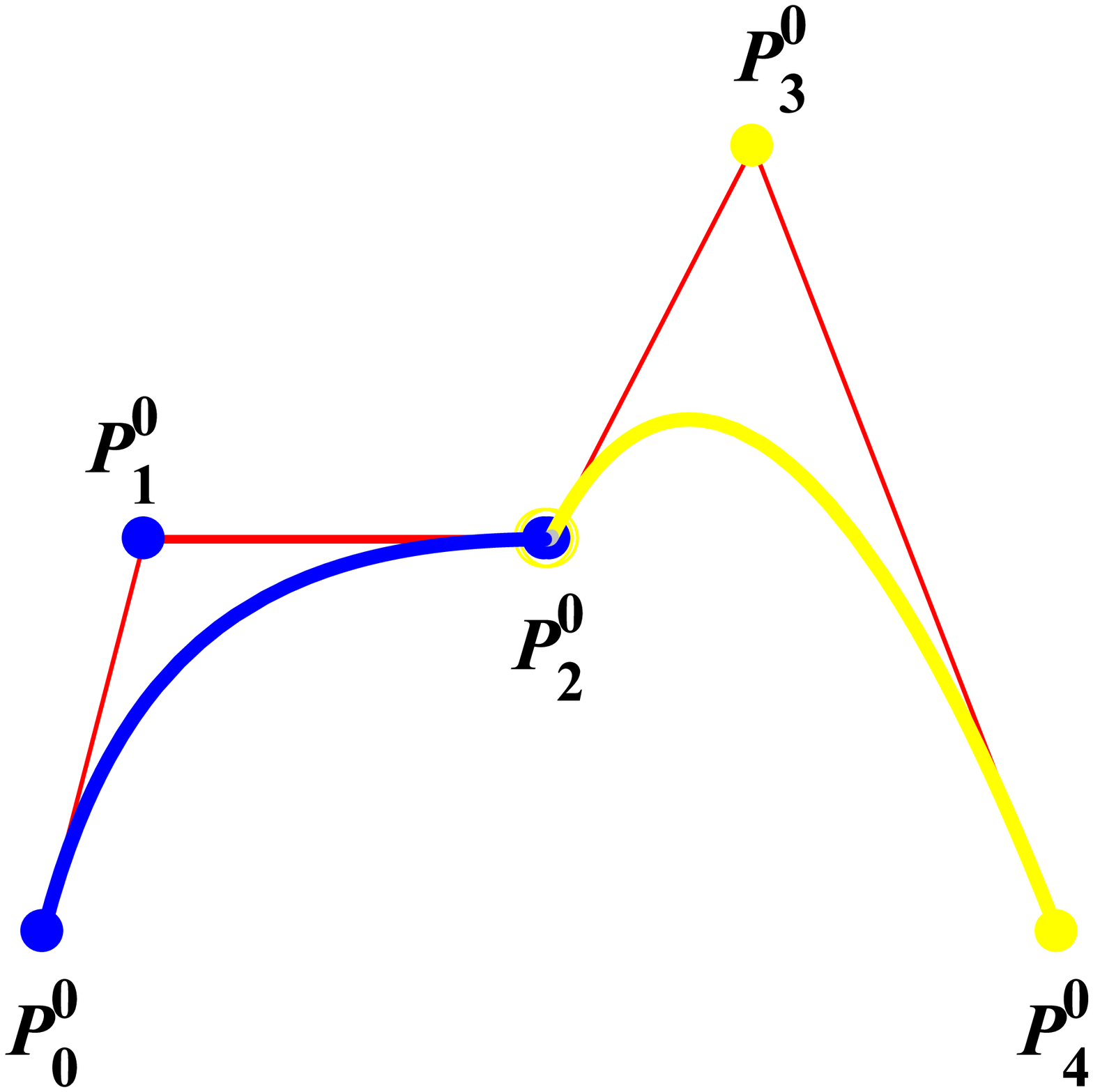}\label{Fig6c}}
\caption{The quadric NURBS curve before and after the knot insertion, and its regular control curve.}
\label{Fig6}
\end{figure*}

\begin{figure*}
\centering
\subfigure[$t=2$] {\includegraphics[height=3cm,width=3cm]{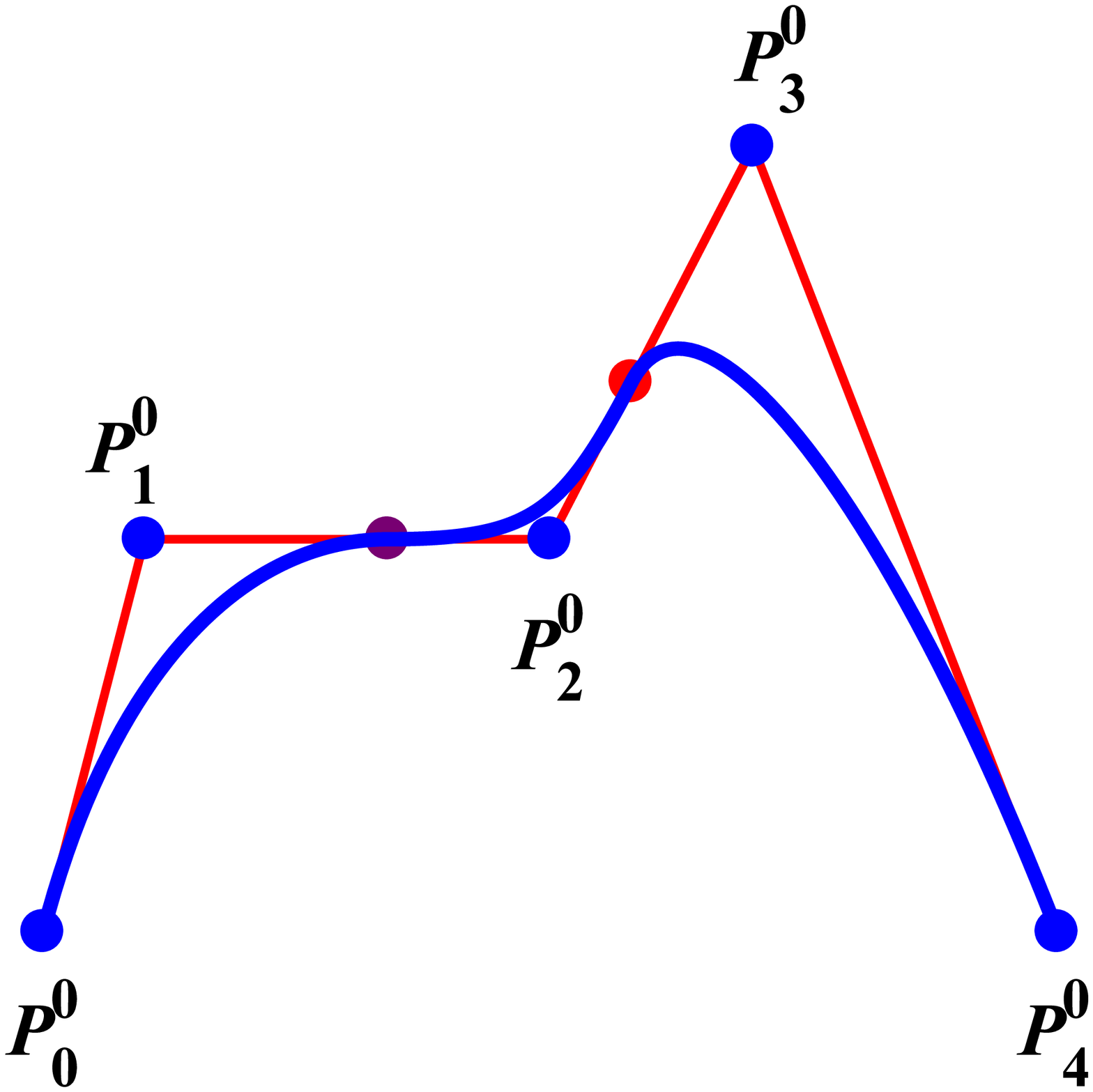}}\hspace{3ex}
\subfigure[$t=3$] {\includegraphics[height=3cm,width=3cm]{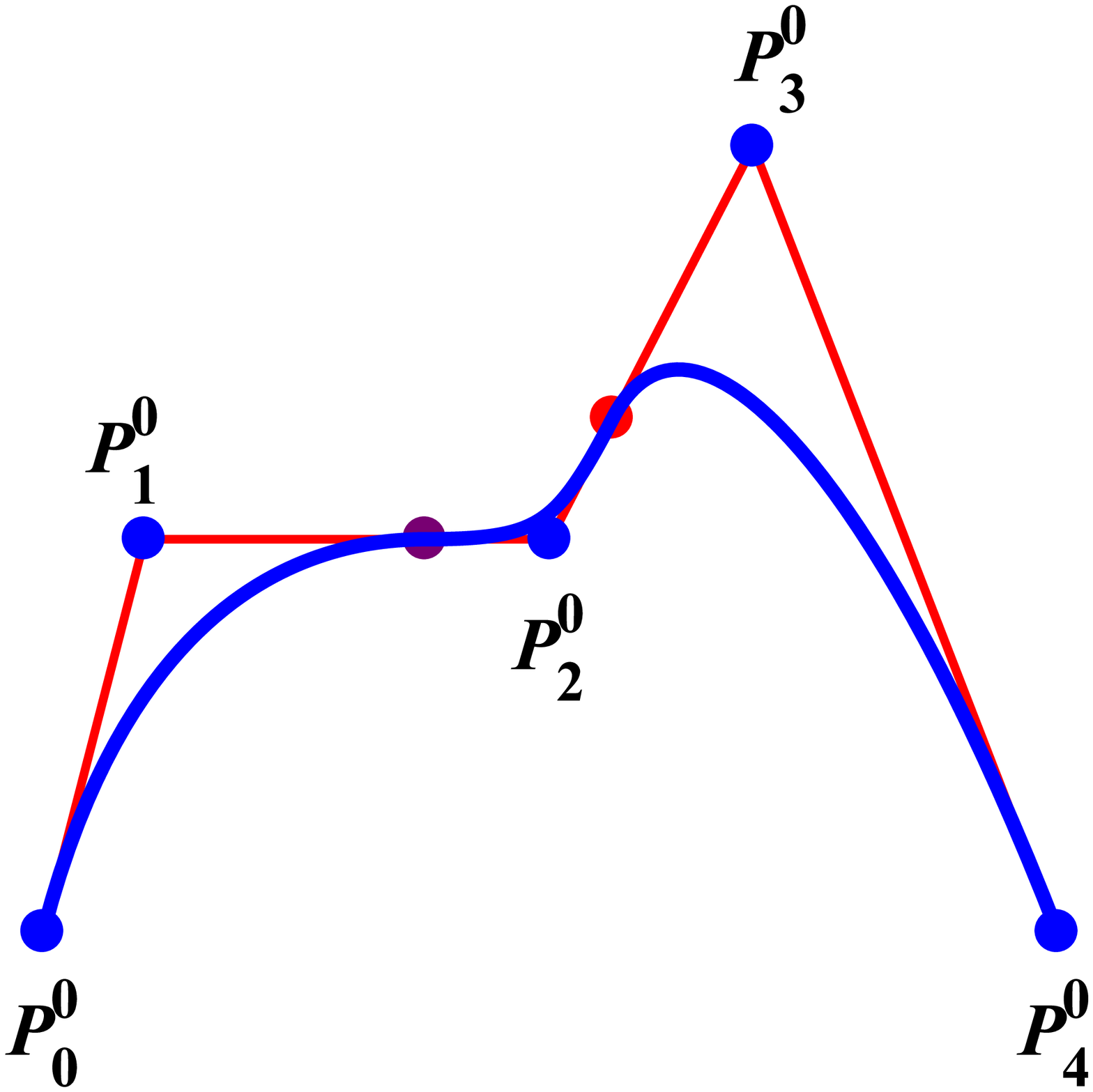}}\hspace{3ex}
\subfigure[$t=5$] {\includegraphics[height=3cm,width=3cm]{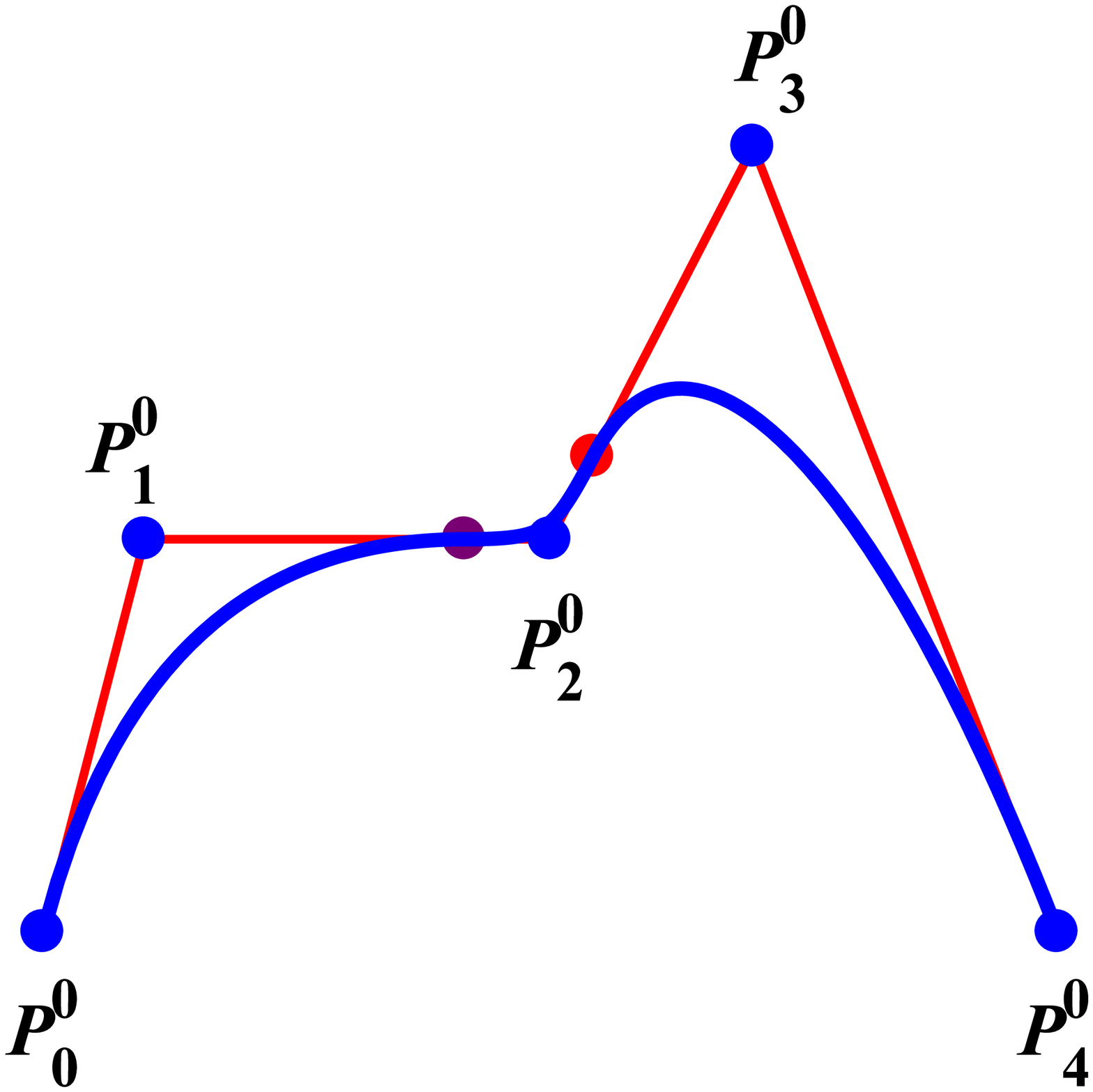}}\hspace{3ex}
\subfigure[$t=10$] {\includegraphics[height=3cm,width=3cm]{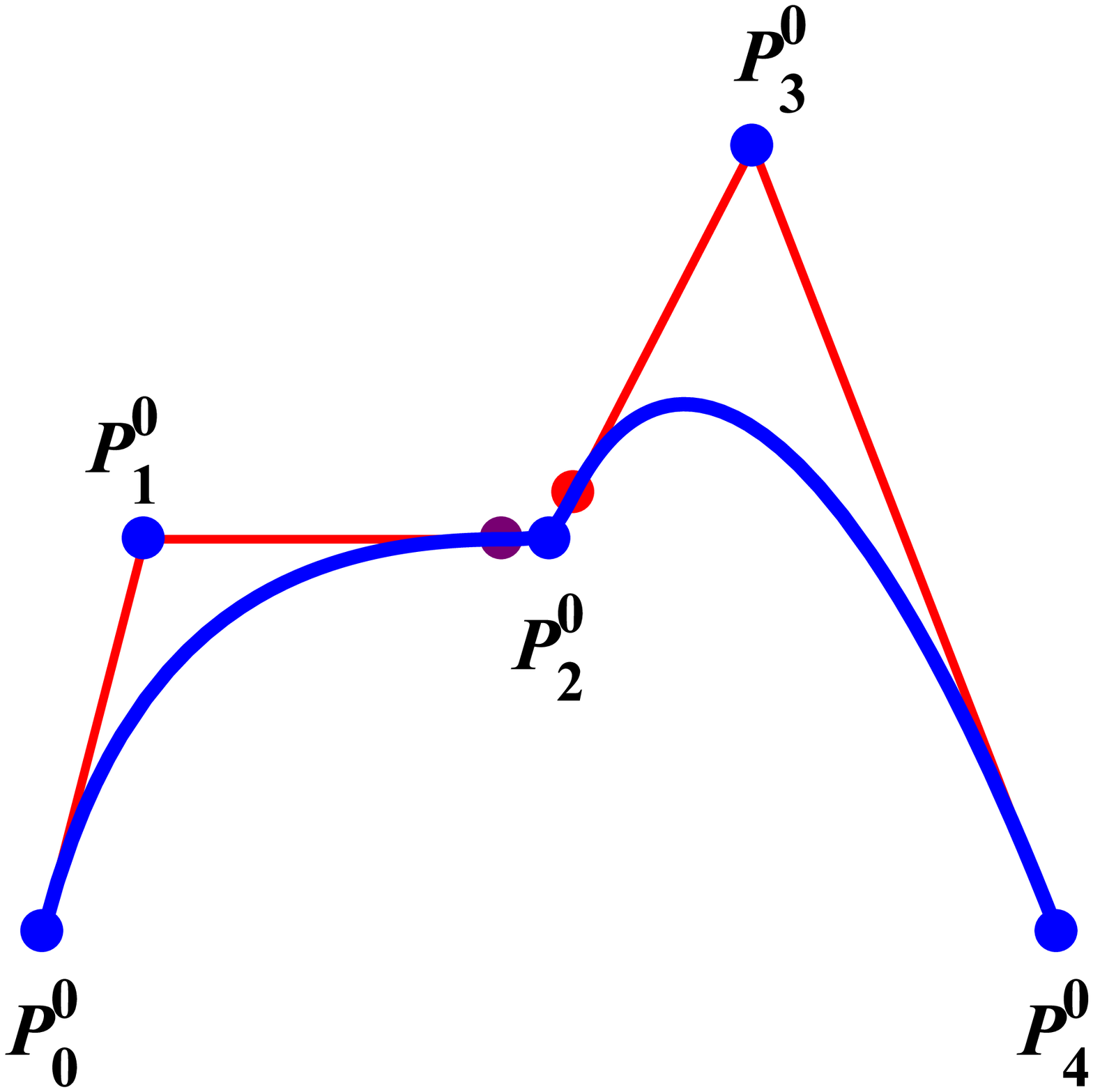}}
\caption{Toric degeneration of the quadric NURBS curve.}
\label{Fig7}
\end{figure*}
\end{example}

\begin{example}
\label{example2}
 \begin{figure}[h!]
\centering
\subfigure[NURBS curve] {\includegraphics[height=3.5cm,width=3.5cm]{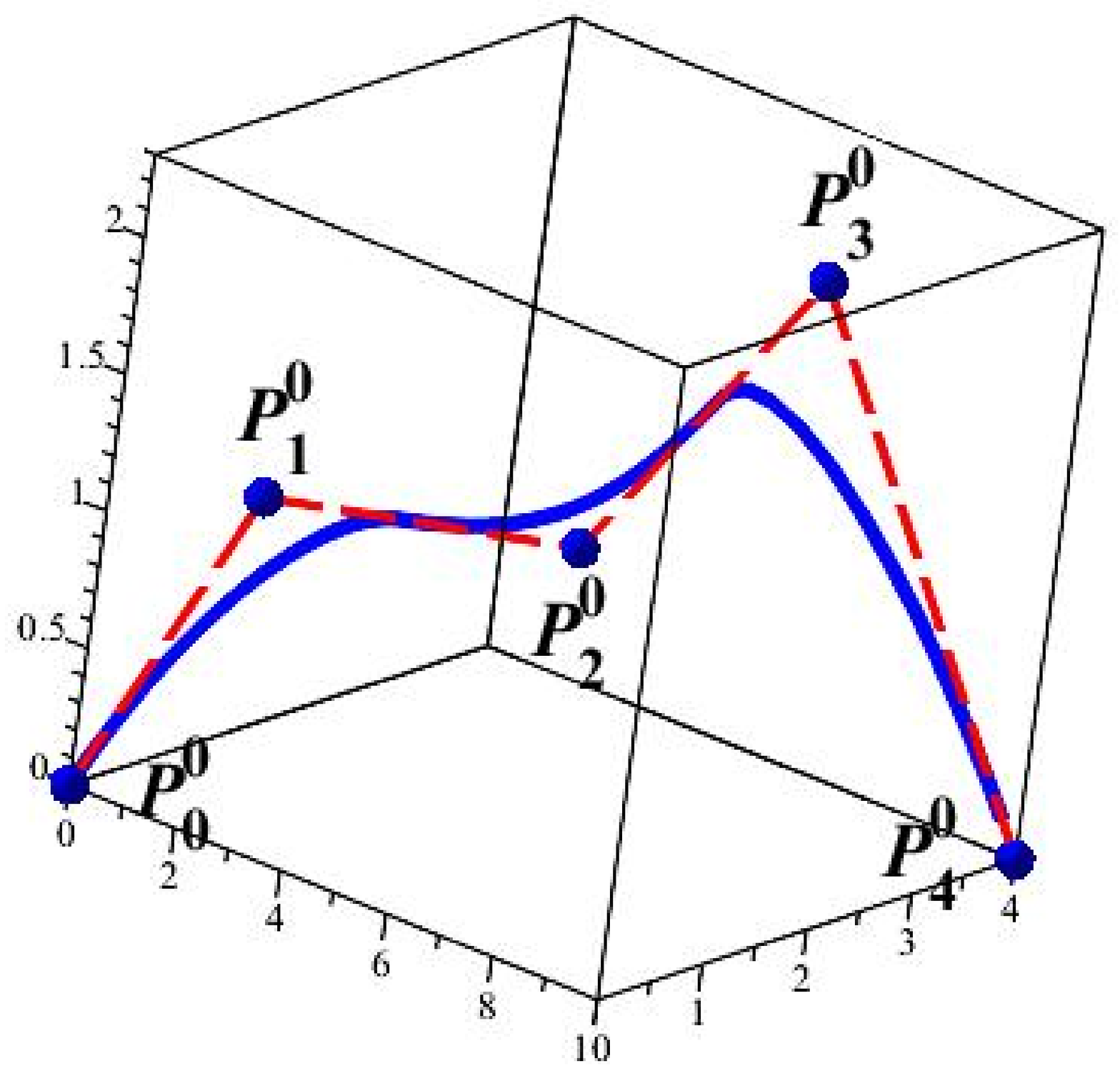}\label{Fig8a}}\hspace{3ex}
\subfigure[Regular control curve] {\includegraphics[height=3.5cm,width=3.5cm]{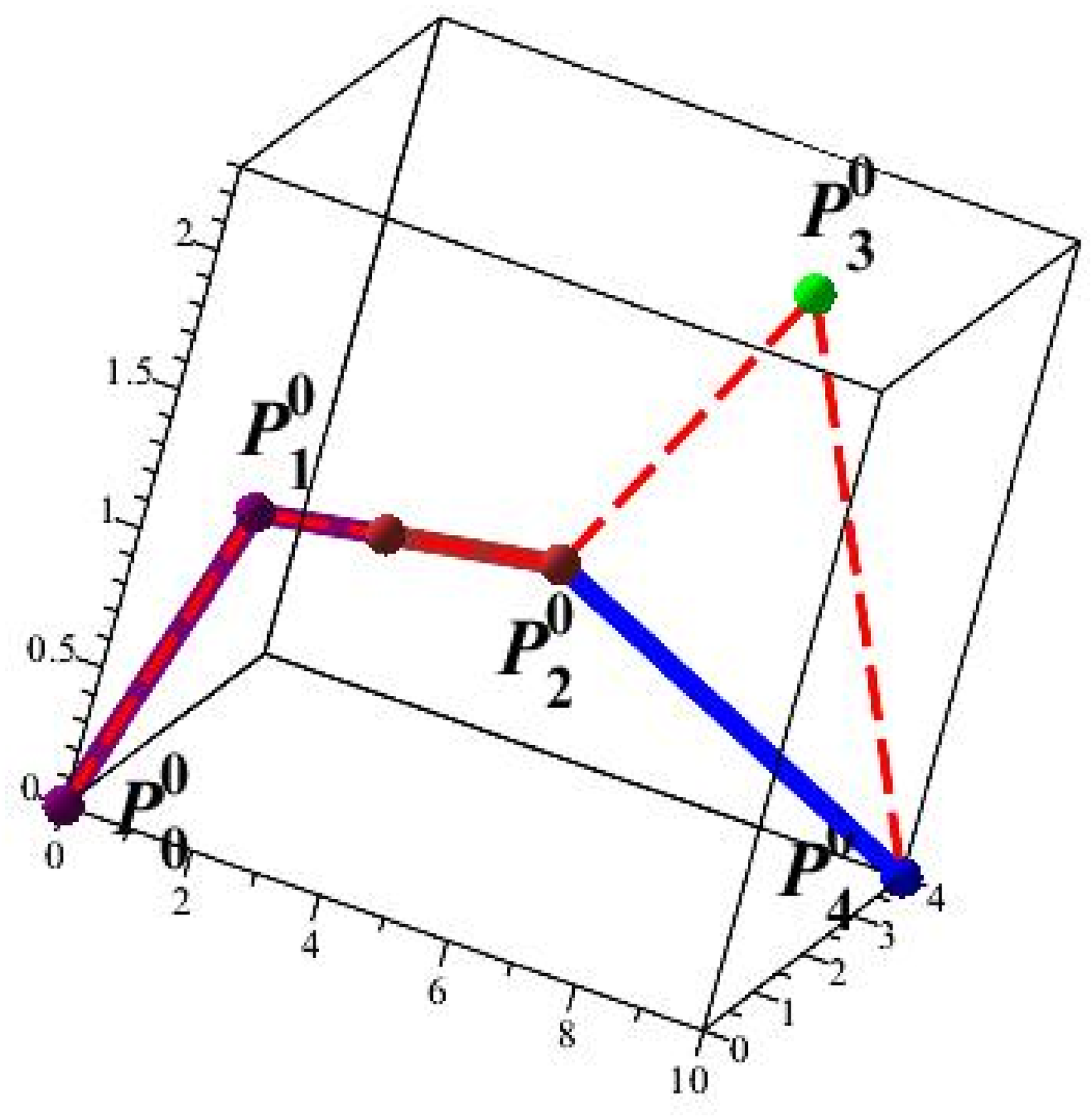}\label{Fig8b}}
\caption{The quadric NURBS curve and its regular control curve.}
\label{Fig8}
\end{figure}
\em{
Consider the quadratic  NURBS curve $\textbf{R}_{\mathcal{A},\omega,\mathcal{B}}$ in  Example $\ref{example1}$ with the control points $\mathcal{B}=\{\textbf{P}^0_0,\textbf{P}^0_1,\textbf{P}^0_2,\textbf{P}^0_3,\textbf{P}^0_4\}$ and weights $\omega=\{3,2,3,2,5\}$ (see Fig.~$\ref{Fig8a}$). Suppose that the lifting function $\lambda$ has the assignments $\{1,4,4,1,1\}$ at the lattice points of $\mathcal{A}=\{0,1,2,3,4\}$.
 After inserting the knots $\frac{1}{4}$ and $\frac{3}{4}$,  $\textbf{R}_{\mathcal{A},\omega_\lambda(t),\mathcal{B}}$ is converted to the union of three pieces of rational quadratic B\'{e}zier curves,   $\textbf{R}_{\bar{\mathcal{A}},\bar{\omega}_\lambda(t),\bar{\mathcal{B}}_\lambda(t)}=\bigcup_{m=1}^{3}\textbf{F}_{\mathcal{A}^m,\omega^m_{\lambda^m}(t),\mathcal{B}^m_{\lambda^m}(t)}$. The lifting function ${\lambda}$ induces the assignments on $\bar{\mathcal{A}}$ by $\{\{1,4,4\},\{4,4,4\},\{4,1,1\}\}$ and then derives a regular decomposition $\bar{S}_{{\lambda}}=\{\{\{0,1\},\{1,2\}\},\{\{2,3,4\}\},\{\{4,6\}\}\}$ of $\bar{\mathcal{A}}$.

Consider the regular control curve of $\textbf{F}_{\mathcal{A}^1,\omega^1_{\lambda^1}(t),\mathcal{B}^1_{\lambda^1}(t)}$ with the control points $\{\textbf{P}^2_0,\textbf{P}^2_1,\textbf{P}^2_2\}$, weights $\{\omega^2_0,\omega^2_1,\omega^2_2\}$ and lifting function $\lambda^1=\{1,4,4\}$.
Since $\lambda(1)=\lambda(2)=4$,  the weights $\omega^2_0=\omega^0_0=3$, $\omega^2_1=\omega^0_1=2$, $\lim_{t\rightarrow \infty }\omega^2_2=\frac{2}{3}\omega _1^0+\frac{1}{3}\omega _2^0=\frac{7}{3}$ and the control points $\textbf{P}^2_0=\textbf{P}^0_0$, $\textbf{P}^2_1=\textbf{P}^0_1$, $\lim_{t\rightarrow \infty }\textbf{P}^2_2=\frac{ \frac{2}{3}\omega _1^0\textbf{P}_1^0+\frac{1}{3}\omega_2^0\textbf{P}_2^0}{\frac{2}{3}\omega _1^0+\frac{1}{3}\omega _2^0}=\frac{ 4}{7}\textbf{P} _1^0+\frac{ 3}{7}\textbf{P} _2^0$.
The lifting function $\lambda^1=\{1,4,4\}$
induces a regular decomposition $S_{\lambda}^1=\{\{0,1\},\{1,2\}\}$ of $\mathcal{A}^1$.
For the subset $\{0,1\}$, we can construct a  linear B\'{e}zier curve by the control points $\{\textbf{P}^2_0,\textbf{P}^2_1\}=\{\textbf{P}^0_0,\textbf{P}^0_1\}$ and weights $\{\omega^2_0,\omega^2_1\}=\{3,2\}$. For the subset $\{1,2\}$, another linear B\'{e}zier curve constructed by the control points $\{\textbf{P}^2_1,\textbf{P}^2_2\}=\{\textbf{P}^0_1,\frac{ 4}{7}\textbf{P} _1^0+\frac{ 3}{7}\textbf{P} _2^0\}$ and weights $\{\omega^2_1,\omega^2_2\}=\{2,\frac{7}{3}\}$ is obtained. We set $\textbf{P}^*=\frac{ 4}{7}\textbf{P} _1^0+\frac{ 3}{7}\textbf{P} _2^0$, then the regular control curve of the first rational B\'{e}zier curve is the union of those two line segments $\overline{\textbf{P}^0_0\textbf{P}^0_1}\cup\overline{\textbf{P}^0_1\textbf{P}^*}$.

Consider the regular control curve of $\textbf{F}_{\mathcal{A}^2,\omega^2_{\lambda^2}(t),\mathcal{B}^2_{\lambda^2}(t)}$ with the control points $\{\textbf{P}^2_2,\textbf{P}^2_3,\textbf{P}^2_4\}$, weights $\{\omega^2_2,\omega^2_3,\omega^2_4\}$ and lifting function $\lambda^2=\{4,4,4\}$. Since $\lambda(1)=\lambda(2)=4$ and $\lambda(2)>\lambda(3)$,
the weights  $\lim_{t\rightarrow \infty }\omega^2_2=\frac{2}{3}\omega _1^0+\frac{1}{3}\omega _2^0=\frac{7}{3},\omega^2_3=\omega^0_2=3,\lim_{t\rightarrow \infty }\omega^2_4=\frac{1}{3}\omega_2^0=1$ and the control points $\lim_{t\rightarrow \infty }\textbf{P}^2_2=\frac{ 4}{7}\textbf{P} _1^0+\frac{ 3}{7}\textbf{P} _2^0$,$\textbf{P}^2_3=\textbf{P}^0_2$,$\lim_{t\rightarrow \infty }\textbf{P}^2_4=\textbf{P}^0_2$.
The lifting function $\lambda^2=\{4,4,4\}$ induces a regular decomposition $S_{\lambda}^2=\{\{2,3,4\}\}$ of $\mathcal{A}^2$. For the subset $\{2,3,4\}$, we can construct a rational quadratic  B\'{e}zier curve by the control points $\{\textbf{P}^2_2,\textbf{P}^2_3,\textbf{P}^2_4\}=\{\frac{ 4}{7}\textbf{P} _1^0+\frac{ 3}{7}\textbf{P} _2^0,\textbf{P}^0_2,\textbf{P}^0_2\}$ and weights $\{\omega^2_2,\omega^2_3,\omega^2_4\}$ $=\{\frac{7}{3},3,1\}$.
Since the control points $\textbf{P}^2_4$ goes to $\textbf{P}^2_3=\textbf{P}^0_2$ while $t$ goes to infinity and we set $\textbf{P}^*=\frac{ 4}{7}\textbf{P} _1^0+\frac{ 3}{7}\textbf{P} _2^0$, then the regular control curve degenerates into a line segment
  $\overline{\textbf{P}^*\textbf{P}^0_2}$.

Consider the regular control curve of $\textbf{F}_{\mathcal{A}^3,\omega^3_{\lambda^3}(t),\mathcal{B}^3_{\lambda^3}(t)}$ with the control points $\{\textbf{P}^2_4,\textbf{P}^2_5,\textbf{P}^2_6\}$, weights $\{\omega^2_4,\omega^2_5,\omega^2_6\}$ and lifting function $\lambda^3=\{4,1,1\}$.  Since $\lambda(2)>\lambda(3)$, the weights $\lim_{t\rightarrow \infty }\omega^2_4$ $=\frac{1}{3}\omega_2^0=1$, $\omega^2_5=\omega^0_3=2$, $\omega^2_6=\omega_4^0=5$ and the control points $\lim_{t\rightarrow \infty }\textbf{P}^2_4$ $=\textbf{P}^0_2$, $\textbf{P}^2_5=\textbf{P}^0_3$, $\textbf{P}^2_6=\textbf{P}^0_4$. The lifting function $\lambda^3=\{4,1,1\}$ induces a regular decomposition $S_{\lambda}^3=\{\{4,6\}\}$ of $\mathcal{A}^3$. For the subset $\{4,6\}$, we can construct a  linear B\'{e}zier curve by the control points $\{\textbf{P}^2_4,\textbf{P}^2_6\}$ $=\{\textbf{P}^0_2,\textbf{P}^0_4\}$ and weights $\{\omega^2_4,\omega^2_6\}$ $=\{1,5\}$.
Then the regular control curve of the first rational B\'{e}zier curve is the  line segment $\overline{\textbf{P}^0_2\textbf{P}^0_4}$.

Since $\textbf{P}^*$ is located in  line segment $\overline{\textbf{P}^0_1\textbf{P}^0_2}$, then the regular control curve of the quadratic  NURBS curve $\textbf{R}_{\mathcal{A},\omega,\mathcal{B}}$ is the union of  three line segments  $\overline{\textbf{P}^0_0\textbf{P}^0_1}\cup\overline{\textbf{P}^0_1\textbf{P}^0_2}\cup\overline{\textbf{P}^0_2\textbf{P}^0_4}$, which is shown in Fig.~$\ref{Fig8b}$.
}
\end{example}
\begin{example}
\label{example3}
\begin{figure}[h!]
\centering
\subfigure[NURBS curve] {\includegraphics[height=3cm,width=3cm]{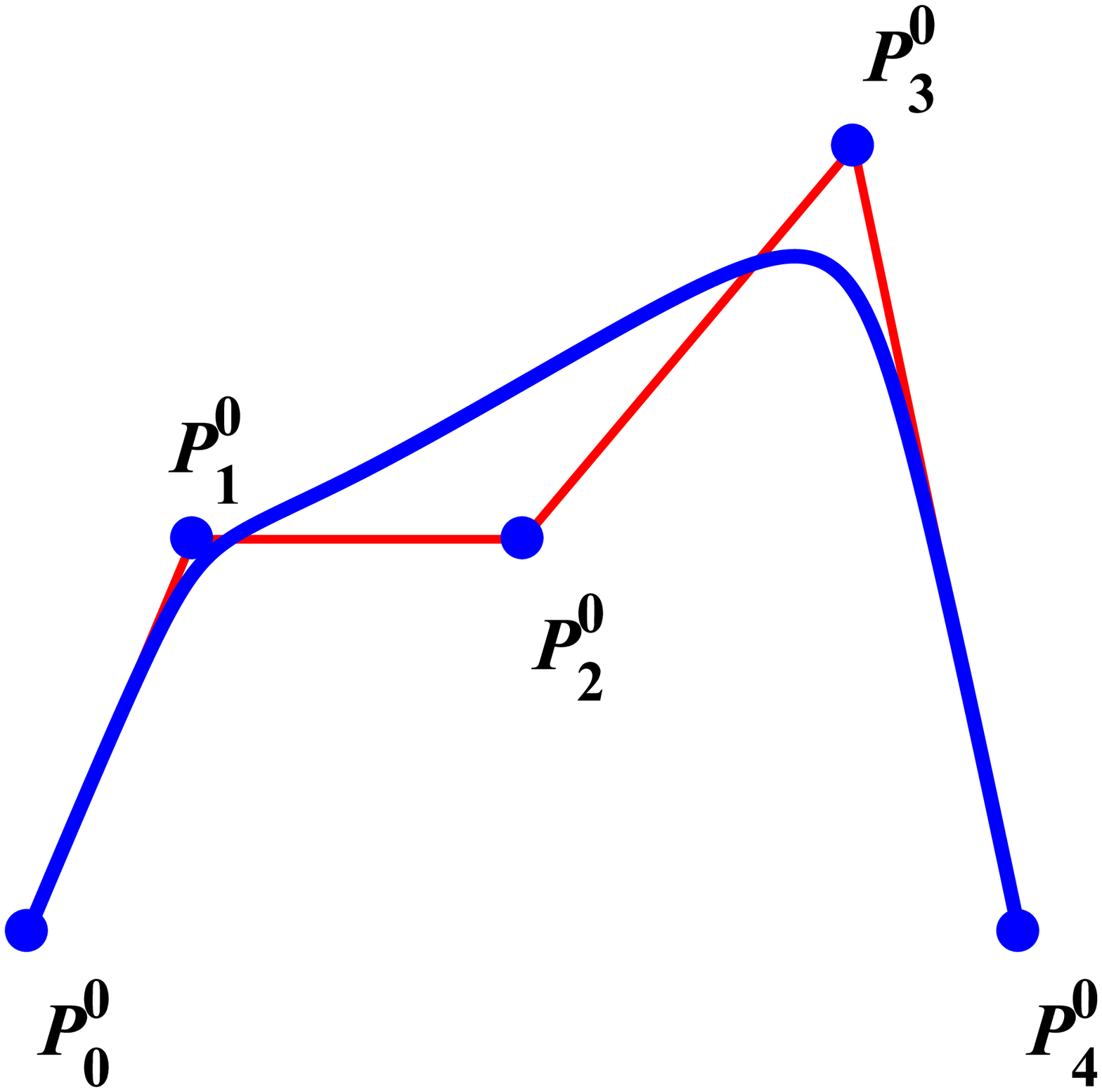}\label{Fig9a}}\hspace{3ex}
\subfigure[Regular control curve] {\includegraphics[height=3cm,width=3cm]{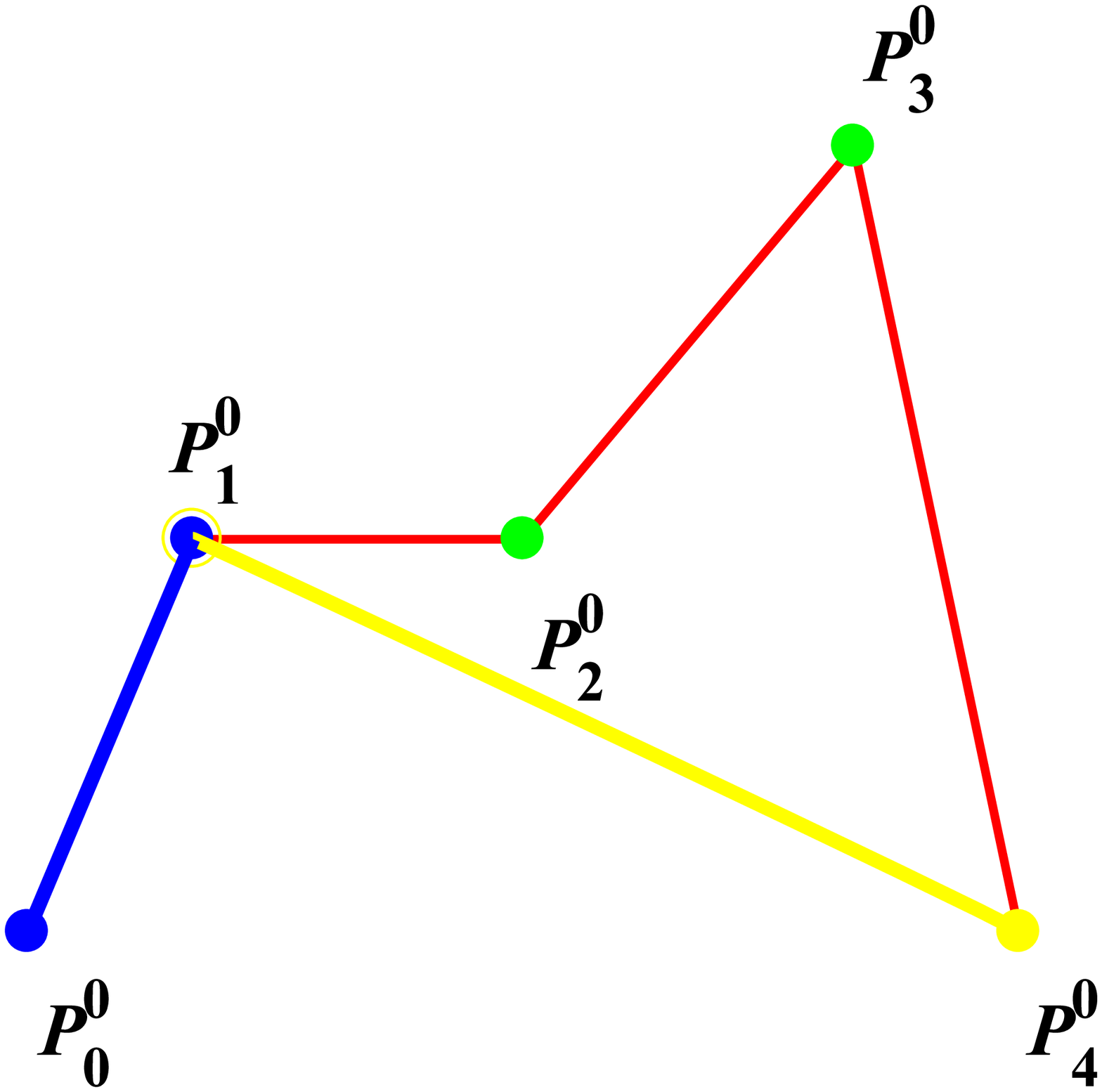}\label{Fig9b}}
\caption{The cubic NURBS curve and its regular control curve.}
\label{Fig9}
\end{figure}
\begin{figure*}
\centering
\subfigure[$t=2$] {\includegraphics[height=3cm,width=3cm]{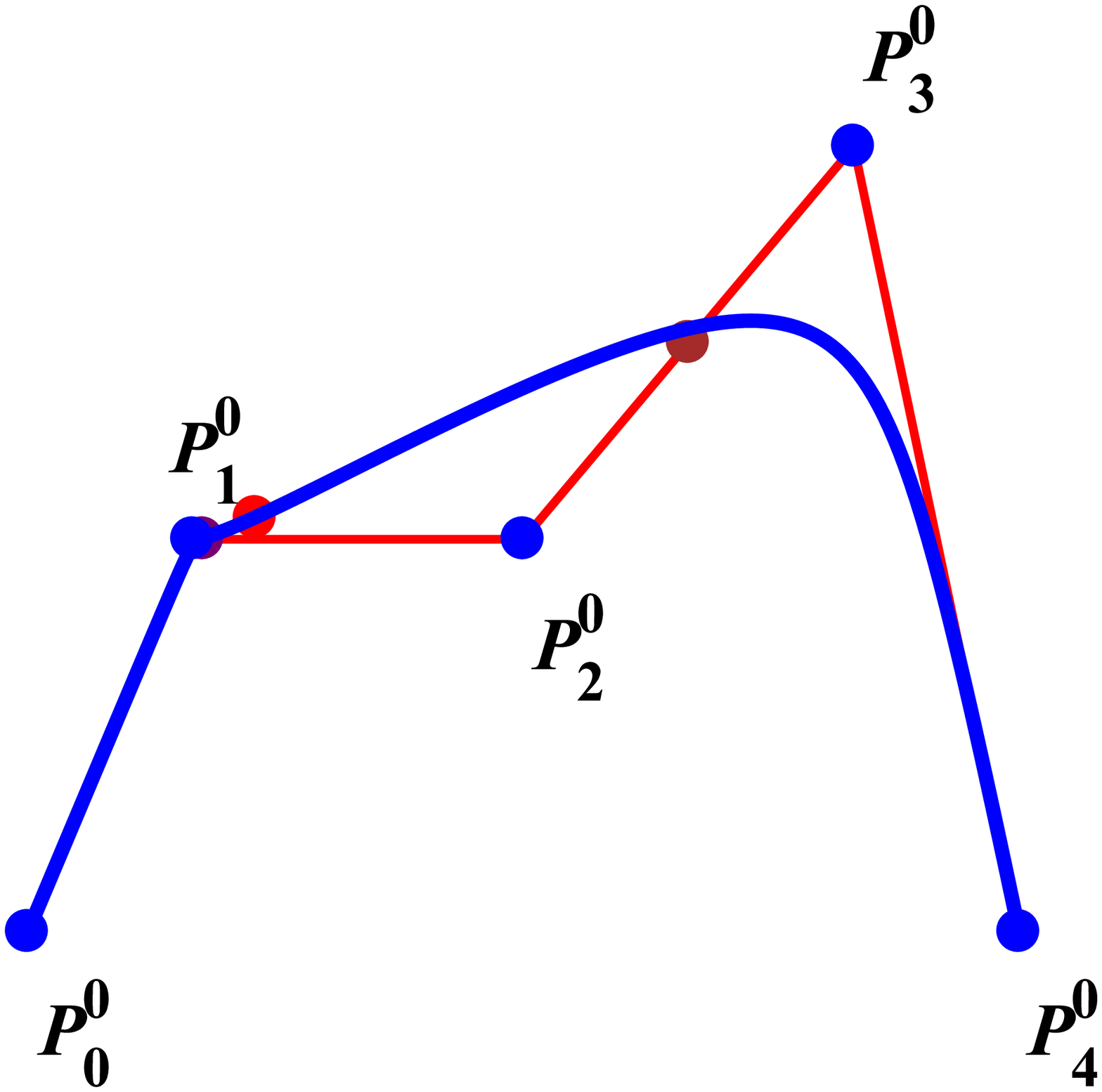}}\hspace{3ex}
\subfigure[$t=10$] {\includegraphics[height=3cm,width=3cm]{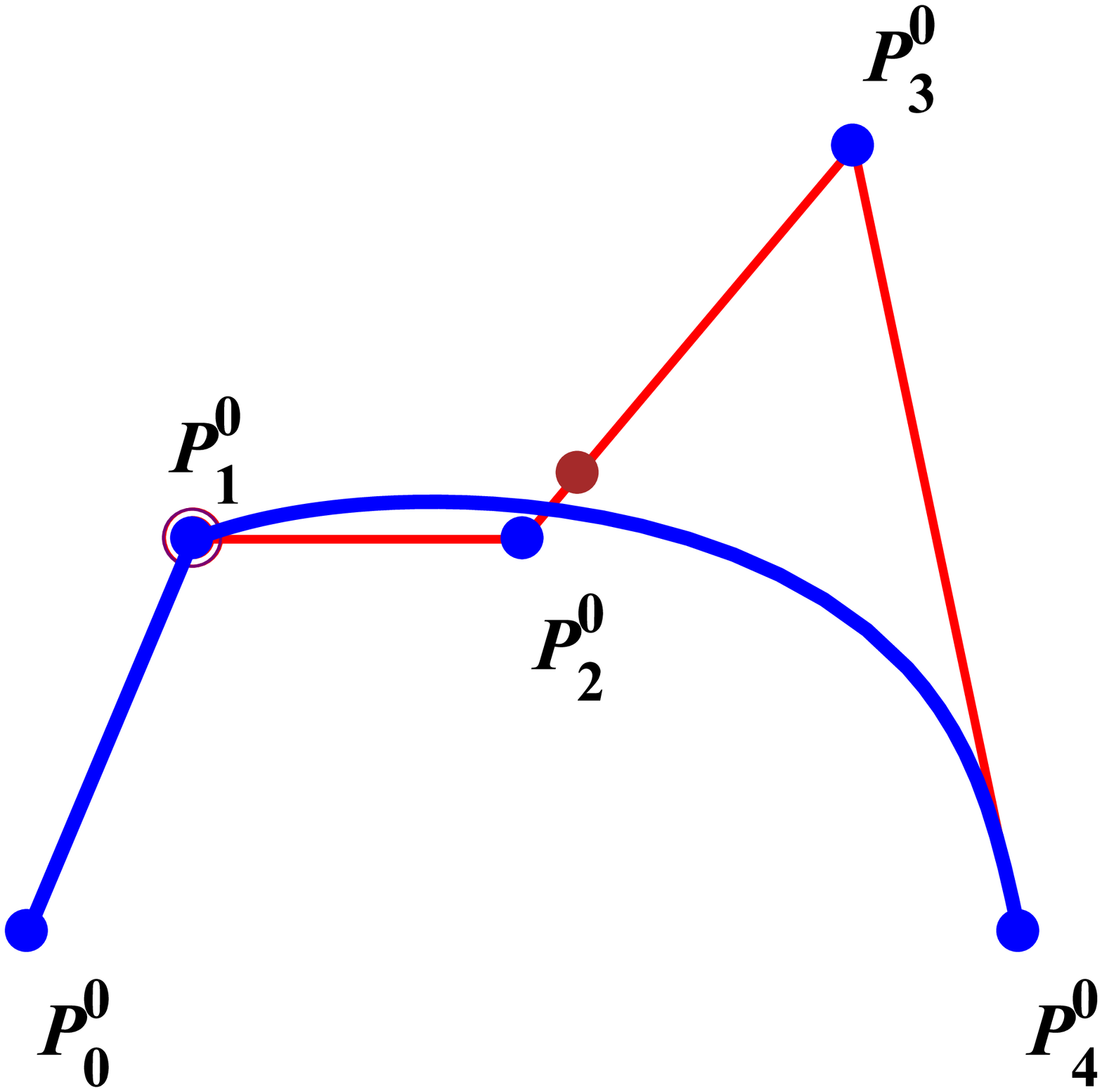}}\hspace{3ex}
\subfigure[$t=20$] {\includegraphics[height=3cm,width=3cm]{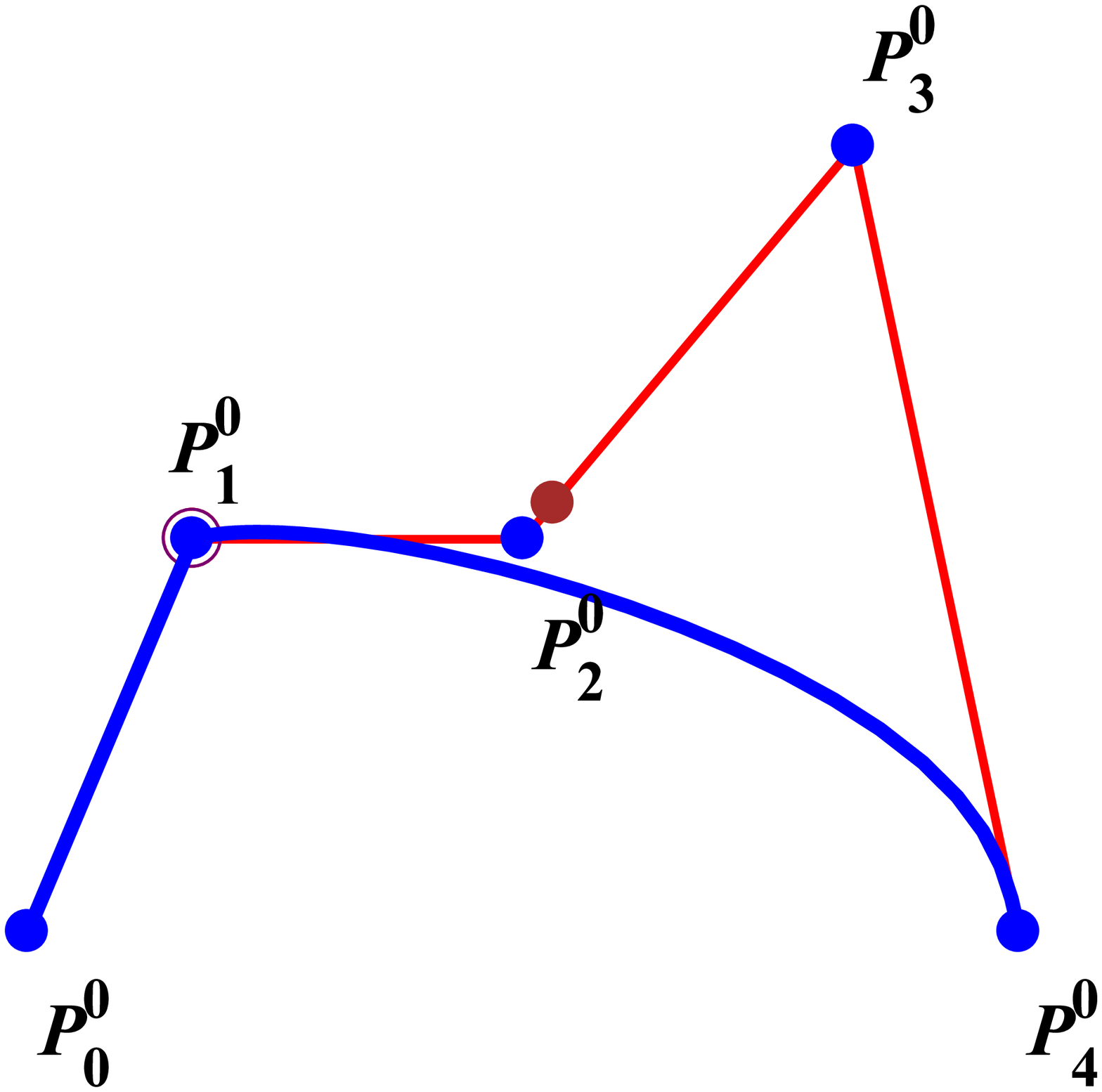}}\hspace{3ex}
\subfigure[$t=30$] {\includegraphics[height=3cm,width=3cm]{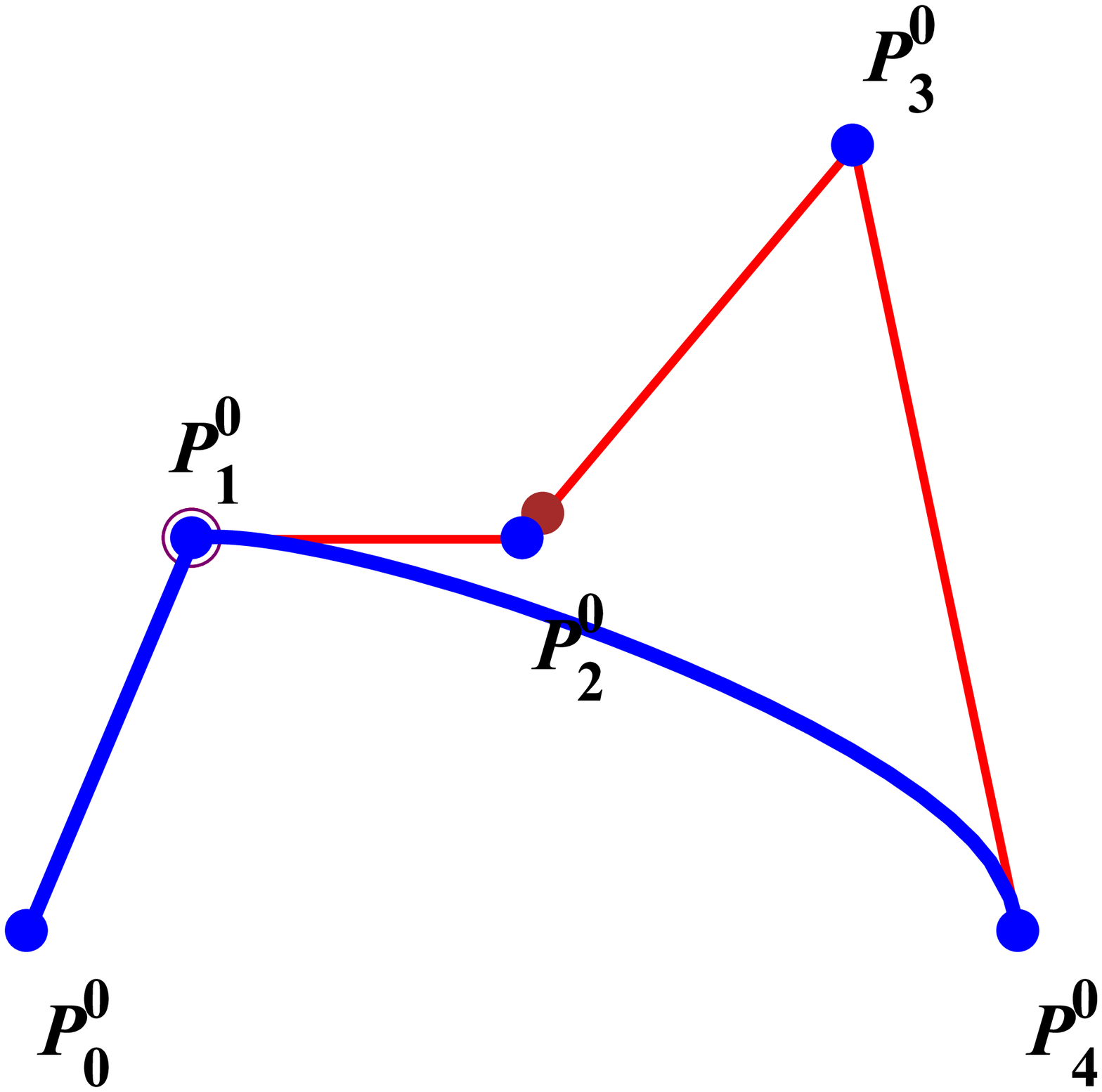}}
\caption{Toric degeneration of the cubic NURBS curve.}
\label{Fig10}
\end{figure*}
\em{
Let
\begin{eqnarray*}
\textbf{R}_{\mathcal{A},\omega,\mathcal{B}}(u)=\frac{\sum_{i=0}^{4}\omega_i^0 \textbf{P}_i^0N_{i,3}(u)}{\sum_{i=0}^{4}\omega_i^0N_{i,3}(u)}, u\in [0,1],
\end{eqnarray*}
be a cubic NURBS curve defined on knot vector $U^0=\{0,0,0,0,\frac{1}{3},1,1,1,1\}$ with the weights $\omega=\{1,4,1,4,1\}$ and  control points $\mathcal{B}=\{\textbf{P}^0_0,\textbf{P}^0_1,\textbf{P}^0_2,\textbf{P}^0_3,\textbf{P}^0_4\}$ (see Fig.~$\ref{Fig9a}$).
Suppose that the lifting function $\lambda$ has the assignments $\{1,4,2,1,1\}$ at the lattice points of $\mathcal{A}=\{0,1,2,3,4\}$. $\textbf{R}_{\mathcal{A},\omega_\lambda(t),\mathcal{B}}$ is converted to the union of two pieces of rational cubic  B\'{e}zier curves,  $\textbf{R}_{\bar{\mathcal{A}},\bar{\omega}_\lambda(t),\bar{\mathcal{B}}_\lambda(t)}=\bigcup_{m=1}^{2}\textbf{F}_{\mathcal{A}^m,\omega^m_{\lambda^m}(t),\mathcal{B}^m_{\lambda^m}(t)}$.
The weights and control points of $\textbf{R}_{\bar{\mathcal{A}},\bar{\omega}_\lambda(t),\bar{\mathcal{B}}_\lambda(t)}$ are
\begin{eqnarray*}
\bar{\omega}_\lambda(t)&&=\{\omega^2_0,\omega^2_1,\omega^2_2,\omega^2_3,\omega^2_4,\omega^2_5,\omega^2_6\}\\&&=\{t^{\lambda(0)}\omega^0_0,t^{\lambda(1)}\omega^0_1,\sum_{i=1}^{2}f_{1;2}^{i;2}t^{\lambda(i)}\omega _i^0,\\&&\sum_{i=1}^{3}f_{1;3}^{i;0}t^{\lambda(i)}\omega _i^0,\sum_{i=2}^{3}f_{2;3}^{i;1}t^{\lambda(i)}\omega _i^0,t^{\lambda(3)}\omega^0_3,t^{\lambda(4)}\omega^0_4\},
\end{eqnarray*}
\begin{eqnarray*}
\bar{\mathcal{B}}_\lambda(t)&&=\{\textbf{P}^2_0,\textbf{P}^2_1,\textbf{P}^2_2,\textbf{P}^2_3,\textbf{P}^2_4,\textbf{P}^2_5,\textbf{P}^2_6\}\\&&=\{\textbf{P}^0_0,\textbf{P}^0_1,\frac{ \sum_{i=1}^{2}f_{1;2}^{i;2}t^{\lambda(i)}\omega _i^0\textbf{P}_i^0}{\sum_{i=1}^{2}f_{1;2}^{i;2}t^{\lambda(i)}\omega _i^0},\\&&\frac{ \sum_{i=1}^{3}f_{1;3}^{i;0}t^{\lambda(i)}\omega _i^0\textbf{P}_i^0}{\sum_{i=1}^{3}f_{1;3}^{i;0}t^{\lambda(i)}\omega _i^0},\frac{ \sum_{i=2}^{3}f_{2;3}^{i;1}t^{\lambda(i)}\omega _i^0P_i^0}{\sum_{i=2}^{3}f_{2;3}^{i;1}t^{\lambda(i)}\omega _i^0},\textbf{P}^0_3,\textbf{P}^0_4\}.
\end{eqnarray*}
The lifting function ${\lambda}$ induces the assignments on $\bar{\mathcal{A}}$ by $\{\{1,4,4,4\},\{4,2,1,1\}\}$, and then derives a regular decomposition $\bar{S}_{{\lambda}}=\{\{\{0,1\},\{1,2,3\}\},\{\{3,6\}\}\}$ of $\bar{\mathcal{A}}$.

Consider the regular control curve of $\textbf{F}_{\mathcal{A}^1,\omega^1_{\lambda^1}(t),\mathcal{B}^1_{\lambda^1}(t)}$ with the control points $\{\textbf{P}^2_0,\textbf{P}^2_1,\textbf{P}^2_2,\textbf{P}^2_3\}$,  weights $\{\omega^2_0,\omega^2_1,\omega^2_2,\omega^2_3\}$ and lifting function $\lambda^1=\{1,4,4,4\}$. The lifting function $\lambda^1$ induces a regular decomposition $S_{\lambda}^1=\{\{0,1\},\{1,2,3\}\}$ of $\mathcal{A}^1=\{0,1,2,3\}$. The regular control curve  is the union of  a linear B\'{e}zier curve  by the control points $\{\textbf{P}^2_0,\textbf{P}^2_1\}$ and a rational quadratic B\'{e}zier curve  by the control points $\{\textbf{P}^2_1,\textbf{P}^2_2,\textbf{P}^2_3\}$ and their corresponding weights. Since $\lambda(1)>\lambda(2)>\lambda(3)$, the control points $\textbf{P}^2_0=\textbf{P}^0_0$, $\textbf{P}^2_1=\lim_{t\rightarrow \infty }\textbf{P}^2_2=\lim_{t\rightarrow \infty }\textbf{P}^2_3=\textbf{P}^0_1$, then the regular control curve degenerates into a line segment $\overline{\textbf{P}^0_0\textbf{P}^0_1}$.

Consider the regular control curve of $\textbf{F}_{\mathcal{A}^2,\omega^2_{\lambda^2}(t),\mathcal{B}^2_{\lambda^2}(t)}$ with the control points $\{\textbf{P}^2_3,\textbf{P}^2_4,\textbf{P}^2_5,\textbf{P}^2_6\}$, weights $\{\omega^2_3,\omega^2_4,\omega^2_5,\omega^2_6\}$ and lifting function $\lambda^2=\{4,2,1,1\}$. The lifting function $\lambda^2$ induces a regular decomposition $S_{\lambda}^2=\{\{3,6\}\}$ of $\mathcal{A}^2=\{3,4,5,6\}$. The regular control curve  is a  linear B\'{e}zier curve  by the control points  $\{\textbf{P}^2_3,\textbf{P}^2_6\}$. Since $\lambda(1)>\lambda(2)>\lambda(3)$, the control points  $\lim_{t\rightarrow \infty }\textbf{P}^2_3=\textbf{P}^0_1$, $\lim_{t\rightarrow \infty }\textbf{P}^2_4=\textbf{P}^0_2$, $\textbf{P}^2_5=\textbf{P}^0_3$, $\textbf{P}^2_6=\textbf{P}^0_4$, then the regular control curve degenerates into a line segment $\overline{\textbf{P}^0_1\textbf{P}^0_4}$.

Then the regular control curve  of the cubic  NURBS curve $\textbf{R}_{\mathcal{A},\omega,\mathcal{B}}$ is the union of two line segments  $\overline{\textbf{P}^0_0\textbf{P}^0_1}\cup \overline{\textbf{P}^0_1\textbf{P}^0_4}$ (see Fig.~$\ref{Fig9b}$). Fig.~$\ref{Fig10}$ shows the degeneration process of the curve with $t=2,10,20,30$, respectively.
}
\end{example}

\begin{example}
\label{example4}
\begin{figure}[h!]
\centering
\subfigure[NURBS curve] {\includegraphics[height=3cm,width=3cm]{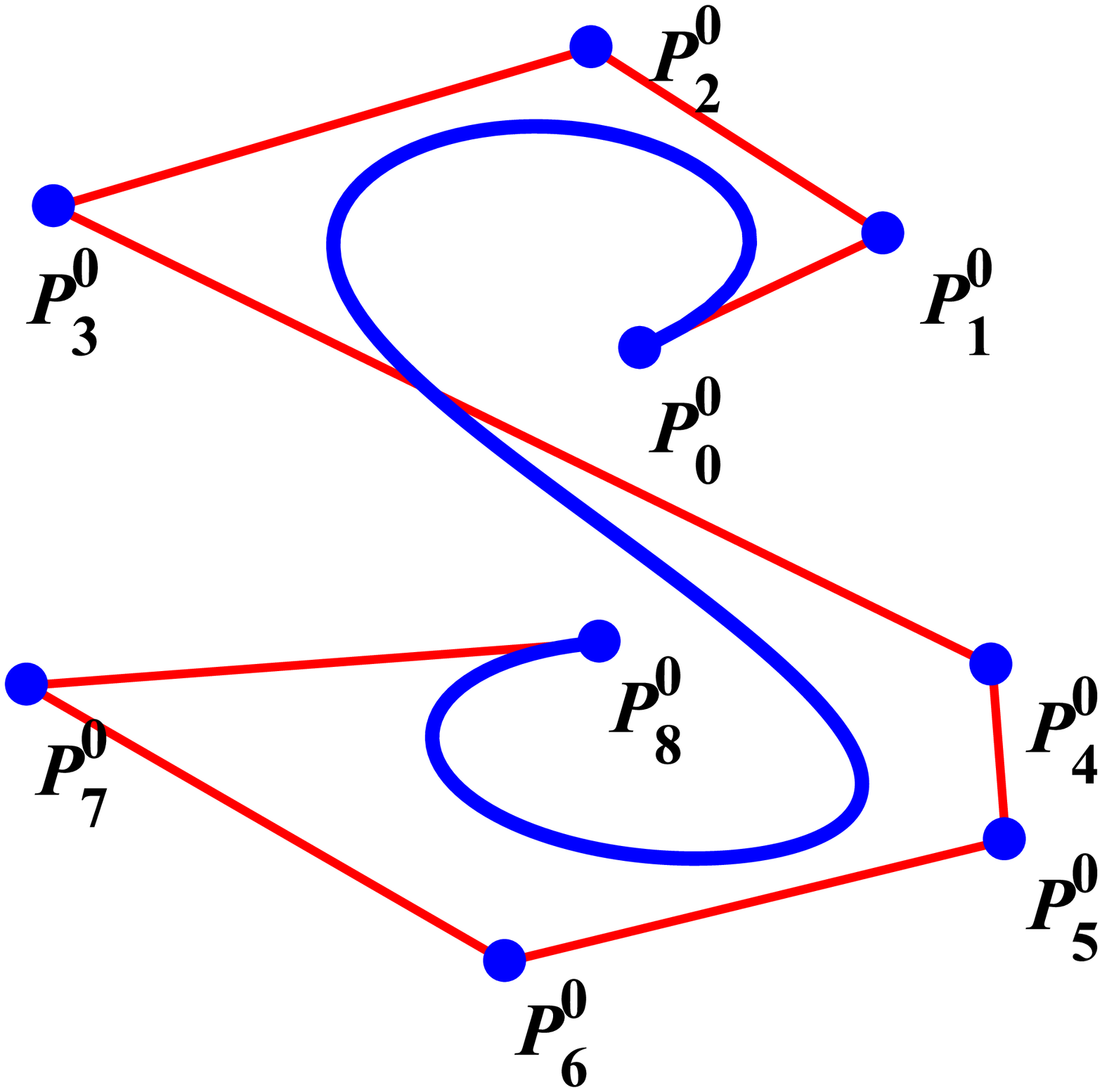}\label{Fig11a}}\hspace{3ex}
\subfigure[Regular control curve] {\includegraphics[height=3cm,width=3cm]{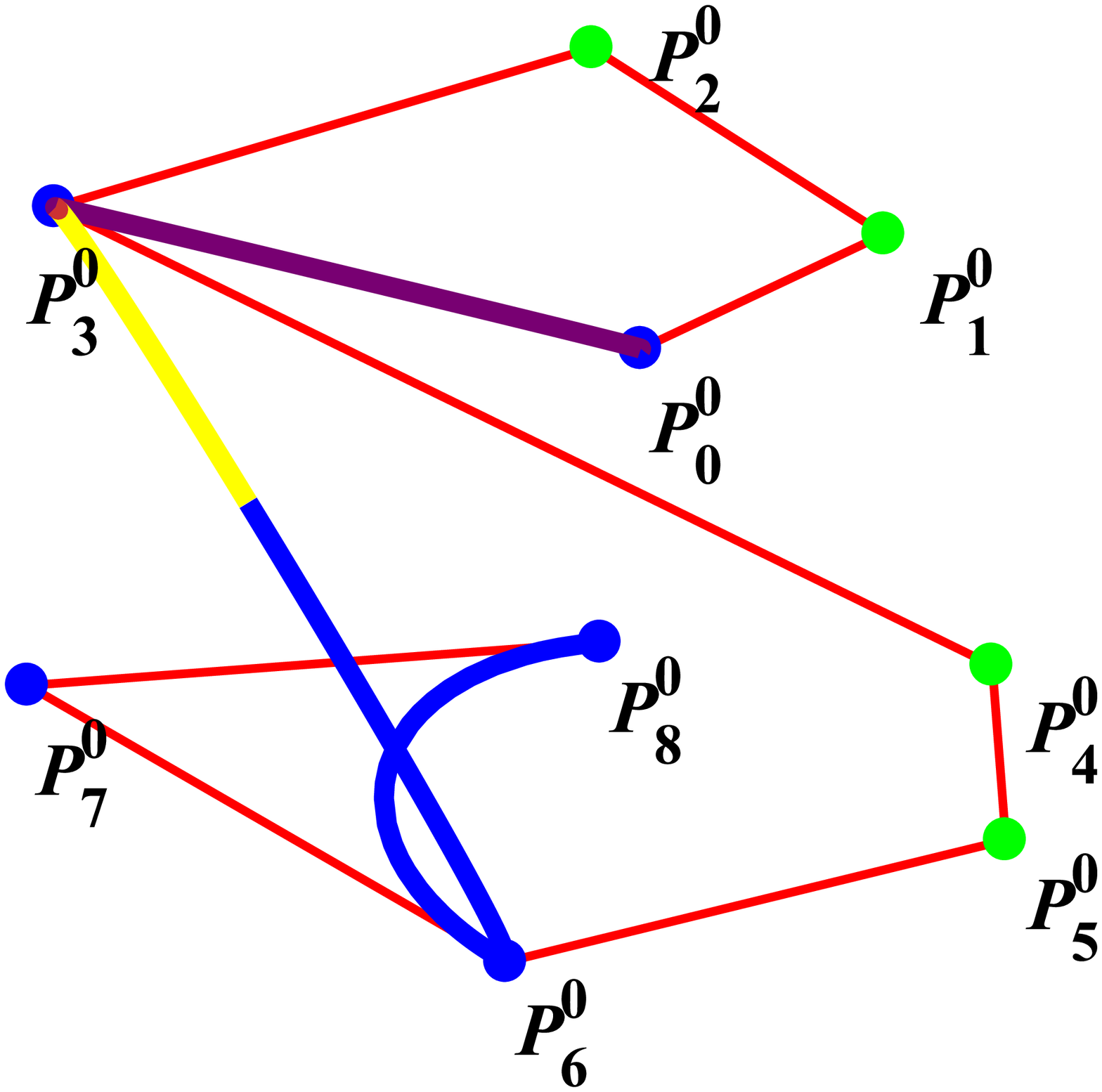}\label{Fig11b}}
\caption{The quintic NURBS curve and its regular control curve.}
\label{Fig11}
\end{figure}
\begin{figure*}
\centering
\subfigure[$t=2$] {\includegraphics[height=3cm,width=3cm]{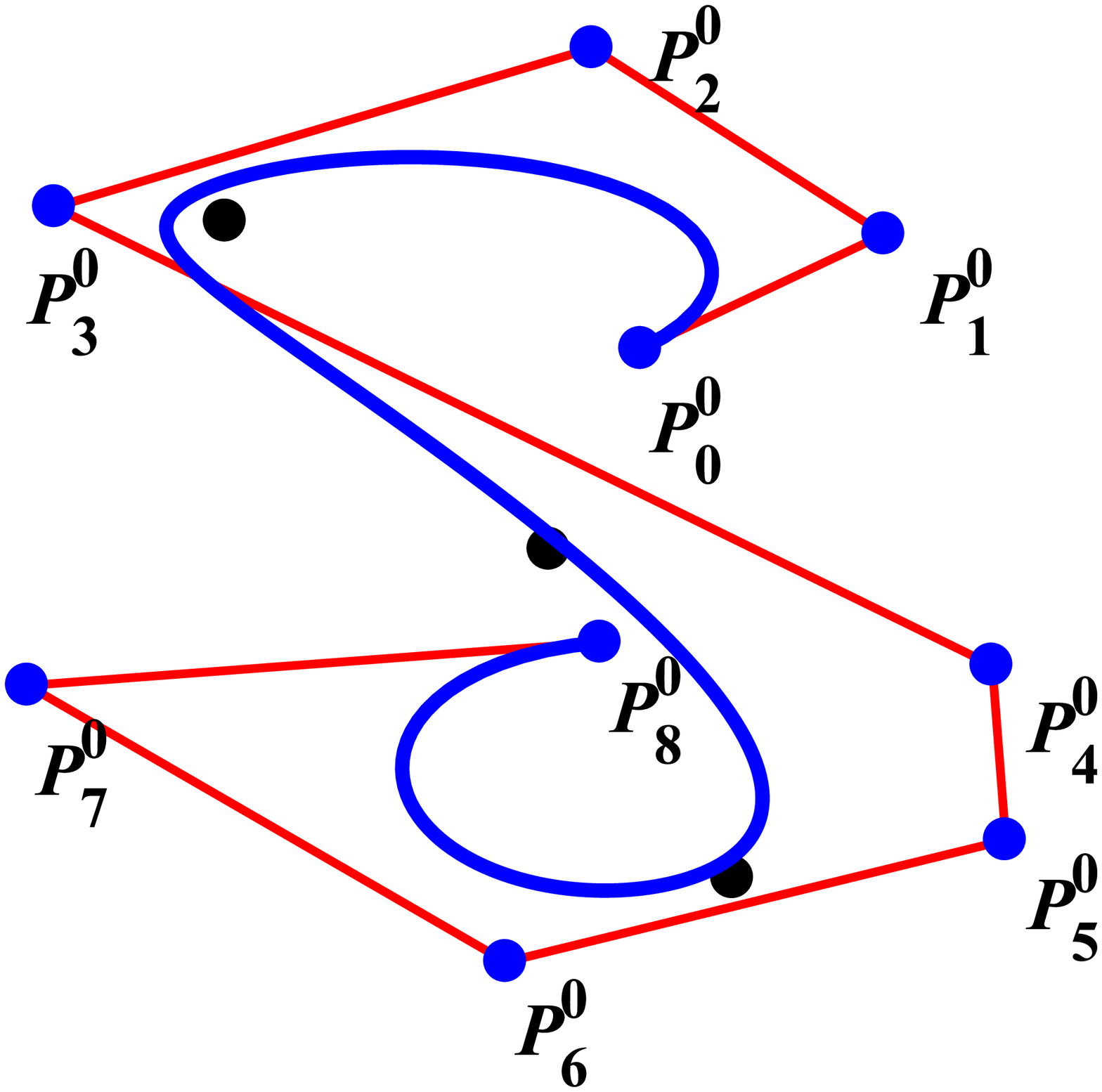}}\hspace{3ex}
\subfigure[$t=5$] {\includegraphics[height=3cm,width=3cm]{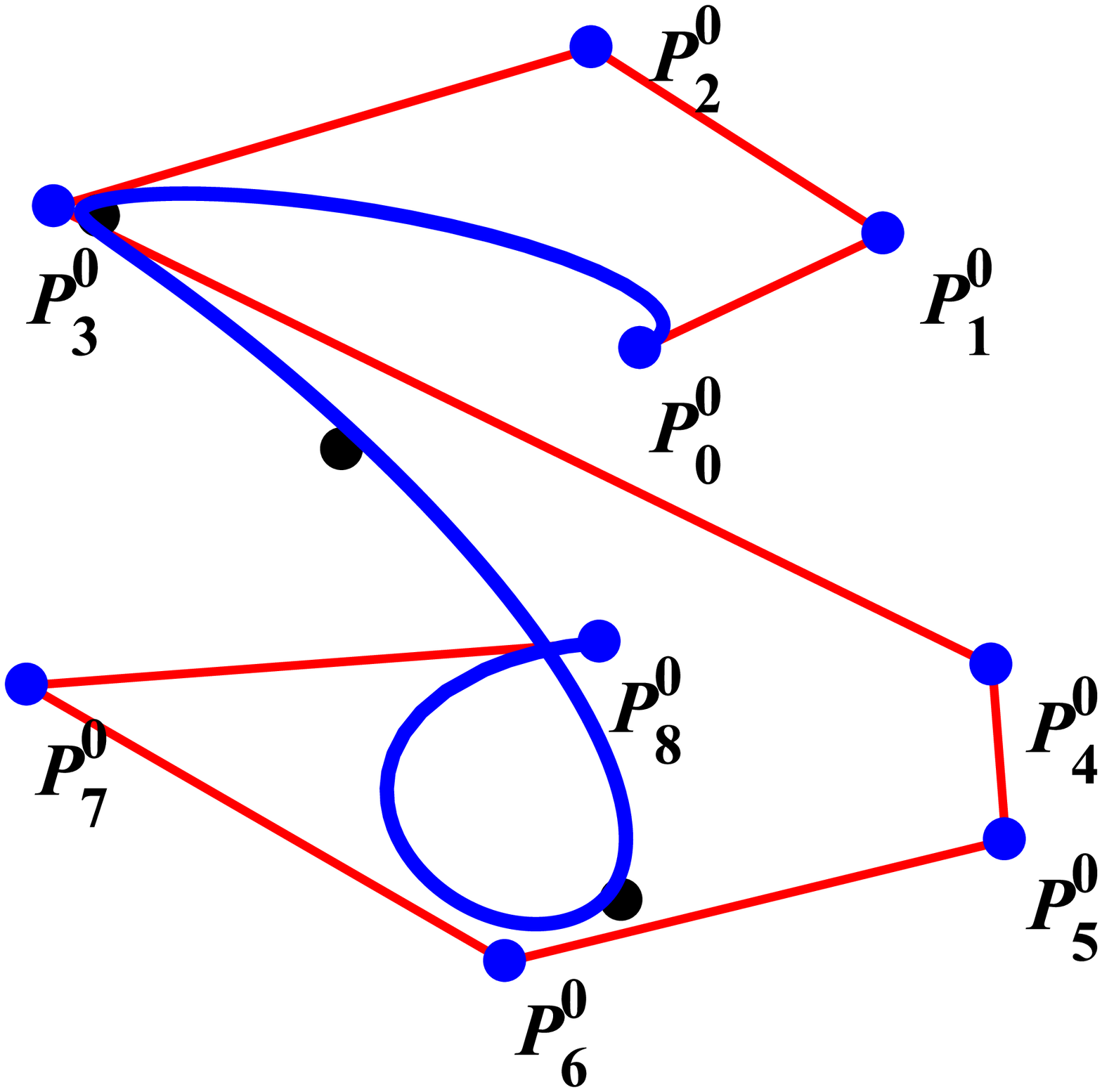}}\hspace{3ex}
\subfigure[$t=10$] {\includegraphics[height=3cm,width=3cm]{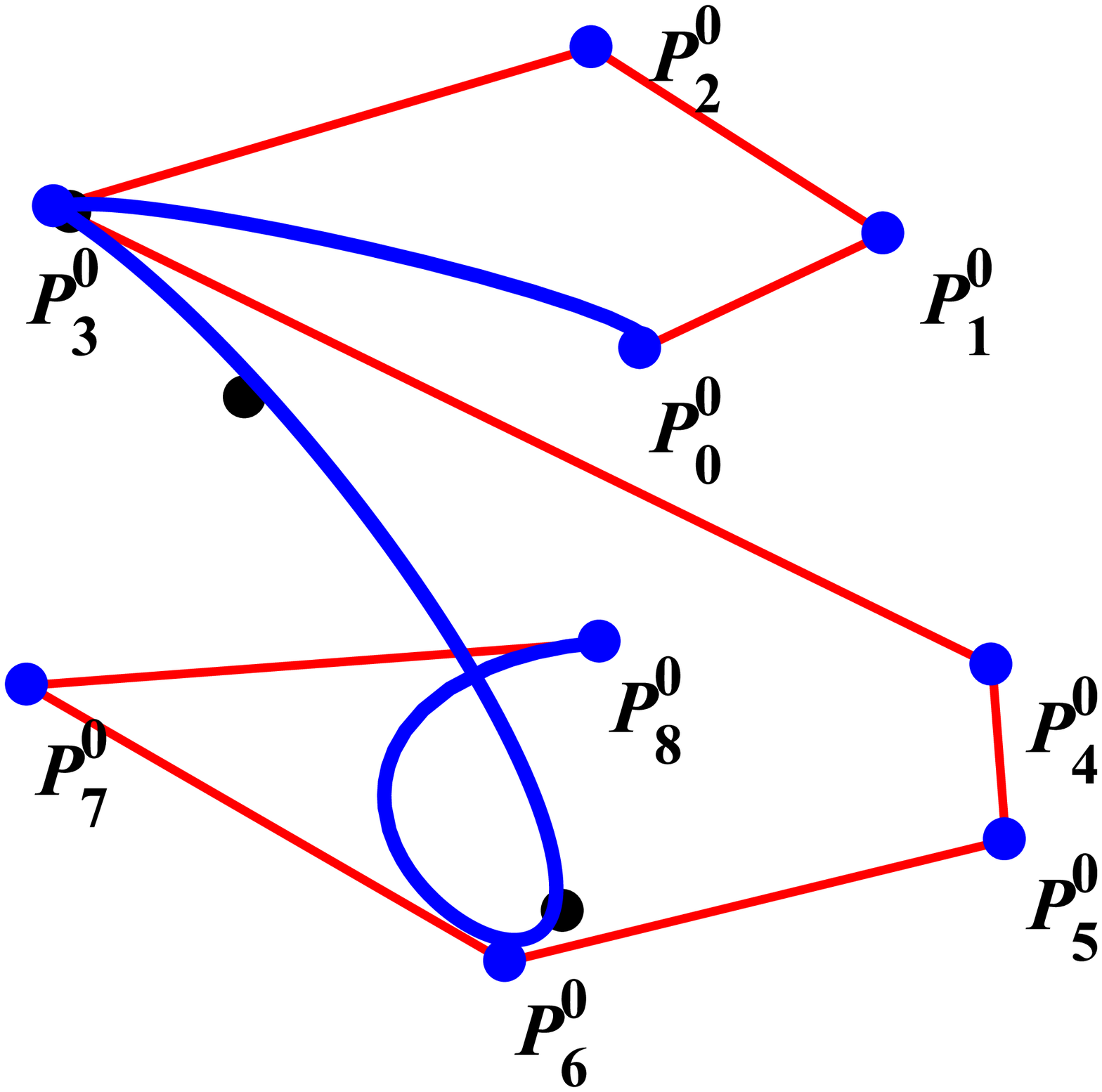}}\hspace{3ex}
\subfigure[$t=20$] {\includegraphics[height=3cm,width=3cm]{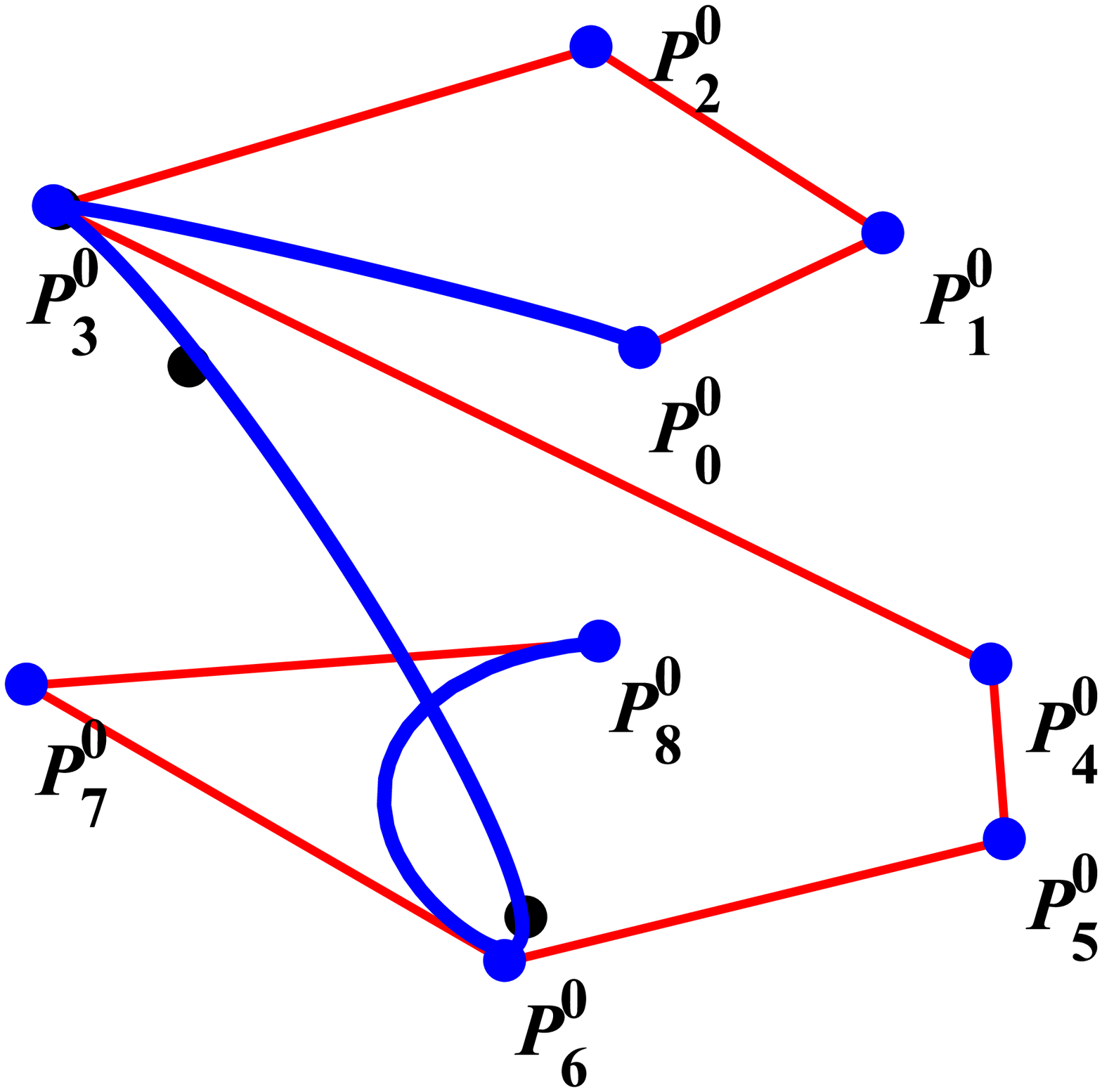}}
\caption{Toric degeneration of the quintic NURBS curve.}
\label{Fig12}
\end{figure*}
\em{
Let
\begin{eqnarray*}
\textbf{R}_{\mathcal{A},\omega,\mathcal{B}}(u)=\frac{\sum_{i=0}^{8}\omega_i^0 \textbf{P}_i^0N_{i,5}(u)}{\sum_{i=0}^{8}\omega_i^0N_{i,5}(u)}, u\in [0,1],
\end{eqnarray*}
be a quintic NURBS curve defined on knot vector
\begin{eqnarray*}
U^0=\{0,0,0,0,0,0,\frac{1}{4},\frac{1}{3},\frac{1}{2},1,1,1,1,1,1\}
\end{eqnarray*}
 with  the weights $\omega=\{1, 2, 3, 2, 1, 3, 2, 1, 2\}$ and control points $\mathcal{B}=\{\textbf{P}^0_0,\textbf{P}^0_1,\textbf{P}^0_2,\textbf{P}^0_3,\textbf{P}^0_4,\textbf{P}^0_5,\textbf{P}^0_6,\textbf{P}^0_7,\textbf{P}^0_8\}$, the  curve is shown in Fig.~$\ref{Fig11a}$.
Suppose that the lifting function $\lambda=\{2, 1, 1, 3, 1, 2, 3, 2, 1\}$,
the regular control curve of the quintic  NURBS curve after the degeneration is shown in Fig.~$\ref{Fig11b}$, which is the union of a rational quadratic B\'{e}zier curve and two line segments. Fig.~$\ref{Fig12}$ shows the degeneration process of the curve with $t=2,5,10,20$, respectively.
}
\end{example}

In the next three examples, we indicate the application of our results for shape deformation. Through the toric degeneration of NURBS curve, if a lifting function is given, then the limit of NURBS curve $\textbf{R}_{\mathcal{A},\omega_\lambda(t),\mathcal{B}}$ is determined. It means that if we choose the lifting functions properly, then the original curve (composed of NURBS curves) can be deformed to the target curve (composed of regular control curves of the NURBS curves).  Furthermore, our results also point out the potential application for computer animation.

\begin{example}\label{example7}
\begin{figure}[h!]
\centering
\subfigure[A wooden club] {\includegraphics[height=3cm,width=3cm]{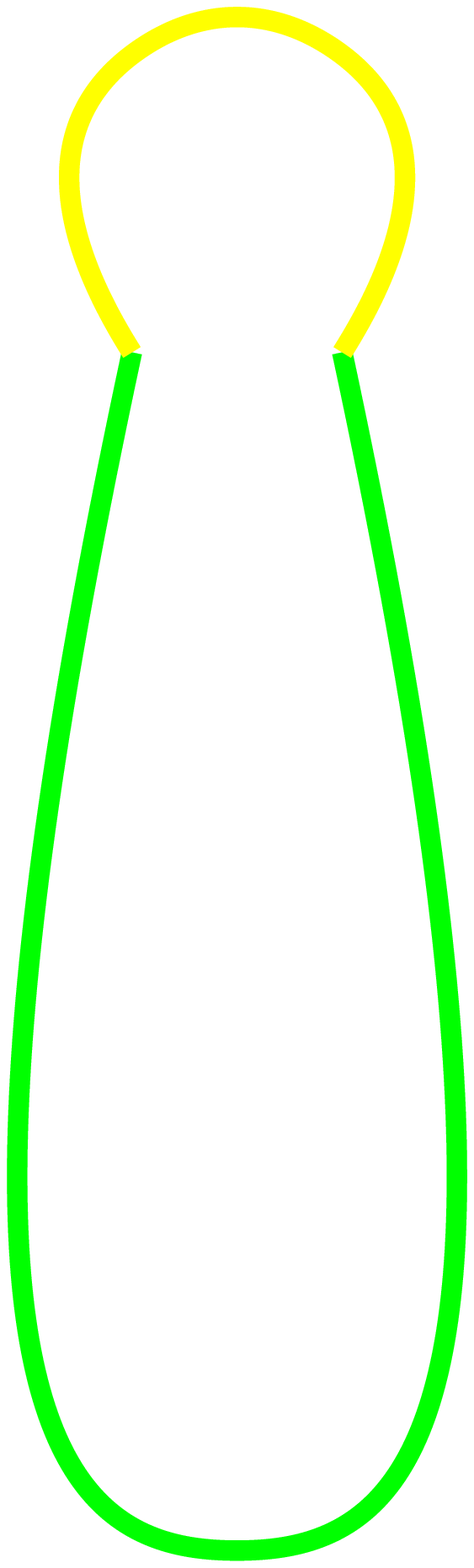}\label{Fig13a}}
\subfigure[$t=5$] {\includegraphics[height=3cm,width=3cm]{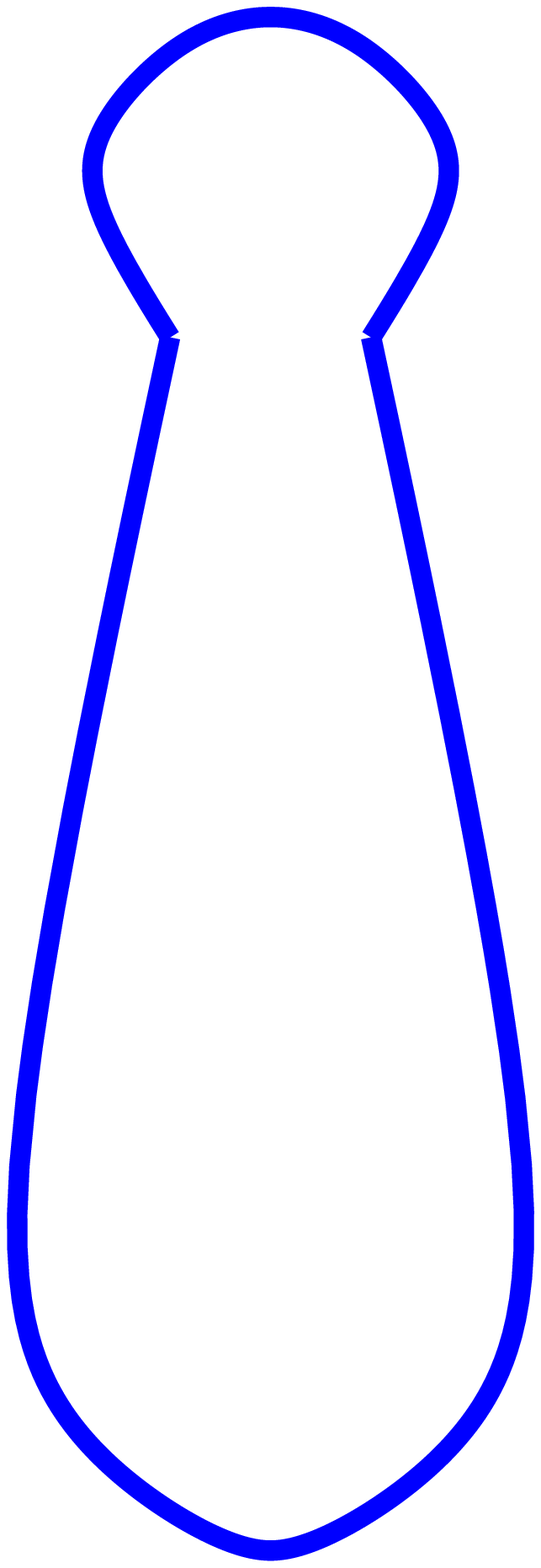}}
\subfigure[$t=10$] {\includegraphics[height=3cm,width=3cm]{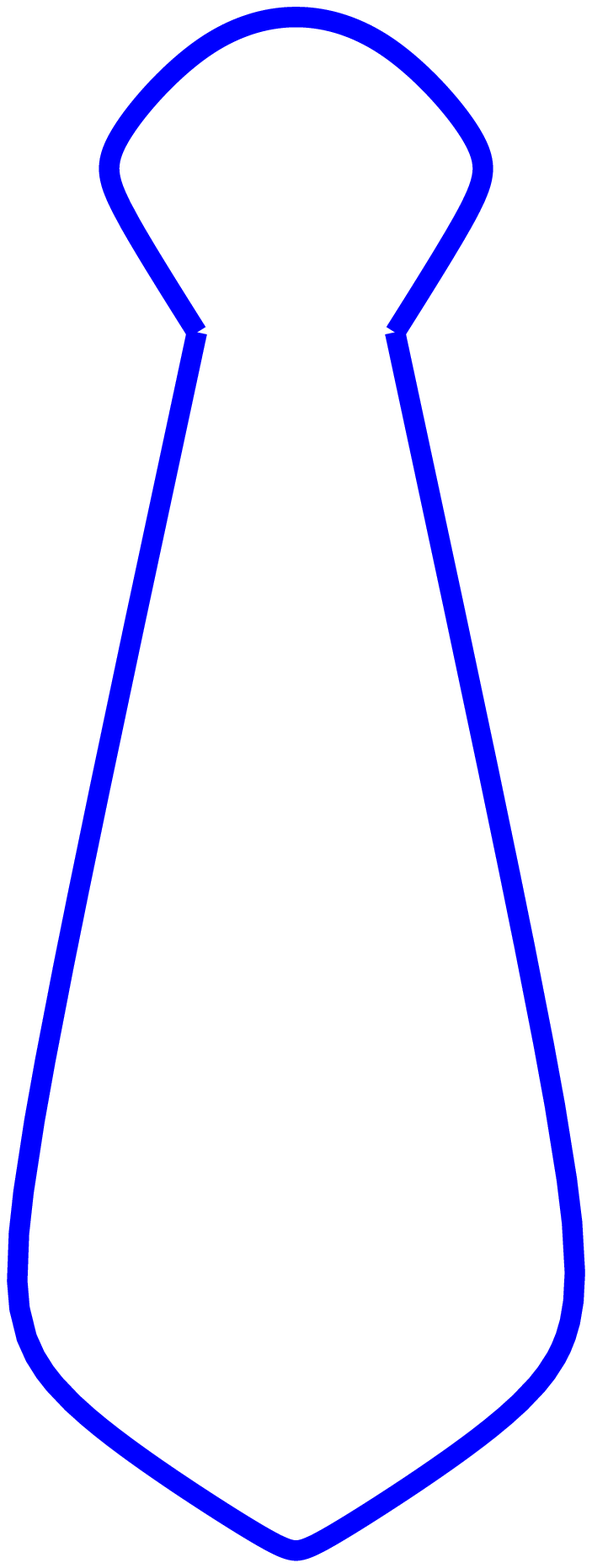}}
\subfigure[$t=20$] {\includegraphics[height=3cm,width=3cm]{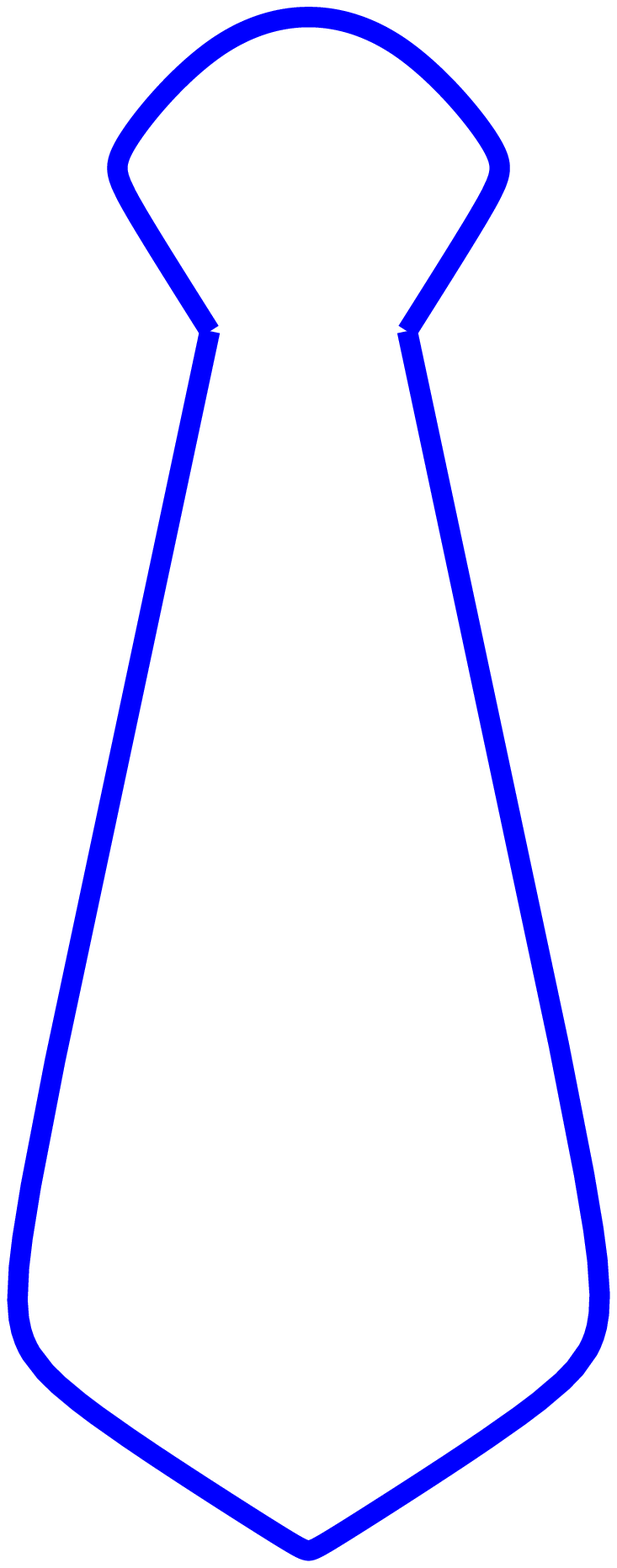}\label{Fig13e}}
\caption{The deformation process of a wooden club.}
\label{Fig13}
\end{figure}
\emph{Figure $\ref{Fig13}$ shows the shape deformation of a wooden club to a tie by using toric degenerations of NURBS curves.
 The wooden club $($see Figure  $\ref{Fig13a})$ is composed of two pieces of NURBS curves on  knot vectors  $\{0,0,0,0,\frac{1}{2},1,1,1,1\}$ and $\{0,0,0,0,\frac{1}{3},\frac{2}{3},1,1,1,1\}$.
The lifting functions correspond to these two pieces of NURBS curves are $\lambda_1=\{1, 2, 2, 2, 2, 1\}$ and $\lambda_2=\{1, 3, 4, 3, 1\}$.
 Then the limit of the wooden club is a tie $($see Figure  $\ref{Fig13e})$.}
\end{example}

\begin{example}\label{example8}
\begin{figure*}
\centering
\subfigure[A vase] {\includegraphics[height=3cm,width=3cm]{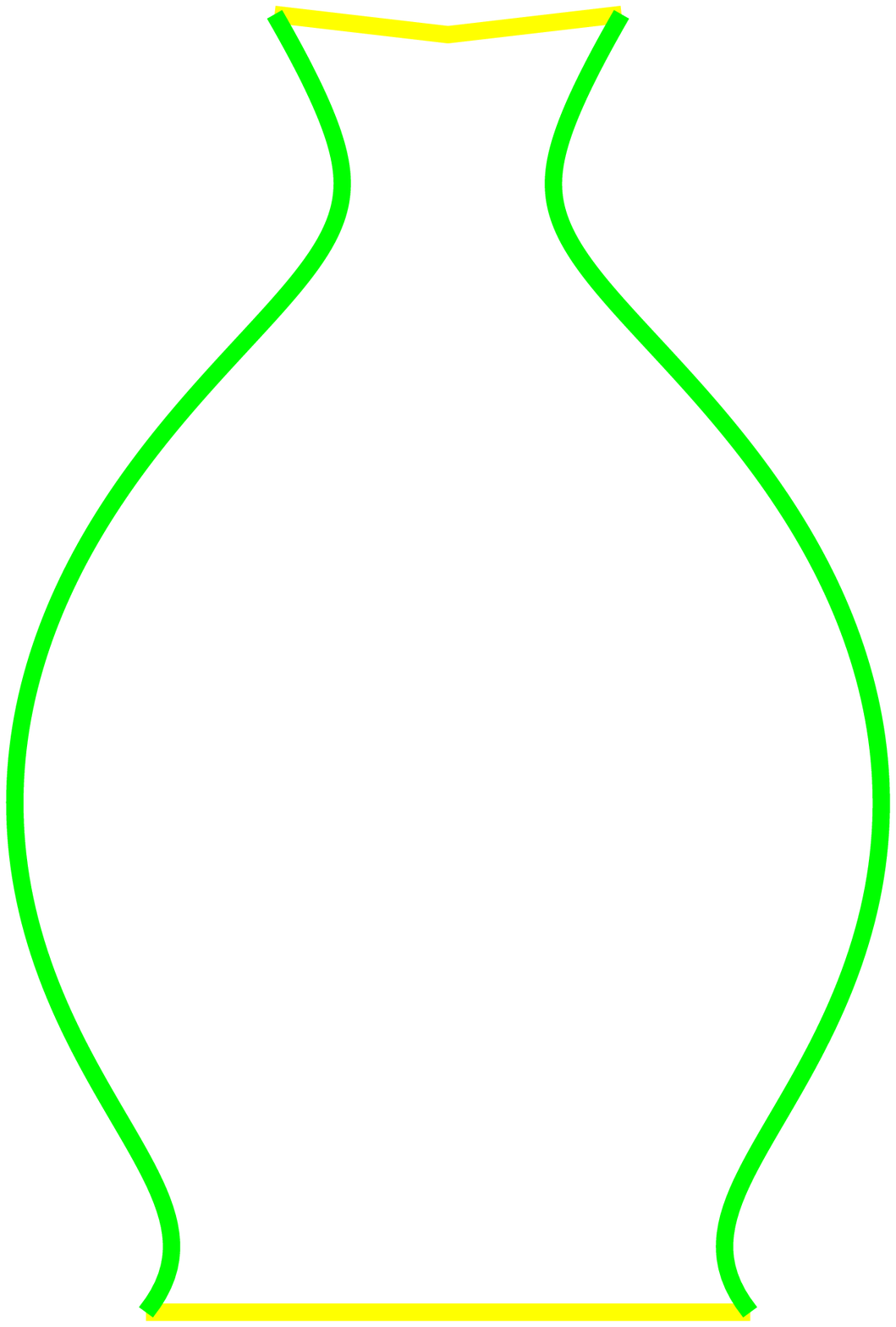}\label{Fig14a}}
\subfigure[$t=0.5$] {\includegraphics[height=3cm,width=3cm]{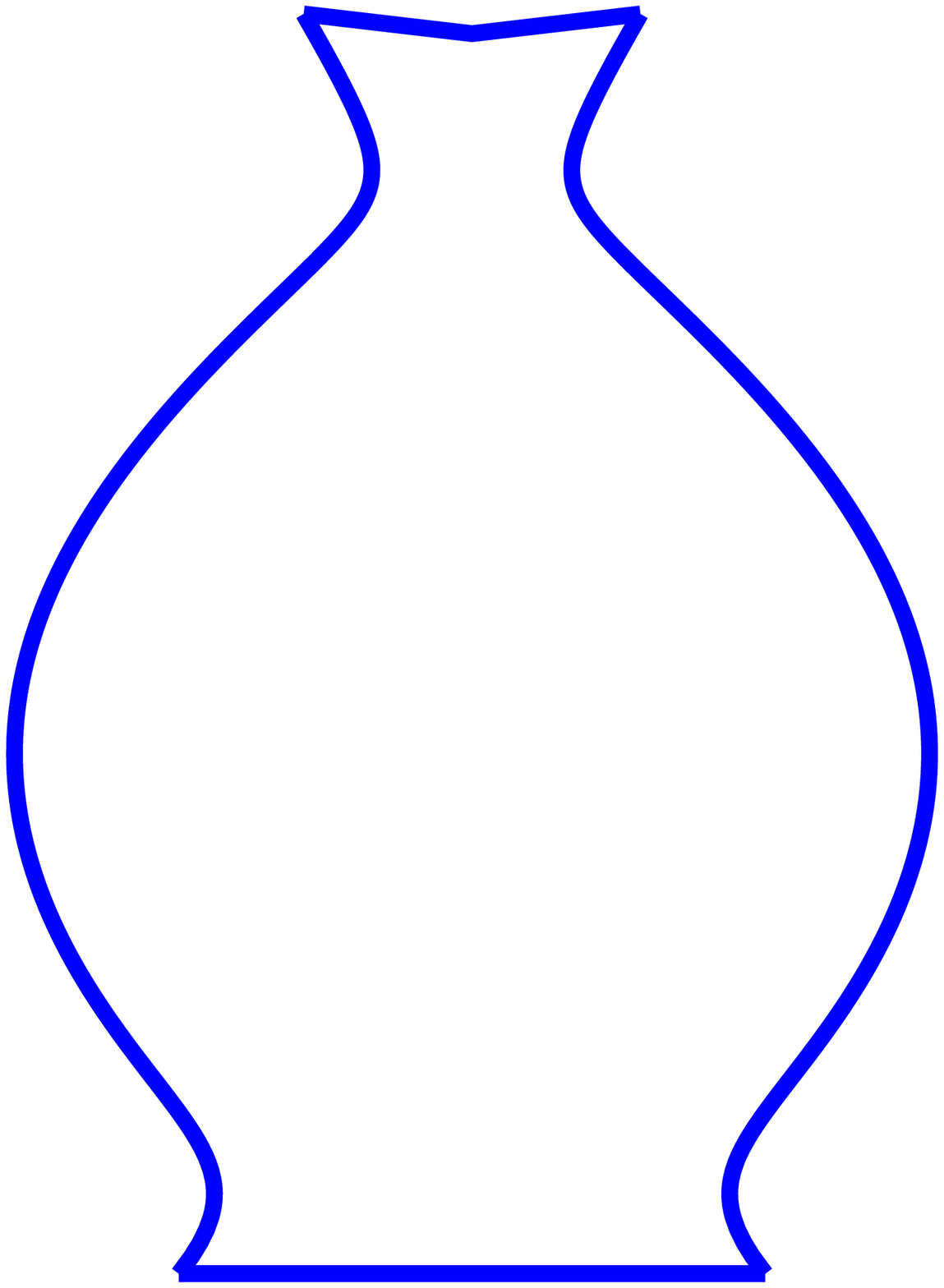}\label{Fig14b}}
\subfigure[$t=2$] {\includegraphics[height=3cm,width=3cm]{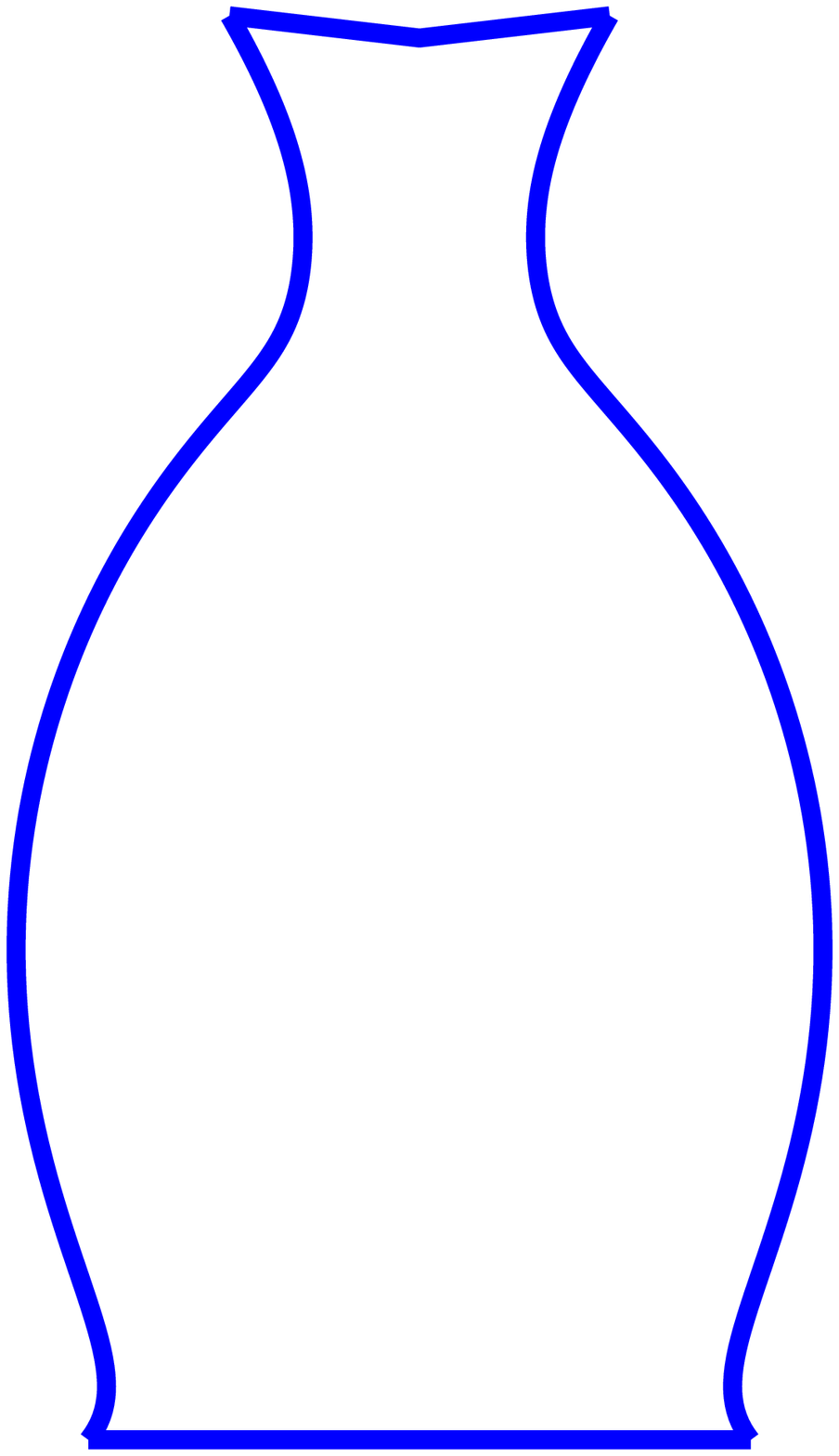}}
\subfigure[$t=4$] {\includegraphics[height=3cm,width=3cm]{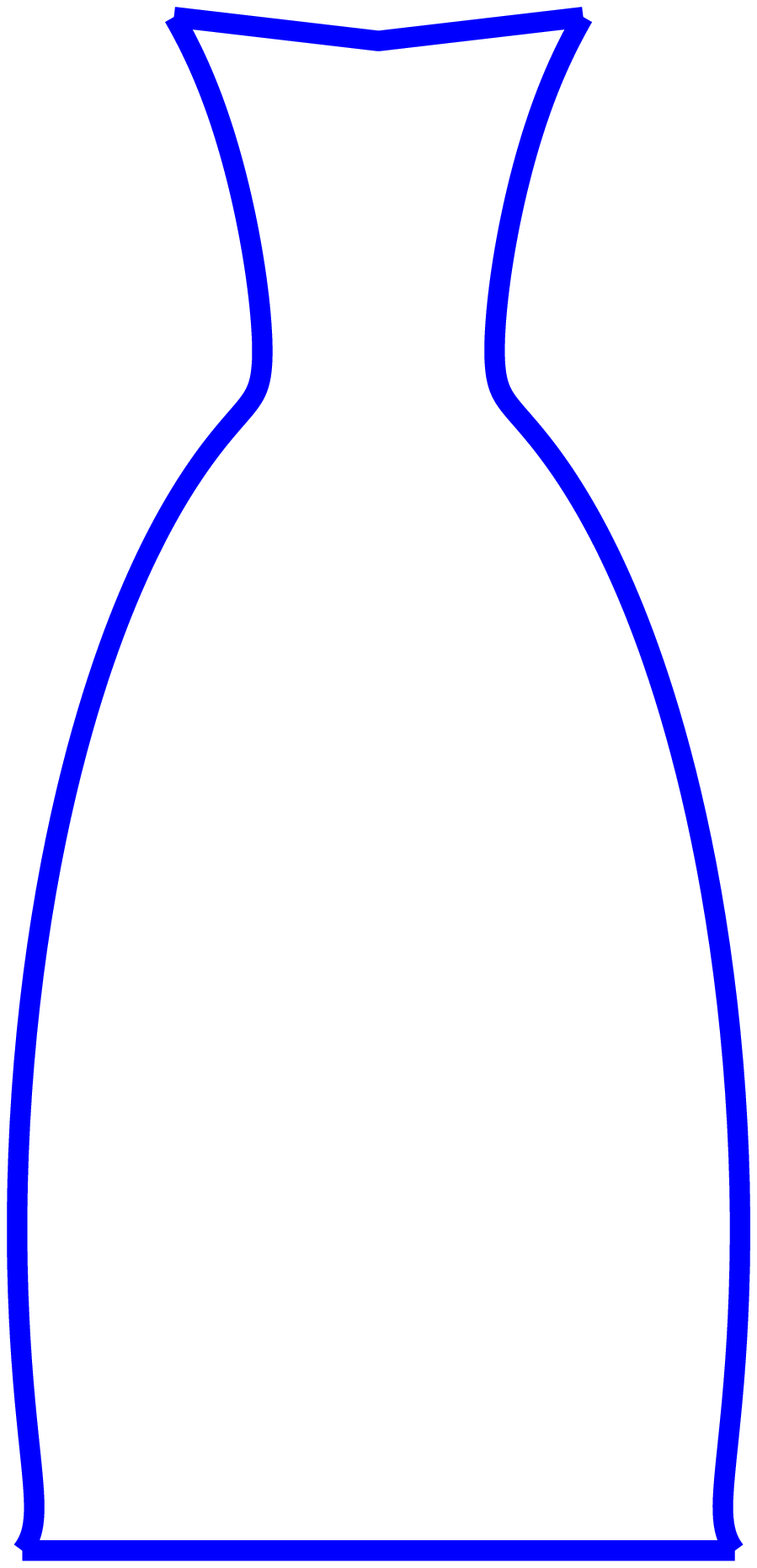}}
\subfigure[$t=8$] {\includegraphics[height=3cm,width=3cm]{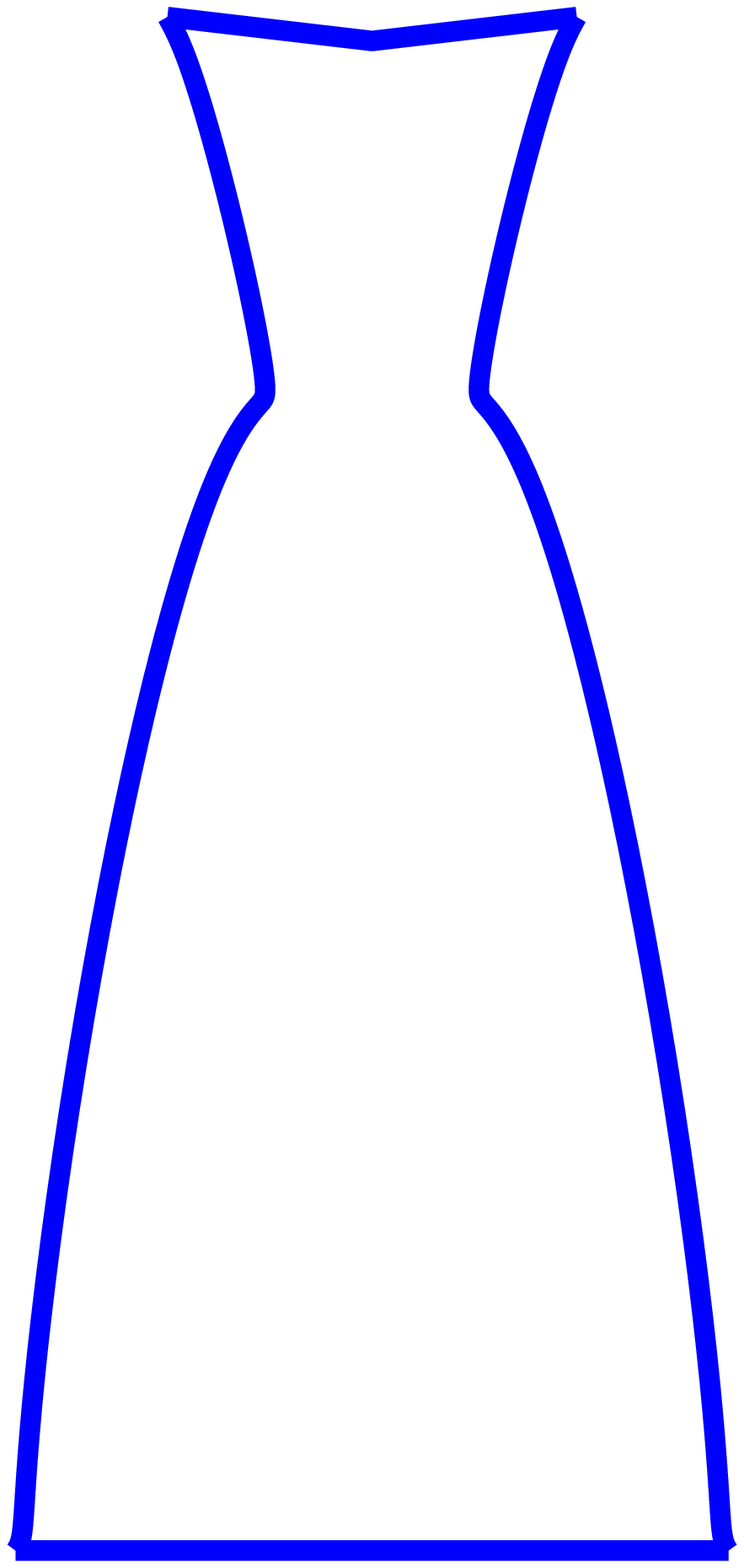}\label{Fig14d}}
\caption{The deformation process of a vase.}
\label{Fig14}
\end{figure*}
\emph{Figure $\ref{Fig14}$ shows the shape deformation processes of a vase. The vase $($see Figure  $\ref{Fig14a})$ is composed of two pieces of NURBS curves on  knot vector  $\{0,0,\frac{1}{2},1,1\}$, and two pieces of NURBS curves on  knot vector $\{0,0,0,0,0,0,0,0,\frac{1}{2},1,1,1,1,1,1,1,1\}$.
The lifting functions correspond to these four pieces of NURBS curves are $\lambda_1=\lambda_2=\{1,1,1\}$ and   $\lambda_3=\lambda_4=\{2,1,1,1,3,1,1,1,2\}$, respectively.
In the degeneration process of NURBS curves,  the fat vase  transforms into a thin vase $($see Figure  $\ref{Fig14d})$.}
\end{example}


\begin{example}\label{example9}
\begin{figure*}
\centering
\subfigure[A  bear's face] {\includegraphics[height=3cm,width=3cm]{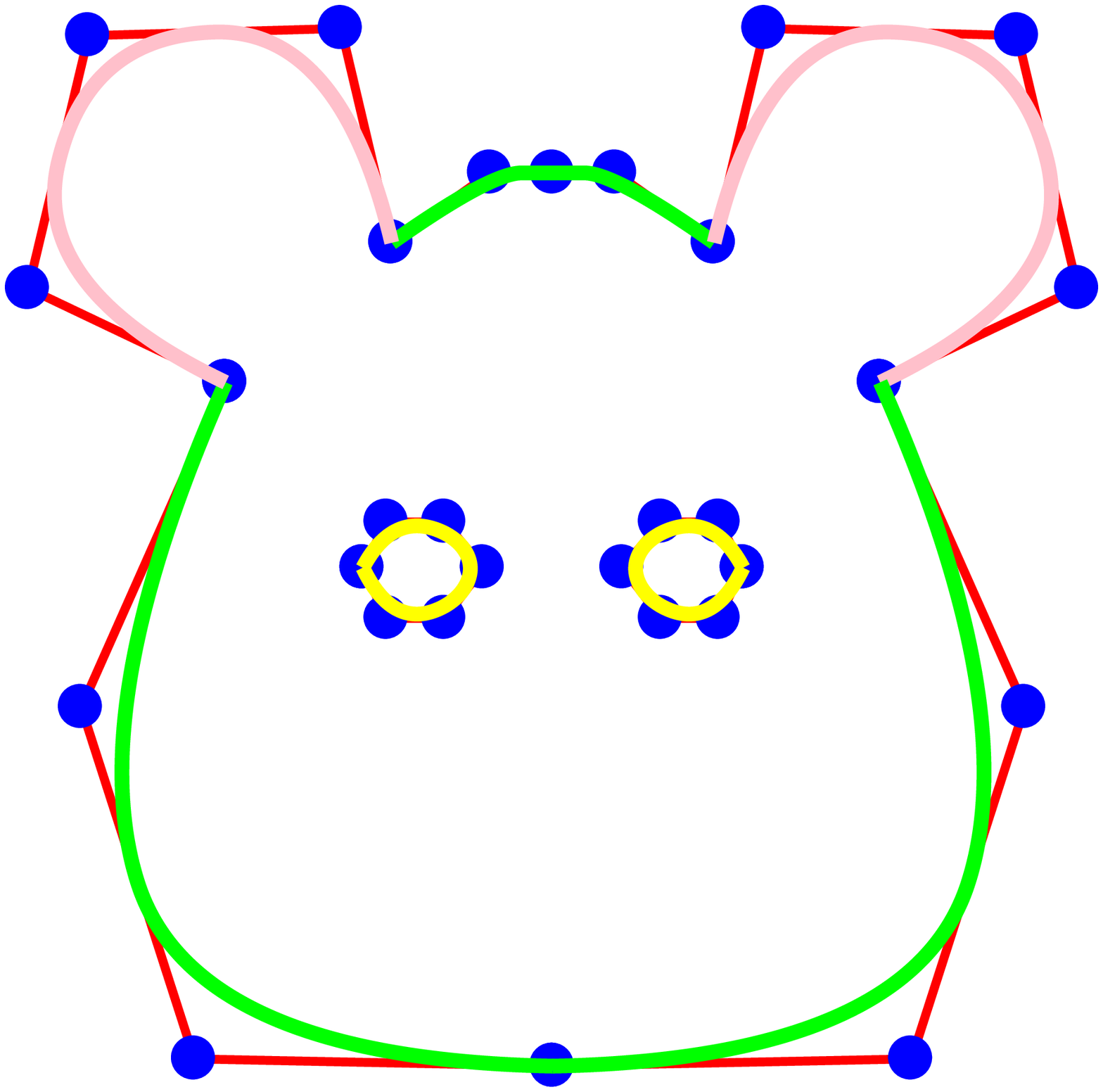}\label{Fig15a}}
\subfigure[$t=1$] {\includegraphics[height=3cm,width=3cm]{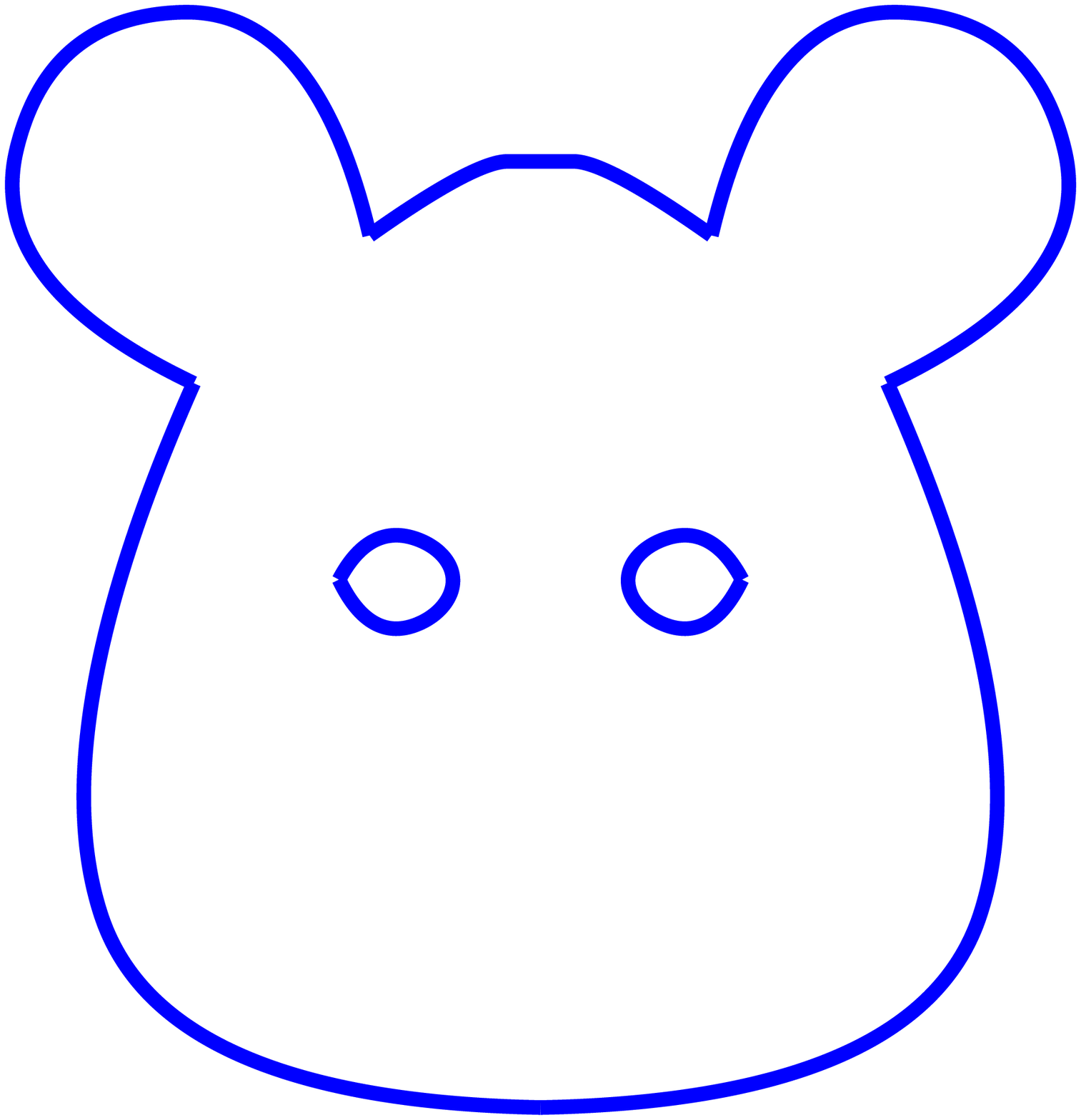}}
\subfigure[$t=4$] {\includegraphics[height=3cm,width=3cm]{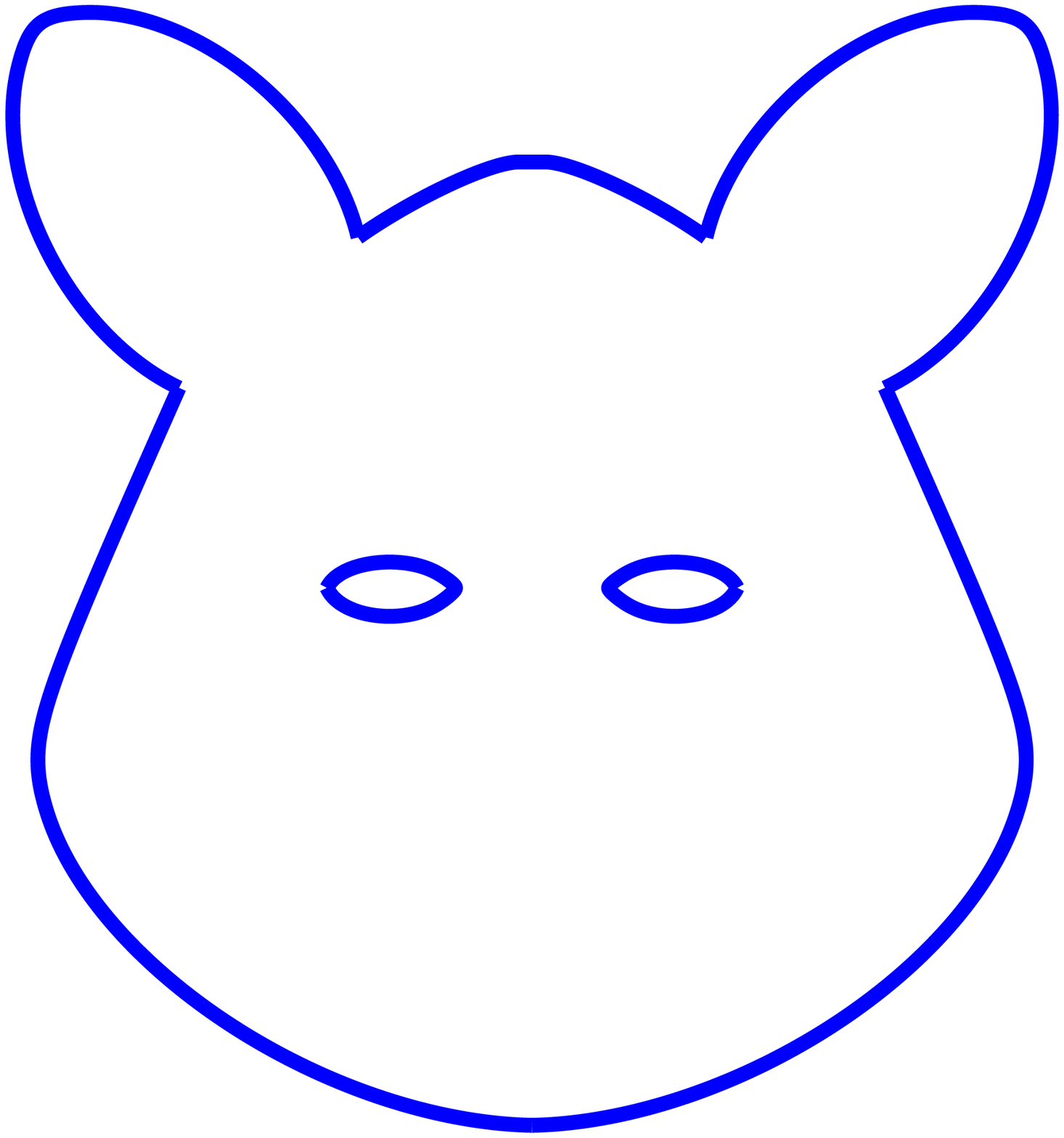}}
\subfigure[$t=12$] {\includegraphics[height=3cm,width=3cm]{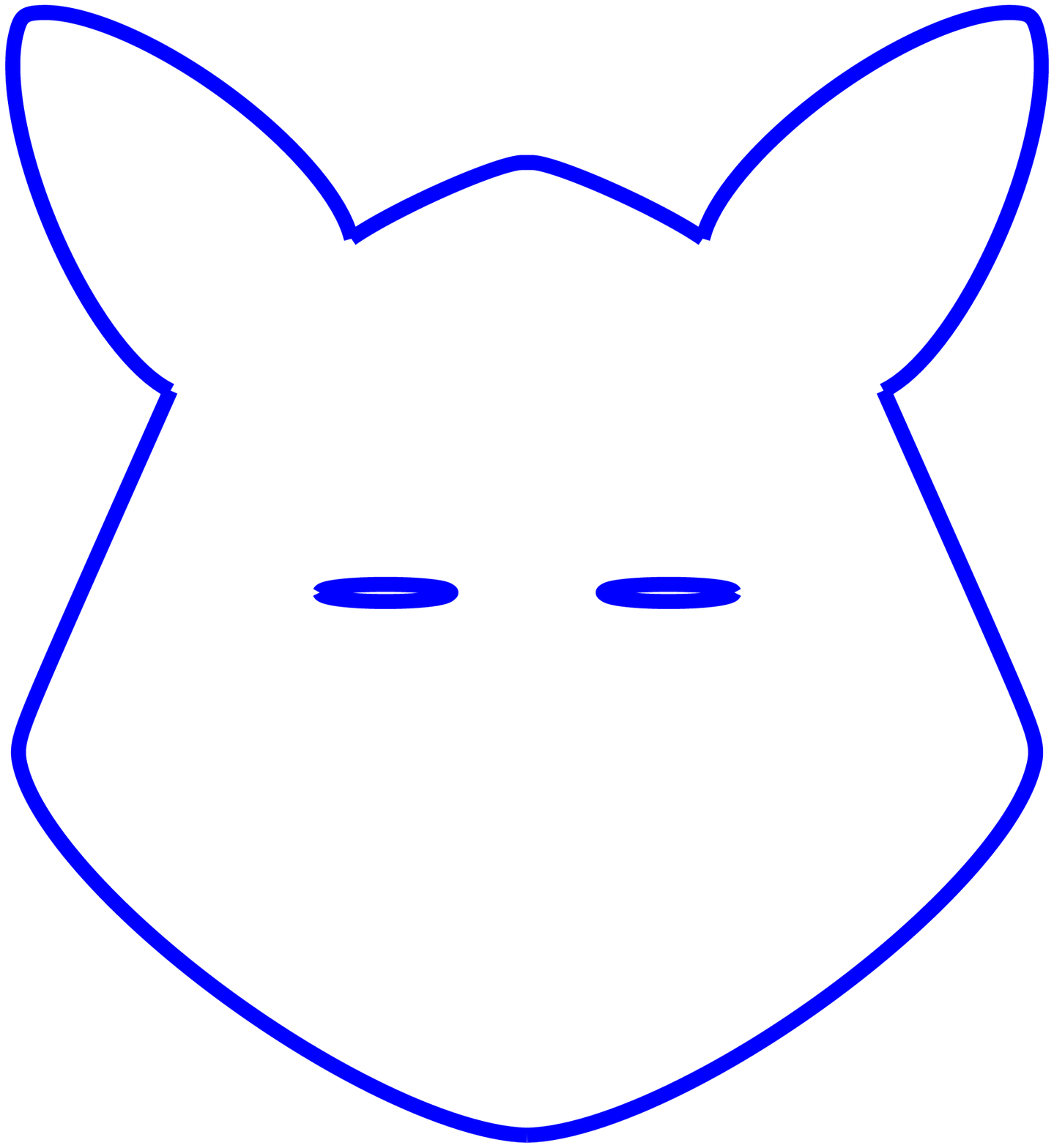}}
\subfigure[$t=20$] {\includegraphics[height=3cm,width=3cm]{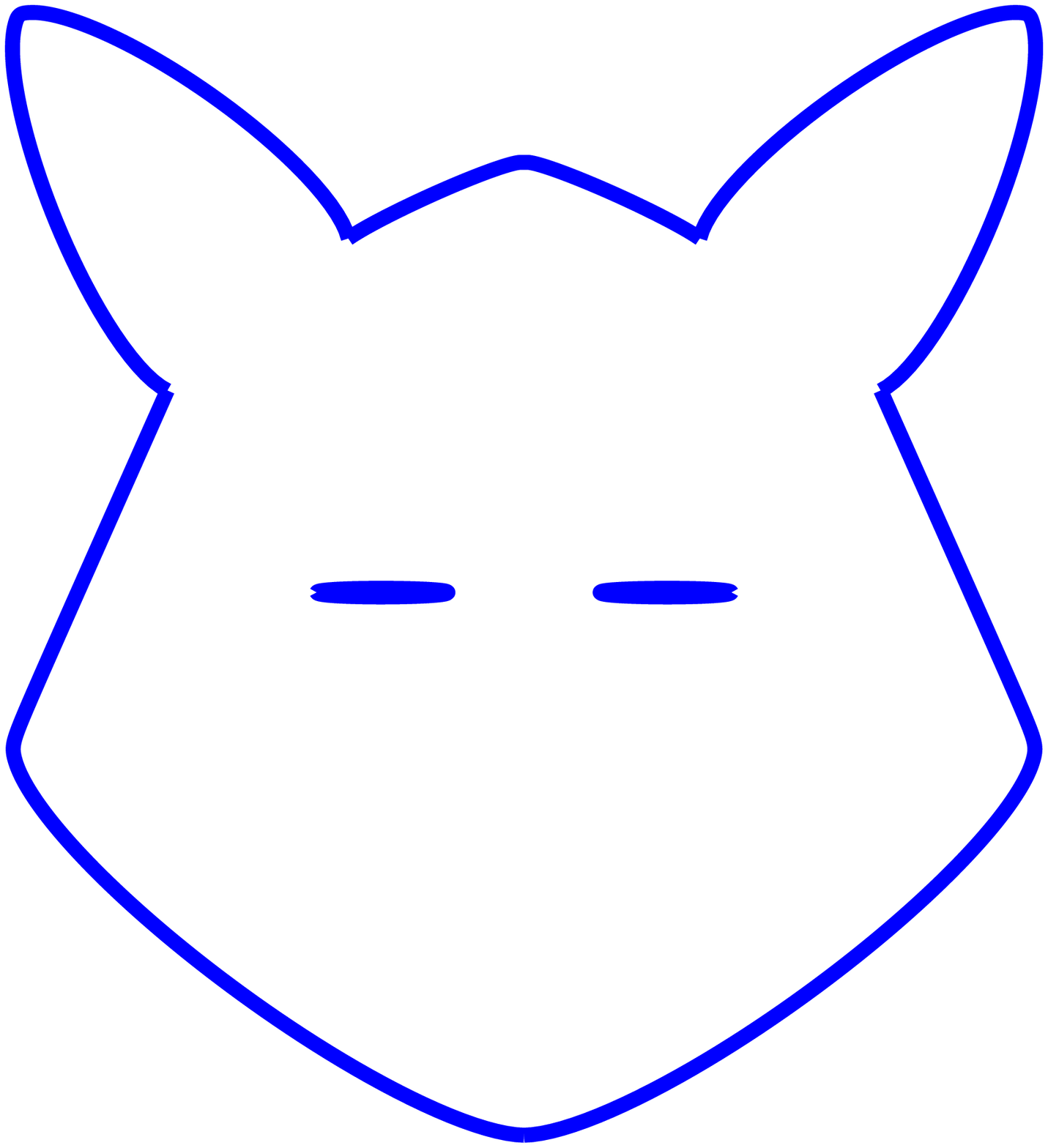}\label{Fig15e}}
\caption{The deformation process of a bear's face.}
\label{Fig15}
\end{figure*}
\emph{Figure $\ref{Fig15}$ shows the shape deformation processes of a bear's face $($see Figure  $\ref{Fig15a})$.
Each ear of the bear is   a quadratic NURBS curve  defined on  knot vector  $\{0,0,0,\frac{1}{2},1,1,1\}$ with the lifting function  $\lambda_1=\{1,2,1,1\}$.
Each eye of the bear is  a cubic NURBS curve defined on  knot vector  $\{0,0,0,0,\frac{1}{4},\frac{1}{2},\frac{3}{4},1,1,1,1\}$ with the lifting function $\lambda_2=\{2,1,1,3,1,1,2\}$.
The rest of the bear's face is composed of  three pieces of NURBS curves  on  knot vector  $\{0,0,0,\frac{1}{3},\frac{2}{3},1,1,1\}$ with the lifting function  $\lambda_3=\{1,1,2,1,1\}$.
After the toric degenerations of NURBS curves, the bear's face degenerates to  a fox's face $($see Figure  $\ref{Fig15e})$.}
\end{example}

\section{Conclusion}\label{conclusion}
In this paper, we define a regular control curve of a NURBS curve by  regular decomposition and propose the geometric meaning of this control curve.  The regular control curve is the limit of a NURBS curve when the control points and weights are fixed but the parameter $t \rightarrow \infty$. Conversely, the regular control curve is also a curve which is the limit of a NURBS curve with control points, but differing weights.
If the regular decompositions induced by different lifting functions, then the limit curves of the NURBS curve (i.e., regular control curves) are different (see Example~\ref{example5}).
The control polygon of a NURBS curve is a regular control curve when the NURBS curve reduced by a certain regular decomposition (see Example~\ref{example5}(2)). This paper also improves the geometric meaning of weights of NURBS curve,  the curve tends to the regular control curve we defined when all of weights approach infinity.
In Example~\ref{example4}, the NURBS curve arises a self-intersection in the toric degeneration process  (see Fig.~$\ref{Fig12}$) and then our work provides possible application for checking the injectivity of NURBS curve.
Moreover, we also provide an idea for shape deformation of NURBS curves by the presented results. We will study the application of the toric degeneration of NURBS curve for animation in future.

\section*{Acknowledgments}
This work is partly supported by the National Natural Science Foundation of China (Nos. 11671068, 11601064, 11271060, 11290143), Fundamental Research of Civil Aircraft (No. MJ-F-2012-04), and the Fundamental Research Funds for the Central Universities (Nos. DUT16LK38, DUT17LK09).

\section*{References}


\end{document}